\documentclass[twocolumn,amsmath,amssymb,showpacs,aps,10pt,prb,superscriptaddress]{revtex4-1}
\pdfoutput=1

\usepackage{dm}

\begin{document}

\title{Magnetic order on a topological insulator surface \texorpdfstring{\\}{} with warping and proximity-induced superconductivity}
\author{Daniel Mendler}
\email{daniel.mendler@kit.edu}
\affiliation{Institut für Theoretische Festkörperphysik, Karlsruhe Institute of Technology, 76131 Karlsruhe, Germany}
\affiliation{Institute of Nanotechnology, Karlsruhe Institute of Technology, 76344 Eggenstein-Leopoldshafen, Germany}
\author{Panagiotis Kotetes}
\affiliation{Institut für Theoretische Festkörperphysik, Karlsruhe Institute of Technology, 76131 Karlsruhe, Germany}
\author{Gerd Schön}
\affiliation{Institut für Theoretische Festkörperphysik, Karlsruhe Institute of Technology, 76131 Karlsruhe, Germany}
\affiliation{Institute of Nanotechnology, Karlsruhe Institute of Technology, 76344 Eggenstein-Leopoldshafen, Germany}

\hypersetup{
  pdfauthor={Daniel Mendler, Panagiotis Kotetes, Gerd Schön},
  pdftitle={Magnetic order on a topological insulator surface with warping and proximity-induced superconductivity}
}

\begin{abstract}
We determine the nature of the magnetic order on the surface of a topological insulator (TI) which develops due to hexagonal warping and the resulting Fermi surface (FS)
nesting in the presence of a repulsive Hubbard interaction. For this purpose we investigate the spin susceptibility and derive a Landau theory to compare the different
accessible phases. For a nearly hexagonal FS and sufficiently strong interaction the magnetic ground state is formed by a skyrmion lattice, i.e., by a superposition of three
helical spin density waves which preserves \group{C_3} symmetry. The magnetic ground state is topologically nontrivial with a nonzero skyrmion charge, which can be
stabilized and controlled by an applied magnetic field. By bringing the TI in proximity to a conventional superconductor one can engineer a \group{C_3}-symmetric topological
superconductor. We explore the modification of the phase diagram as well as the mutual influence between the skyrmion structure and a multipolar distribution of supercurrents,
which can provide information about the underlying skyrmion charge.
\end{abstract}


\maketitle

\section{Introduction}

The recent predictions \cite{BHZ,Fu-Kane,HJZhang} and discovery \cite{QSHE,Hsieh1,*Hsieh2,Chen,Xia,BrueneQHEBulkHgTe,Xia,AndoSpinTexture} of topological insulators (TIs) have
brought about novel concepts and applications relying on the presence of topologically protected surface states
\cite{KaneMele1,*KaneMele2,*Moore,*ZhangTQFT,*Roy,KaneHasan,ZhangQi,Ando}. For three-dimensional TIs the electronic surface states are cha\-racte\-rized by spin-momentum
locking, which in its simplest form yields a helical Dirac cone energy dispersion. As long as time-reversal (\OpT) symmetry is preserved, the surface states are massless and
protected against elastic backscattering. Their robustness is particu\-larly useful when engineering topological superconductors (TSCs) by bringing the topo\-lo\-gi\-cal
surface in proximity to a conventional SC \cite{AkhmerovMFDetection,Linder,FuKane2008,AliceaReview,BeenakkerReview,KotetesClassi}. TSCs have the property that Majorana bound
states may occur at domain walls, vortices or other topological defects. A number of promising experi\-men\-tal steps towards the implementation of TSCs have already been
reported \cite{Morpurgo,*Analytis,*Brinkman,*Lu,*Du,*Molenkamp,*Yacoby}.

The topologically nontrivial properties rely on the presence of a single Dirac point in the surface state energy dispersion \cite{KaneHasan,Hsieh2}. Away from the Dirac
point the dispersion can become distorted due to material-specific effects. For example, for Bi-based topological insulators the Fermi surface (FS) exhibits \group{C_{3v}}-symmetric
warping, which modifies the FS such that it approaches the form of a hexagon \cite{FuHex,ModelHamTI,STMofHex,GedikHex,Chulkov,KurodaHexWarp1sArpes}. The consequences of
warping on magnetic \cite{FuHex,JiangStripe,BaumStern1,*BaumStern2} and transport pro\-per\-ties \cite{CarbotteCond1,*CarbotteCond2,*OptCond} have been discussed extensively
in the literature. The most important ones arise from the nesting of the hexagonal FS, characterized by three nesting vectors $\vec{Q}_{1,2,3}$. It renders the system
susceptible to the spontaneous deve\-lop\-ment of helical magnetic order. The possible ground states involve single- or triple-\vec{Q} helical magnetism, with the latter also
inclu\-ding the possibility of skyrmion lattice phases \cite{BaumStern2}. The question of which one of these phases constitutes the magnetic ground state and how it depends on
various control fields, has not yet been fully resolved.

In the present work we elucidate this question and determine the properties of the magnetic ground state of the TI surface under the influence of warping and a repulsive
Hubbard interaction. For this purpose we first analyze the spin susceptibility, from which we extract the magnetic instabilities as a function of the chemical potential. It
can be varied by chemical doping, which allows tuning the shape of the FS from convex via nearly hexagonal to snowflake-like. To identify the magnetic ground state, we derive
a Landau theory for the magnetic order parameter. At fourth order it shows that for a he\-xa\-go\-nal FS the single-\vec{Q} (stripe) magnetic order \cite{JiangStripe} is less
favorable than two possible triple-\vec{Q} phases which transform according to the $A_1$ and $A_2$ irreducible representations (IRs) of the relevant \group{C_{3v}} point group.
These two phases differ in the value of the topological skyrmion charge $\OpC=0$ and $\OpC=\pm1$. The sixth-order Landau expansion shows that the $A_2$ phase corresponding to a
topo\-lo\-gi\-cal\-ly nontrivial skyrmion lattice is favored.

The development of helical magnetism on the TI surface opens perspectives for engineering TSCs, namely by bringing the TI in proximity to a conventional SC, without the need for
vortices \cite{FuKane2008,AliceaReview,BeenakkerReview,KotetesClassi,Choy,*Nakosai2013,*Ojanen,*Bena}. In such systems Majorana fermions can be trapped at defects where the
energy difference between the magnetic order parameter and the proximity-induced gap changes sign \cite{FuKane2008,AliceaReview,KSG,*Pekker}. For this reason, we determine the
magnetic phases in the presence of a spin singlet, proximity-induced superconducting gap. Our analysis shows that a \group{C_3}-symmetric TSC can be rea\-lized in these hybrid
systems.

In the final part of the present work we propose means to control the magnetic order on the TI surface and investigate signatures of it. Specifically, we show that an
external perpendicular magnetic field can stabilize the $A_2$ phase and select the sign of the skyrmion charge $\OpC=\pm1$. An in-plane magnetic field or imposed supercurrent
breaks the \group{C_{3v}} symmetry and modifies the phase dia\-gram. In addition, we note that the magnetic skyrmion order may induce a multipolar supercurrent distribution in
the SC in pro\-xi\-mi\-ty to the warped TI, reflecting the nonzero skyrmion charge on the surface.

This paper is organized as follows: In \secref{warping} we investigate the effects of warping on the TI surface states and their energy dispersion. Next we exa\-mi\-ne in
\secref{magnetic} the instability of the TI surface states towards the development of a magnetic phase. To distinguish different com\-pe\-ting phases we need to derive a
sixth-order Landau theory. In \secref{proximity} we examine the consequences of a proximity-induced superconducting gap. In \secref{external} we suggest me\-thods on how to
manipulate the magnetic phases, and point out that an induced supercurrent could provide information about the skyrmion charge. We conclude in \secref{conclusion}
with a summary. Se\-ve\-ral extensions are discussed in the appendices. In order to estimate the consequences of the imperfect nesting we analyze in
\appendixref{susc-high-symmetry} the susceptibility in more detail. In \appendixref{landau} we present details on the
derivation of the Landau theory. Finally, in \appendixref{classification} we perform a symmetry-based classification of the magnetic order parameters.

\section{Topological insulator with warping}\label{sec:warping}

In this section we introduce the Hamiltonian describing the surface states of a TI with hexagonal war\-ping. We determine the eigenvalue and eigenstates, and show that warping
leads to nesting of the resulting Fermi surface.

\begin{figure}[t]
\includegraphics{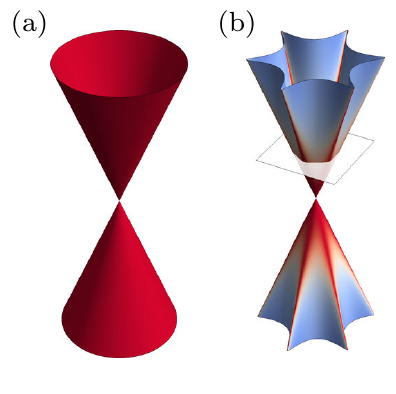}
\hskip0.1cm
\includegraphics{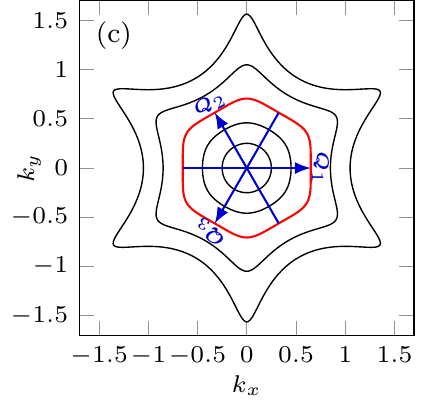}
\caption{(a) Ordinary  Dirac cone ($\gamma = 0$) and (b) warped ($\gamma \ne 0$) energy dispersions for the surface states of a 3D topological insulator. The color coding
denotes the direction-resolved density of states. (c) Warped Fermi surfaces for different values of the chemical potential. For $\mu=\mu_{\text{hex}}$ the FS is closest to
hexagonal, leading to nearly perfect nesting with three nesting vectors $\vec{Q}_{1,2,3}$. The plane in (b) corresponds to this optimal value $E=\mu_{\text{hex}}$.}
\label{fig:cone}
\end{figure}

\subsection{The Hamiltonian of a warped TI}\label{sec:warping:symmetries}

The surface states (in the $xy$ plane) of a warped TI follow from the modified Dirac Hamiltonian \cite{FuHex},
\begin{eqnarray}
\HH_0(\vec{k})=v\left(k_x\sigma_y-k_y\sigma_x\right)+\frac{\gamma}{2}(k_+^3+k_-^3)\sigma_z+\epsilon_0(\vec{k})\, .\quad\label{eq:Hamiltonian0}
\end{eqnarray}

\noindent
The \vec{\sigma} Pauli matrices act on the $\ket{\up},\ket{\down}$ eigenstates of the $z$ component of the total angular momentum ope\-ra\-tor, which for the materials under
investigation almost coincides with the spin operator \vec{S} \cite{He,*VallaFullSpinPol}. Here and throughout we set $\hbar=1$. The first term of the Hamiltonian is the
Rashba spin-orbit coupling  with Dirac ``velocity'' $v$. It is invariant under arbitrary rotations about the $z$ axis and mirror operations. With only this term present, we
obtain a Dirac cone spectrum typical for relativistic mas\-sless particles with a constant density of states (DOS) as shown in \figref{cone,a}.

The second term in the Hamiltonian
\begin{eqnarray}
  \frac{\gamma}{2}(k_+^3+k_-^3)\sigma_z = \gamma k_x(k_x^2-3 k_y^2)\sigma_z
\end{eqnarray}

\noindent with $k_\pm = k_x\pm \imath k_y = k e^{\pm \imath \theta_{\vec{k}}}$ and strength $\gamma$ introduces warping effects. It reduces the continuous rotational symmetry
to a discrete \group{C_3} subgroup, generated by a $2\pi/3$ counterclockwise rotation ($C_3$) of the system about the $z$ axis. The \group{C_3} symmetry is supplemented by
the invariance of the Hamiltonian under the mirror operation at the $yz$ plane $\sigma_v\colon x\mapsto -x$, leading to a \group{C_{3v}} point group symmetry for the TI
surface. The character table of the point group is presented in \tableref{c3v_characters}.

The third term in the Hamiltonian is assumed to be inva\-riant under all symmetry operations. In general it includes particle-hole asymmetric contributions, such as a chemical
potential $\mu$ or a quadratic kinetic term $\propto\vec{k}^2$. For the rest of our discussion we will only consider the effect of the chemical potential, which can be
experi\-mentally controlled by doping or gating the topological insulator \cite{HsiehTunableTI}; i.e., we set $\epsilon_0(\vec{k})=-\mu$. We note that the Hamiltonian
respects \OpT symmetry.

\begin{table}[b]
  \caption{Character table of \group{C_{3v}} and examples of quantities transforming according to its irreducible representations (IRs).}
  \label{table:c3v_characters}
  \begin{ruledtabular}
    \begin{tabular}{>{$}l<{$}|>{$}r<{$}>{$}r<{$}>{$}r<{$}|>{$}l<{$}|>{$}l<{$}}
    \text{IR} & \II & 2C_3 & 3\sigma_v & \text{Linear} & \text{Higher order}\\
    \hline
    A_1&       1 &    \phantom{-}1 &    \phantom{-}1 & z & x^2+y^2,\, z^2,\, z^3,\, y(y^2-3x^2)\\
    A_2&       1 &    \phantom{-}1 &              -1 & S_z & x(x^2 - 3y^2)\\
    E &        2 &              -1 &               0 & (x,y),\, (S_x,S_y) & (2xy,x^2-y^2),\, (xz,yz)\\
    \end{tabular}
  \end{ruledtabular}
\end{table}

For convenience, we introduce dimensionless forms for the wave vector \vec{k} and Hamiltonian by scaling them with $k_s=\sqrt{v/\gamma}$ and $E_s = v k_s$, respectively.
Representative va\-lues for the material parameters $v$ and $\gamma$ can be inferred from angular resolved photoemission spectroscopy (see for instance \citeref{JiangStripe}
and re\-fe\-rences therein). As an example we consider \chem{Bi_2Te_3} throughout the text with $k_s= \SI{0.1}{\angstrom^{-1}}$ and $E_s=\SI{0.26}{eV}$.

\subsection{Energy dispersion and eigenstates}\label{sec:warping:eigenstates}

For our further analysis we rewrite the Hamiltonian in the
form $\HH_0(\vec{k})=\vec{g}(\vec{k})\cdot\vec{\sigma} - \mu$, introducing the vector
\begin{eqnarray}
  \vec{g}(\vec{k}) =
  \begin{pmatrix}
    -k_y\\
    k_x\\
    k_x(k_x^2-3 k_y^2)
  \end{pmatrix}
  =
  |\vec{g}(\vec{k})|
  \begin{pmatrix}
    \sin \vartheta_{\vec{k}} \cos \varphi_{\vec{k}}\\
    \sin \vartheta_{\vec{k}} \sin \varphi_{\vec{k}}\\
    \cos \vartheta_{\vec{k}}
  \end{pmatrix}\quad
\end{eqnarray}
with  polar angles given by $\cot \vartheta_{\vec{k}}= \sin\varphi_{\vec{k}}(k_x^2-3k_y^2)$, $\tan\varphi_{\vec{k}}=-k_x/k_y$ and $\varphi_{\vec{k}}=\theta_{\vec{k}}+\pi/2$.
The resulting spectrum of the Hamiltonian under the influence of warping is $\epsilon_{\vec{k},\pm} = \pm |\vec{g}(k,\theta_{\vec{k}})| - \mu = \pm k\sqrt{1+ k^4
\cos^2(3\theta_{\vec{k}})}-\mu$. It is depicted in \figref{cone,b} for different values of the chemical potential $\mu$. The corresponding eigenstates, commonly called
helicity eigenstates, are
\begin{eqnarray}
  \!\ket{\vec{k},\!+}
  \!=\!
  \begin{pmatrix}
    e^{-\frac{\imath\varphi_{\vec{k}}}{2}} \cos \frac{\vartheta_{\vec k}}{2}\\
    e^{+\frac{\imath\varphi_{\vec{k}}}{2}} \sin \frac{\vartheta_{\vec k}}{2}
  \end{pmatrix}\!,\;
  \ket{\vec{k},\!-}
  \!=\!
  \begin{pmatrix}
    -e^{-\frac{\imath\varphi_{\vec{k}}}{2}} \sin \frac{\vartheta_{\vec k}}{2}\\
    \phantom{-}e^{+\frac{\imath\varphi_{\vec{k}}}{2}} \cos \frac{\vartheta_{\vec k}}{2}
  \end{pmatrix}.\quad
\end{eqnarray}

For the following discussions it is useful to list the transformation properties of \vec{k}, the Pauli matrices \vec{\sigma}, $\vec{g}(\vec{k})$, and the helicity eigenstates
under the action of the \group{C_{3v}} generators $C_3$ and $\sigma_v$. We use the convention that ($\HD^-_\OpG$) $\HD^+_\OpG$ defines the representation of the group element
\OpG acting on (axial) vectors. The representation $\HD^+_{C_3}=\HD^-_{C_3}$ of the rotation $C_3$ is given by the rotation matrix about the $z$ axis by $2\pi/3$,
ac\-ting like $C_3\vec{k} = \HD_{C_3}^+\vec{k}$ and $C_3\vec{\sigma} = \HD_{C_3}^-\vec{\sigma}$. The mirror ope\-ra\-tion yields $\sigma_v\vec{k} = \HD_{\sigma_v}^+\vec{k} =
\diag(-1,1,1)\vec{k}$ and $\sigma_v\vec{\sigma} = \diag(1,-1,-1)\vec{\sigma} = \HD^-_{\sigma_v}\vec{\sigma}=-\HD^+_{\sigma_v}\vec{\sigma}$.

Since the Hamiltonian is invariant under the operation of any element $\OpG\in\group{C_{3v}}$, we have
\begin{eqnarray}
 \vec{g}(\OpG\vec{k})\cdot\OpG\vec{\sigma}=\vec{g}(\vec{k})\cdot\vec{\sigma} \implies \vec{g}(\OpG\vec{k})=\HD^{-}_\OpG\vec{g}(\vec{k})\,.
\end{eqnarray}

\noindent As a result, we also obtain the transformation properties of the helicity eigenstates,
\begin{eqnarray}
\hspace*{-1.6mm}\ket{C_3\vec{k},\!\pm}=e^{-\imath\pi\sigma_z/3}\ket{\vec{k},\pm}
\text{ and } \ket{\sigma_v\vec{k},\pm}=\imath \sigma_x\ket{\vec{k},\pm}.\quad
\end{eqnarray}

\subsection{Warping-induced Fermi surface nesting}\label{sec:warping:nesting}

As illustrated in \figref{cone,c} the shape of the FS can be modified by tuning the chemical
potential $\mu$ \cite{FuHex}. It evolves from roughly circular for small values of
$\mu \lesssim 0.5$ to more hexagonal for larger values and finally to snowflake-like for $\mu\gtrsim 1$. The shape of the FS is defined by the cubic
equation $|\vec{g}(k,\theta_{\vec{k}})|=\mu$. For $\cos(3\theta) = 0$ it is determined by $k(\theta,\mu) = \mu$. In general it is given by
\begin{eqnarray}
  \hspace*{-6mm}k(\theta,\mu) &=& \sqrt{t(\theta,\mu) - \dfrac{1}{3\cos^2(3\theta)t(\theta,\mu)}}\,,\label{eq:fermi}\\
  \hspace*{-6mm}t(\theta,\mu) &=& \sqrt[3]{\frac{\mu^2}{2\cos^2(3\theta)}\!+\!\sqrt{\frac{\mu^4}{4\cos^4(3\theta)}\!+\!\frac{1}{27\cos^6(3\theta)}}}\,. \nonumber
\end{eqnarray}
In the following we focus on the situation where the FS becomes nearly perfectly hexagonal with strong (but not perfect) FS nesting. The latter occurs for
$\mu=\mu_{\text{hex}}=0.725$. As shown in \figref{cone,c} the nesting wave vectors $\pm \vec{Q}_{1,2,3}$ are
\begin{eqnarray}
\vec{Q}_1&\equiv&2k_0(1,\, 0)\,, \nonumber\\
\vec{Q}_2&\equiv&C_3\vec{Q}_1 = 2k_0(-1/2,\, +\sqrt{3}/2)\,, \nonumber\\
\vec{Q}_3&\equiv&C_3\vec{Q}_2 = 2k_0(-1/2,\, -\sqrt{3}/2)
\end{eqnarray}

\noindent with Fermi wave vector $k_0$. The relation $\epsilon_{k_0,+} = 0$ implies $\mu = |\vec{g}(k_0,0)| = k_0\sqrt{1+k_0^4}$.

\section{Magnetic instability}\label{sec:magnetic}

In this section we investigate the tendency of the surface states to spontaneously form a magnetic ground state. In \secref{magnetic:interaction} we discuss the magnetic order
parameter developing for a repulsive Hubbard interaction at the mean-field level. In \secref{magnetic:landau} we derive a Landau theory for the surface magnetization, which we
analyze up to second order of the Landau expansion in \secref{magnetic:landau2}, up to fourth order in \secref{magnetic:landau4}, and up to sixth order in
\secref{magnetic:landau6}. At second order we can only determine the tendency of the system to develop a magnetic order parameter; at fourth order we can infer the single- or
triple-\vec{Q} character of the magnetic phase. We need the sixth order to completely determine the structure of the order parameter in the triple-\vec{Q} phase. Two phase
degrees of freedom, which correspond to Goldstone modes, remain open at any order of the Landau theory.

\subsection{Magnetic interaction and order parameter}\label{sec:magnetic:interaction}

We assume a repulsive Hubbard interaction ($U>0$) of the form
\begin{eqnarray}
\OpH_U\nohc&=& U\int \difv{r} n_\up(\vec{r})\,n_\down(\vec{r})\nonumber\\
&=&U\int\difv{r}\left[\frac{\rho^2(\vec{r})}{4}-\frac{\vec{S}^2(\vec{r})}{3}\right]\,,
\end{eqnarray}

\noindent with particle and spin density operators $\rho(\vec{r})=\hpsi\hc(\vec{r})\II\hpsi(\vec{r})$ and $\vec{S}(\vec{r})=\hpsi\hc(\vec{r})(\vec{\sigma}/2)\hpsi(\vec{r})$.
Since we do not expect charge density-wave instabilities we will neglect the nonmagnetic contribution to the interaction. We proceed
with a mean-field decoupling of the spin interaction by introducing the related order parameter $\vec{M}(\vec{r})=-U\left<\vec{S}(\vec{r})\right>$. Thus the spin-dependent
part of the interaction becomes
\begin{eqnarray}
\OpH_{\text{mag}}\nohc=\int \difv{r}\frac{\vec{M}^2(\vec{r})}{U}+\int \difv{r}\hpsi\hc(\vec{r})\vec{M}(\vec{r})\cdot\vec{\sigma}\hpsi(\vec{r})\,.\qquad
\end{eqnarray}

\noindent After introducing the Fourier transforms, e.g.,
\begin{eqnarray}
  \!\vec{M}(\vec{r})=\int\difk{q} e^{\imath\vec{q}\cdot\vec{r}}\vec{M}_{\vec{q}}\nohc
\end{eqnarray}

\noindent we have
\begin{eqnarray}
  \OpH_{\text{mag}}\nohc &=& \int\difk{q}\Bigg[\frac{|\vec{M}_{\vec{q}}|^2}{U} \nonumber\\
  &&+\int \difk{k}\hpsi_{\vec{k}+\vec{q}/2}\hc\vec{M}_{\vec{q}}\nohc\cdot\vec{\sigma} \hpsi_{\vec{k}-\vec{q}/2}\nohc\Bigg]\,,
\end{eqnarray}

\noindent and the self-consistency relation reduces to
\begin{eqnarray}
  \vec{M}_{\vec{q}}\nohc
  &=&-U\int\difv{r}e^{-\imath\vec{q}\cdot\vec{r}}\braket{\vec{S}(\vec{r})}\nonumber\\
  &=&-\frac{U}{2}\int\difk{k}\braket{\hpsi\hc_{\vec{k}-\vec{q}/2}\vec{\sigma}\hpsi_{\vec{k}+\vec{q}/2}\nohc}\,.
\end{eqnarray}

The order parameter has the symmetry property $\vec{M}\nohc_{-\vec{q}}=\vec{M}^*_{\vec{q}}$ and transforms under a group operation \OpG of the point group \group{C_{3v}}
according to $\OpG \vec{M}_{\vec{q}} \equiv \HD_{\OpG}^-\vec{M}_{\OpG^{-1}\vec{q}}$. A complete classification of the possible magnetic order parameters with wave vectors
\vec{q}, $C_3\vec{q}$ and $C_3^2\vec{q}$ under the \group{C_{3v}} point group is presented in \appendixref{classification}.

\subsection{Free energy functional of the topological insulator}
\label{sec:magnetic:landau}

To obtain information about the dominant magnetic instability and preferred wave vectors we derive a Landau free energy functional for magnetism. For this purpose we integrate
out the electronic degrees of freedom in the frame of the path-integral formalism in terms of Grassman fields. The free energy $\OpF =-\ln \OpZ/\beta$ depends on the partition
function
\begin{eqnarray}
  \OpZ = \int \mathrm{D}[\bar{\psi},\psi]\, e^{-\OpS[\bar{\psi},\psi]}\,,
\end{eqnarray}

\noindent with the action
\begin{eqnarray}
\OpS[\bar{\psi},\psi]&=&\sum_{k,q}\bar{\psi}_{k+q}\left\{-\left[\imath k_n - \HH_0(k)\right]\delta_{q,0} + \vec{M}_q\cdot\vec{\sigma}\right\}\psi_k\nohc\nonumber\\
&\equiv& \sum_{k,q} \bar{\psi}_{k+q}\left[-\HG_0^{-1}(k)\delta_{q,0} + \HatOpV(k+q,k)\right] \psi_k\nohc\nonumber\\
&\equiv&\bar{\psi}\left(-\HG_0^{-1}+\HatOpV\right)\psi\,.\label{eq:action}
\end{eqnarray}

\noindent Here we introduced the fermionic and bosonic (2+1)-vectors $k=(\vec{k},k_n)$ and $q=(\vec{q}, \omega_n)$, with the fermionic and bosonic Matsubara
frequencies $k_n=(2n+1)\pi/\beta$ and $\omega_n=2n\pi/\beta$. Furthermore, we introduced the shorthand notation $\sum_{k} \equiv \sum_{k_n}\int\difv{k}/(2\pi)^2$ for the
integration over the (2+1)-vectors and the Kronecker delta $\delta_{q,q'}=(2\pi)^2\delta(\vec{q}-\vec{q}')\delta_{\omega_n\nohc,\omega'_n}$.
The noninteracting Green's function is $\HG_0(k)=[\imath k_n-\HH_0(k)]^{-1}$. In the last line of \eqref{action}, $\OpS$ is written in a basis-independent form, with
operators $\HG_0$ and $\HatOpV$ defined by the matrix e\-le\-ments $\HG_0(k)\equiv\braket{k|\HG_0|k}$ and $\HatOpV(k+q,k) \equiv \braket{k+q|\HatOpV|k}$.
Here we are interested in the static magnetization and concentrate on the zero-frequency component of $\vec{M}_{\vec{q},0}$ and the static magnetic potential $\HatOpV(k+q,k)
\equiv \HatOpV(\vec{q})\delta_{\omega_n,0}$.

Integration over the Grassmann variables yields the fermionic determinant
\begin{eqnarray}
  \OpZ=\det \beta \left(-\HG^{-1}_0 + \HatOpV \right)
  = e^{\trln\left[\beta \left(-\HG^{-1}_0 + \HatOpV \right)\right]}
\end{eqnarray}

\noindent and the related free energy
\begin{eqnarray}
  \OpF=-\frac{1}{\beta} \ln \OpZ
  =\OpF_0-\frac{1}{\beta} \trln\left(\II - \HG_0\HatOpV \right)\,.
\end{eqnarray}

\noindent A series expansion of the loga\-rithm in powers of the magnetic potential, complemented by the quadratic term  emerging from the mean-field decoupling,
yields the effective free energy for the magnetic order parameter
\begin{eqnarray}
\OpF_{\text{mag}}=\frac{1}{\beta} \sum_{\nu=1}^\infty\frac{\tr \big(\HG_0\HatOpV\big)^\nu}{\nu}+\int\difk{q}\frac{|\vec{M}_{\vec{q}}|^2}{U}\,,\quad
\end{eqnarray}

\noindent where
\begin{eqnarray}
\HG_0(k)=\frac{\imath k_n+\mu + \vec{g}(\vec{k})\cdot \vec{\sigma}}{\left(\imath k_n+\mu\right)^2-|\vec{g}(\vec{k})|^2}
=\sum_{s=\pm}\frac{P_s(\vec{k})}{\imath k_n-\epsilon_{\vec{k},s}}\,.\quad
\end{eqnarray}

\noindent In the last form we introduced the projectors $P_\pm(\vec{k})=\left[\II \pm \uvec{g}(\vec{k})\cdot \vec{\sigma} \right]/2$ with  unit vector
$\uvec{g}(\vec{k})\equiv\vec{g}(\vec{k})/|\vec{g}(\vec{k})|$.

\subsection{Spin susceptibility and magnetic instability -- Landau theory at quadratic order}\label{sec:magnetic:landau2}

\begin{figure}[t]
\includegraphics{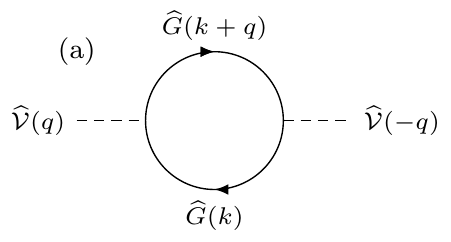}
\hspace*{1em}
\includegraphics{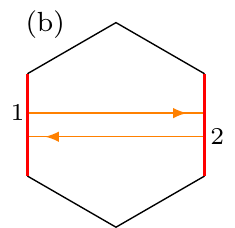}
\caption{(a) Loop diagram for the quadratic term of the Landau expansion in the magnetic order parameters. Solid lines represent the fermionic propagators. (b) Representative
magnetic scattering process for momentum transfers with one of the FS nesting wave vectors, leading to strong enhancement of the spin susceptibility.}
\label{fig:scattering}
\end{figure}

The leading magnetic instability is determined by the quadratic term of the Landau functional,
\begin{eqnarray}
\OpF^{(2)}&=&\frac{1}{2}\int\difk{q}M^a_{-\vec{q}}\left(\frac{2}{U}\delta^{ab}-\chi^{ab}_{\vec{q}}\right)M^b_{\vec{q}}\,,
\end{eqnarray}

\noindent which depends on the spin susceptibility $\chi^{ab}_{\vec{q}}=\sum_{s,s'=\pm} \chi^{ab}_{\vec{q},s,s'}$. A summation over repeated indices $a,b=x,y,z$ is implied.
In \figref{scattering,a} we show the loop diagram corresponding to the susceptibility and the do\-mi\-nant second-order scattering between nested sides of the FS. The spin
susceptibility is a sum of contributions involving upper and lower helicity bands, $s,s'=\pm$,
\begin{eqnarray}
\chi^{ab}_{\vec{q},s,s'}&=&-\int \difk{k}
\frac{n_F(\epsilon_{\vec{k},s})-n_F(\epsilon_{\vec{k}+\vec{q},s'})}{\epsilon_{\vec{k},s}-\epsilon_{\vec{k}+\vec{q},s'}}\nonumber\\
&&\times\braket{\vec{k},s|\sigma^a|\vec{k}+\vec{q},s'}\braket{\vec{k}+\vec{q},s'| \sigma^b|\vec{k},s}\,.\quad
\label{eq:susceptibility}
\end{eqnarray}

\begin{figure}[t]
\includegraphics{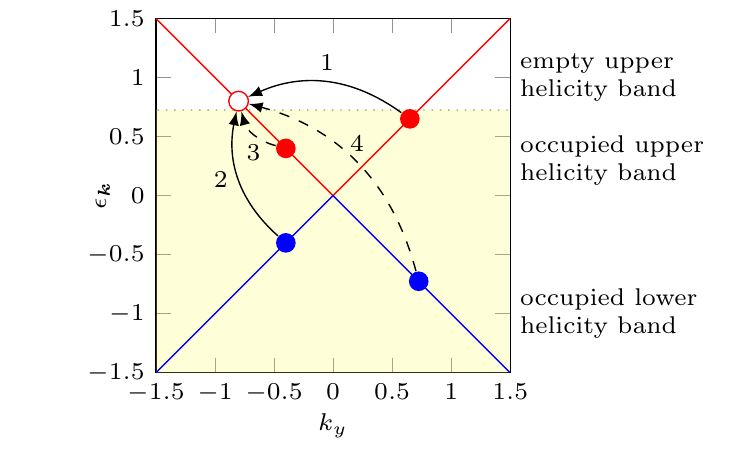}
\caption{Energy dispersions versus $k_y$ for $k_x=0$. Scatte\-ring processes from occupied states below $\mu$ to empty states are illustrated. For a well-nested FS, process 1
yields the dominant contribution to the upper helicity band spin susceptibility, while process 2 is the most relevant contribution from interband scattering. Processes 3 and
4 are suppressed at low temperatures.}
\label{fig:bandstructure}
\end{figure}

In \figref{bandstructure} we illustrate the different scatte\-ring processes contributing to the spin susceptibility. For $\mu>0$ and $T=0$ the lower helicity band
$\ket{\vec{k},-}$ is located deep below the Fermi energy and fully occupied; i.e., $n_F(\epsilon_{\vec{k},-})=1$. Therefore, intraband scattering in the lower band is
suppressed. Only intra- and interband scattering invol\-ving the upper helicity band $\ket{\vec{k},+}$ are relevant. Processes of the type 1  of \figref{bandstructure},
i.e., intraband scattering in the upper helicity band with large wave-vector transfer are the most dominant ones. The interband processes 2 and 4 enter the susceptibility
with a small factor $\sim1/(\epsilon_{\vec{k},-}-\epsilon_{\vec{k},+})<1/\mu$. Processes of the type 3, i.e., intraband scattering in the upper helicity band with small
wave-vector transfer, can be neglected since the occupation of the two states remains practically unchanged.

For an almost hexagonal FS the dominant scatte\-ring process contributing to the spin susceptibility for a wave vector $\vec{Q}_1$ (similarly for the rest) involves nested
parts of the FS as shown in \figref{scattering,b}. In strictly 1D systems \cite{GruenerSpin}, where nesting is perfect, the susceptibility might diverge, which would signal
the onset of a magnetic instability. In the present problem, where nesting is not perfect, the susceptibility does not diverge, but it is strongly enhanced for the wave vectors
$\pm \vec{Q}_i$. In combination with a sufficiently strong interaction this may be sufficient to lead to a magnetic instability.

Diagonalizing the spin susceptibility matrix for each wave vector \vec{q} yields three eigenvalues and unit eigenvectors,
$\hchi_{\vec{q}}\Uvec{M}_{\vec{q}}^i=\chi_{\vec{q}}^i\Uvec{M}_{\vec{q}}^i$, with $i=1,2,3$. The largest eigenvalue corresponds to the leading magnetic in\-sta\-bi\-li\-ty.
In the model considered here there are no other competing instabilities. Instead all potential instabilities are triggered by the same interaction potential $U$. The critical
interaction then follows from the Stoner criterion $\det[(2/U_{\text{crit}})\II -\hchi_{\vec{q}}]=0$.

\begin{figure*}
\includegraphics{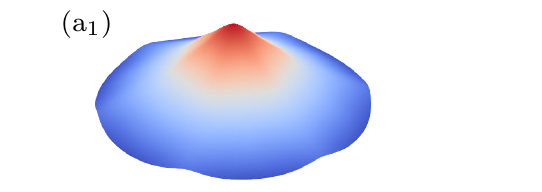}
\includegraphics{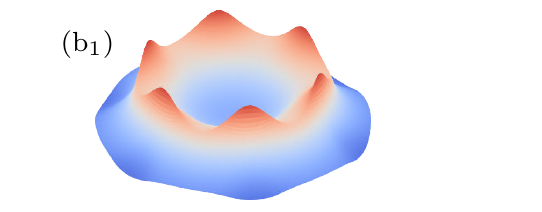}
\includegraphics{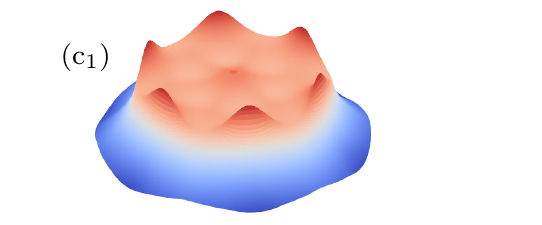}
\includegraphics{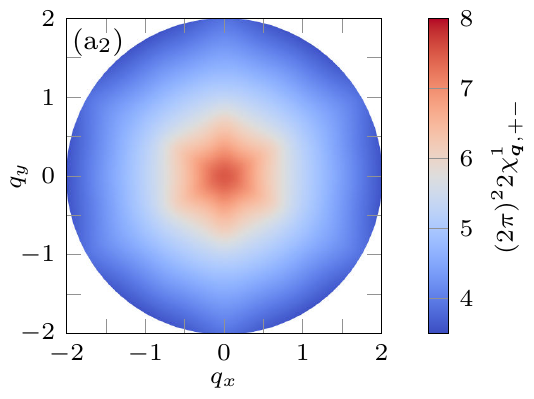}
\includegraphics{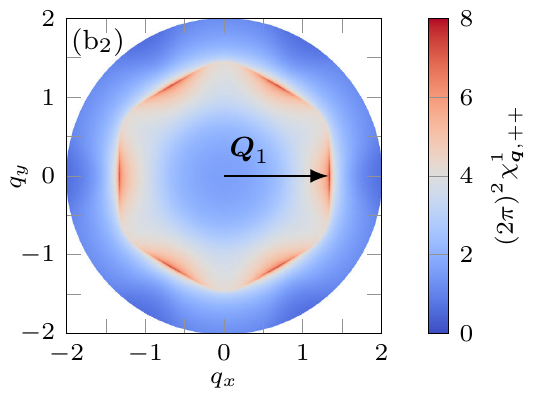}
\includegraphics{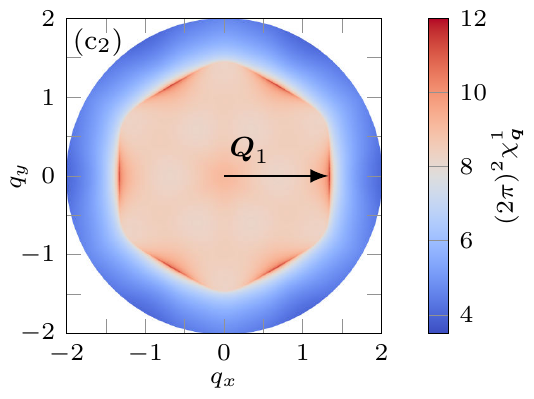}
\caption{Full view [(a$_1$),(b$_1$),(c$_1$)] and top view [(a$_2$),(b$_2$),(c$_2$)] plots of the largest eigenvalue
of the spin susceptibility versus the wave vector $\vec{q}=(q_x,q_y)$.
The parameters are $\mu=\mu_{\text{hex}}=0.725$ and $T=0.01$ in (a$_1$), (b$_1$), (c$_1$)
and $T=0$ in (a$_2$), (b$_2$), (c$_2$).
(a) Interband contribution.
(b) Intraband contribution from the upper helicity band.
(c) Total susceptibility [sum of (a) and (b)]. The nesting vector $\vec{Q}_1$ connects two sides of the FS. The remaining nesting vectors $\vec{Q}_{2,3}$ are related by
$C_3$ rotations.}
\label{fig:polarsusceptibility}
\end{figure*}

In \figref{polarsusceptibility} we plot the largest eigenvalue of the spin susceptibility, $\chi^1_{\vec{q}}$, for different wave vectors $\vec{q}=(q_x,q_y)$. We consider
$T=0$ and focus on the case with the most perfectly hexagonally warped FS, i.e., $\mu=\mu_{\text{hex}}$. We decompose the susceptibility into the inter- and intraband
contributions of the two helicity bands. The major contribution of the interband scattering, shown in \figref{polarsusceptibility,a}, comes from $\vec{q}\approx\vec{0}$ with
only a weak signature of the hexagonal warping. A large value of $\chi^1_{\vec{0}}$ would indicate a tendency to a ferromagnetic ground state. The upper helicity intraband
contribution, plotted in \figref{polarsusceptibility,b}, peaks for nesting wave vectors $\pm\vec{Q}_{1,2,3}$. For $T=0$ the intraband contribution of the lower helicity band
va\-ni\-shes. \figref{polarsusceptibility,c} depicts the total susceptibility, i.e., the sum of the interband and upper helicity intraband contributions. The picture
persists qualitatively unchanged as long as the dimensionless tem\-pe\-ra\-ture is low, i.e. $T \lesssim 0.05$. In \appendixref{susc-high-symmetry} we present further results for
different values of the chemical potential and momentum transfers along high-symmetry lines.

The largest eigenvalue of the susceptibility is $\chi^1_{\vec{Q}_1} \approx 12 / (2\pi)^2$. This implies that a magnetic instability arises when the interaction is stronger
than a critical value, which at $T=0$ is $U_{\text{crit}}=2/\chi_{\vec{Q}_1}^1\approx 6.58$. For \chem{Bi_2Te_3} this corresponds to $U \approx \SI{1.71}{eV}$. If we take into
account only the upper helicity band contribution, we find a higher value of the critical interaction, $U \approx \SI{2.57}{eV}$. Since the two results differ only slightly,
we will concentrate in the following qualitative discussions on the contribution from the upper helicity band only.

If the interaction is sufficiently strong, the system develops the magnetic phase corresponding to the largest eigenvalue. Once this is established, the remaining two
instabilities related to $\chi^{2,3}_{\vec{q}}$ are suppressed since most of the FS is gapped. Therefore we will consider only the emergence of the leading magnetic
instability. This is justified for all temperatures and is confirmed by our numerical results, since the eigenvalues of the remaining instabilities are very small. Our results
reported so far are in agreement with previous studies \cite{JiangStripe,BaumStern2}.

The nature of the order parameter is determined by the unit eigenvector $\Uvec{M}_{\vec{q}}^1$ correspon\-ding to the largest eigenvalue of the spin
susceptibility $\chi_{\vec{q}}^1$. Since we only consider this eigenvalue in the following we omit from here on the index $1$ for convenience; i.e., we write $\Uvec{M}_{\vec{q}}$.
The complex unit eigenvector $\Uvec{M}_{\vec{Q}_1}$ will be parametrized by spherical coordinates with the angles $\zeta$ and $\upsilon$:
\begin{eqnarray}
\Uvec{M}_{\vec{Q}_1}(\zeta,\upsilon)=
\begin{pmatrix}
\cos\zeta\\
\imath\sin\zeta\sin\upsilon\\
\imath\sin\zeta\cos\upsilon
\end{pmatrix}\,.
\end{eqnarray}

\noindent We observe that the $x$ component of the eigenvector exhibits a $\pi/2$ phase shift compared to the $y$ and $z$ components. The particular form implies that, as
a direct consequence of the spin-momentum locking of the surface states, the leading instability corresponds to a helical magnetic phase.

The orientation of the magnetic order can be understood directly from the structure of the Hamiltonian. As we argued above the upper helicity band contribution plays the
dominant role. If we consider the projection of the magnetic term onto this band only we obtain
\begin{eqnarray}
  &&\OpH_{\text{mag}}^+= \int\difk{q}\Bigg[\frac{|\vec{M}_{\vec{q}}|^2}{U} \nonumber\\
  &&+\int\difk{k}\psi_{\vec{k}+\vec{q}/2,+}\hc\vec{M}_{\vec{q}}\nohc\cdot\vec{\sigma}^+_{\vec{q}}(\vec{k})\psi_{\vec{k}-\vec{q}/2,+}\nohc\Bigg]
  \,.\qquad
\end{eqnarray}
Here we introduced the matrix element
\begin{eqnarray}
\vec{\sigma}^+_{\vec{q}}(\vec{k})&\equiv&\braket{\vec{k}+\vec{q}/2,+|\vec{\sigma}|\vec{k}-\vec{q}/2,+}\,,
\end{eqnarray}

\noindent which is related to the effective magnetic moment of the upper helicity band (with details presented in \appendixref{landau}).
We note that $\vec{\sigma}^+_{\vec{q}}(\vec{0})$ assumes the simple form
\begin{eqnarray}
  \vec{\sigma}^+_{\vec{q}}(\vec{0})&=&\begin{pmatrix}
    \cos\vartheta_{\vec{q}/2}\cos\varphi_{\vec{q}/2}+\imath\sin\varphi_{\vec{q}/2}\\
    \cos\vartheta_{\vec{q}/2}\sin\varphi_{\vec{q}/2}-\imath\cos\varphi_{\vec{q}/2}\\
    -\sin\vartheta_{\vec{q}/2}
    \end{pmatrix}\,.
\end{eqnarray}

\noindent Specifically for $\vec{q}=\vec{Q}_1$ we find
\begin{eqnarray}
\vec{\sigma}^+_{\vec{Q}_1}(\vec{0})&=&\begin{pmatrix}\imath\\k_0^3/\mu_{\text{hex}}\\-k_0/\mu_{\text{hex}}\end{pmatrix}\,.
\end{eqnarray}

\noindent The maximum magnetic gap at the he\-xa\-go\-nal FS is achieved when the corresponding moduli $|\vec{M}_{\vec{Q}_i} \cdot \vec{\sigma}^+_{\vec{Q}_i}(\vec{k})|$
($i=1,2,3$) become maximized. Since in the present si\-tua\-tion the FS mainly consists of nearly flat parts nested by the wave vectors $\vec{q}=\vec{Q}_{1,2,3}$, the above
moduli will be maximized when $\Uvec{M}_{\vec{Q}_{1,2,3}}$ are pa\-ral\-lel to $\vec{\sigma}^+_{\vec{Q}_{1,2,3}}(\vec{0})$, respectively. The latter condition is satisfied when
\begin{eqnarray}
\Uvec{M}_{\vec{Q}_1}&\approx&\frac{1}{\sqrt{2}}\begin{pmatrix}1\\\imath k_0^3/\mu_{\text{hex}}\\-\imath k_0/\mu_{\text{hex}}\end{pmatrix}\,.\label{eq:m-vector}
\end{eqnarray}

\noindent The remaining eigenvectors $\Uvec{M}_{\pm\vec{Q}_{2,3}}$ can be obtained from the transformation proper\-ties of the susceptibi\-li\-ty matrix under the group
elements $\OpG\in\group{C_{3v}}$; i.e., $\hchi_{\OpG \vec{q}} = \HD^-_\OpG\hchi_{\vec{q}} \left(\HD^-_\OpG\right)\trans$. This implies $\Uvec{M}_{\OpG\vec{q}} = \HD^-_\OpG
\Uvec{M}_{\vec{q}}$. As a result of the magnetic order a gap opens at the well-nested parts of the FS; however, there is no guarantee for a gap opening at the remaining FS.
This behavior, which is ty\-pi\-cal for two-dimensional systems with imperfect nesting \cite{Schulz}, is indeed found in the present problem, as we will show in
\secref{proximity:self-consistent}.

Numerically we find $\zeta \approx \pi/4$ and $\upsilon \approx 0.9\pi$, in good agreement with the approximate \eqref{m-vector} since
$\sin(\zeta) \approx \cos(\zeta) \approx 1/\sqrt{2}$, $\sin(\upsilon) \approx k_0^3/\mu_{\text{hex}}$, and $\cos(\upsilon) \approx -k_0/\mu_{\text{hex}}$. The diffe\-ren\-ces
between the numerical and approximate result is due to the contributions from those parts of the FS which have poor nesting, such as the rounded corners of the FS. Further
details on the numerically extracted pa\-ra\-me\-ters, including their remarkably weak temperature dependence, are presented in \appendixref{susc-high-symmetry}.

\subsection{Landau theory up to quartic order: Single- versus triple-\texorpdfstring{\vec{Q}}{Q} magnetic phase}
\label{sec:magnetic:landau4}

Previous work on hexagonally warped surface states focused on the spin susceptibility or equivalently on the quadratic part of a Landau expansion. Single-\vec{Q} (i.e.,
stripes) or triple-\vec{Q} magnetic phases were found and described. However, as already noted in \citeref{FuHex}, on this level there remains a large degree of degeneracy
among the potential ground states. Therefore, it is necessary to determine the quartic and, as it will turn out, even the sixth-order coefficients of the Landau expansion. In
this section we show that the quartic order decides in favor of a triple-\vec{Q} order parameter.

To proceed we focus on the dominant nesting vectors $\pm\vec{Q}_i$ and split the favored magnetization vectors $\vec{M}_{\vec{Q}_i} = M_i \Uvec{M}_i$ into a complex amplitude
$M_i = e^{\imath\varPhi_i} |M_i|$ and a unit vector $\Uvec{M}_i=\Uvec{M}_{\vec{Q}_i}$. The unit vectors have been introduced in \secref{magnetic:landau2}.
Note that $\vec{M}_{-\vec{Q}_i} = M_i^* \Uvec{M}_{-\vec{Q}_i}$. After having fixed the orien\-tation of the magnetic order parameters, the remaining 6 degrees of freedom arise
from the moduli $|M_i|$ and the phases $\varPhi_i$. At quartic order the Landau expansion involves only the moduli $|M_i|$ and reads
\begin{eqnarray}
&&\OpF^{(4)} = \alpha \sum_{i=1}^3 |M_i|^2 + \frac{\beta_1}{2} \sum_{i=1}^3 |M_i|^4 + \beta_2 \sum_{i<j} |M_i|^2 |M_j|^2\nonumber\\
&&= \alpha \sum_{i=1}^3 |M_i|^2 + \beta_{\text{S}} \left(\sum_{i=1}^3 |M_i|^2\right)^2 + \beta_{\text{A}} \sum_{i=1}^3 |M_i|^4\,.\quad
\label{eq:landau4}
\end{eqnarray}

\begin{figure}[t]
\includegraphics{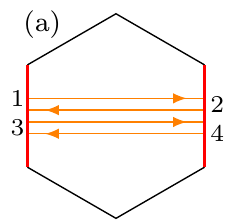}
\includegraphics{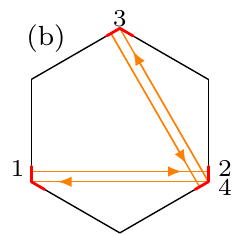}
\includegraphics{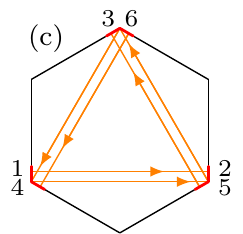}
\caption{Different types of scattering processes between points on the hexagonal FS yield the dominant contribution to the coefficients. Type (a) determines $\beta_1$, type
(b) determines $\beta_2$, and the sixth-order type (c) determines $\eta$. Due to momentum conservation the scattering phase space (indicated in red) which determines
$\beta_2$ and $\eta$ is much smaller than the phase space for $\alpha$ and $\beta_1$.}
\label{fig:scattering-higher}
\end{figure}

\noindent The coefficient $\alpha=2/U-\chi_{\vec{Q}_1}^1$ vanishes at the onset of the instability. The coefficients $\beta_1$ and $\beta_2$ ori\-gi\-na\-te from the
scattering processes shown in \figref{scattering-higher,a} and \hyperref[fig:scattering-higher]{(b)}, respectively. Their values are obtained from four-leg loop diagrams as described
in detail in \appendixref{landau}.

The second-order term in the Landau expansion exhibits U(3) symmetry \cite{FuHex}, which suggests expressing the three complex scalar order parameters as a vector with angles
$\omega$ and $\rho$,
\begin{eqnarray}
  \begin{pmatrix}
    M_1\\
    M_2\\
    M_3
  \end{pmatrix} =
  |M|
  \begin{pmatrix}
    e^{\imath \varPhi_1} \sin \omega \cos \rho\\
    e^{\imath \varPhi_2} \sin \omega \sin \rho\\
    e^{\imath \varPhi_3} \cos \omega
  \end{pmatrix}\,.
\end{eqnarray}

\noindent In general the U(3) symmetry is broken by higher order terms. We show this explicitly by splitting the fourth-order terms in the second line of \eqref{landau4} into
a U(3)-symmetric term with coefficient $\beta_{\text{S}} = \beta_2/2$ and an anisotropic one with coefficient $\beta_{\text{A}} = (\beta_1-\beta_2)/2$.

The relation between $\beta_1$ and $\beta_2$ (or equivalently between $\beta_{\text{S}}$ and $\beta_{\text{A}}$) determines whether the order pa\-ra\-me\-ters of different
nesting vectors compete or coexist. We first consider the case $\beta_S>0$. Then it is sufficient to minimize the anisotropic term, which reads
\begin{eqnarray}
  \beta_{\text{A}}|M|^4 \left[\sin^4 \omega (\cos^4 \rho + \sin^4 \rho) + \cos^4 \omega\right]\,.\quad
\end{eqnarray}

\noindent For $\beta_{\text{A}}<0$ the minima are located at $\omega = 0$, independent of $\rho$ and at $\omega=\pm \pi/2$ and $\rho=0, \pm \pi/2$. This case corresponds
to a magnetization with a single wave vector $\vec{Q}_i$. On the other hand, for $\beta_{\text{A}}>0$ the minimum is found at $\rho = \pm \pi/4$ and $\omega = \arctan
\sqrt{2}$, which implies that $|M_{1,2,3}|$ are equal.

In the case $\beta_S<0$ the favored value for the mo\-du\-lus $|M|$ is obtained by minimizing the  Landau functional up to quartic order. It reduces to the
conditions
\begin{eqnarray}
|M_i| \Big(\alpha + \beta_1 |M_i|^2 + \beta_2 \sum_{j\ne i}|M_j|^2\Big)=0\,.
\end{eqnarray}

\noindent This system of equations allows three solutions: (i) a nonmagnetic phase with $|M_i|=0$, (ii) a single-\vec{Q} phase and (iii) a triple-\vec{Q} phase. In the
single-\vec{Q} case with, e.g., $|M_1|\ne 0$ and $|M_{2,3}|=0$, we obtain $|M_1|^2 =|\alpha|/\beta_1$ and the free energy value $\OpF= -\alpha^2/(2\beta_1)$. For the
triple-\vec{Q} phase we obtain $|M_i|^2=|\alpha|/(\beta_1+2\beta_2)$ for $i=1,2,3$ and the free energy $\OpF = -3\alpha^2/[2(\beta_1 + 2 \beta_2)]$. By com\-pa\-ri\-son we
find that the triple-\vec{Q} phase is stabilized when $\beta_1>\beta_2$ or $\beta_1>-2\beta_2$. Note that a double-\vec{Q} phase cannot appear, since whenever two order
parameters appear together, due to the $\beta_2$ coupling, they act as sources for the remaining third-order parameter leading to a triple-\vec{Q} phase.

The coefficients $\beta_{1,2}$ are obtained by evaluating appropriate four-leg loop diagrams similarly to \secref{magnetic:landau2} (see \appendixref{landau} for details), with
results presented in \figref{landau-higher}. In the whole relevant temperature regime we find $\beta_1 \gg 2|\beta_2|$. This implies that the triple-\vec{Q} phase with equal
values of $|M_i|$ for $i=1,2,3$ is favored over the stripe phase. The stabilization of this triple-\vec{Q} phase renders the system \group{C_3}-symmetric. We want to add, that
although the stripe phase appears unfavored within the model under consideration, it can become relevant in cases where external fields or structural defects break the
\group{C_3} symmetry.

According to the above analysis, the magnitude of the order parameter is fixed at the quartic order of the Landau theory. However, the three phase degrees of freedom remain
undetermined. Two of the phase degrees of freedom, $\varPhi_x=2\varPhi_1-\varPhi_2-\varPhi_3$ and $\varPhi_y=\sqrt{3}(\varPhi_2-\varPhi_3)$, constitute Goldstone modes, i.e.,
phasons related to the broken translational symmetry in the two-dimensional coordinate space. These phases transform according to the two-dimensional IR of
\group{C_{3v}}. The remaining ``center of mass'' phase $\varPhi_z=\varPhi_1+\varPhi_2+\varPhi_3$, which transforms according to the $A_2$ IR of \group{C_{3v}}, will be
determined by a sixth-order term of the Landau expansion, as we will show in the next section.

\subsection{Landau theory at sixth order: Phase locking for the triple-\texorpdfstring{\vec{Q}}{Q} order parameter}\label{sec:magnetic:landau6}

At sixth order of the Landau expansion we find terms of the form $|M_i|^6$, $|M_i|^2|M_j|^4$ with $i\ne j$ and $|M_1M_2M_3|^2$. They are made up of products of moduli
similar to the lower order terms and yield only quantitative corrections. Accordingly, they cannot provide information about the ``center of mass'' phase $\varPhi_z$. This
information is
contained in a new type of term of the form
\begin{eqnarray}
\OpF^{(6)}_{\varPhi_z} &=& \eta (M_1M_2M_3)^2 + \eta (M_1^*M_2^*M_3^*)^2 \nonumber\\
                       &=&2\eta|M_1M_2M_3|^2\cos\left(2\varPhi_z\right)\,,
\end{eqnarray}

\noindent which is allowed at sixth order due to the relation $\vec{Q}_1+\vec{Q}_2+\vec{Q}_3=\vec{0}$. The most relevant underlying scattering process is shown in
\figref{scattering-higher,c}, where the particle is scattered in momentum space twice around the triangular path. The sign of the pa\-ra\-me\-ter $\eta$
is suf\-ficient to fix $\varPhi_z$. If $\eta<0$ ($\eta>0$) the free energy is minimized for $\varPhi_z=0,\pi$ ($\varPhi_z=\pm\pi/2$).

\begin{figure}[t]
\includegraphics{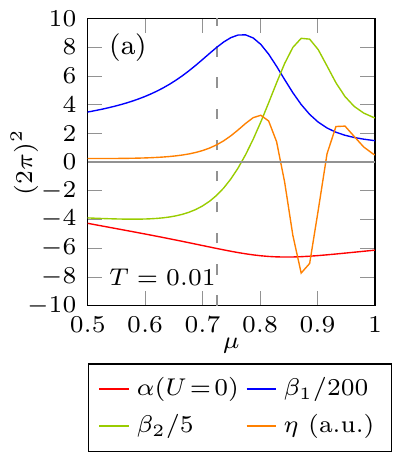}
\includegraphics{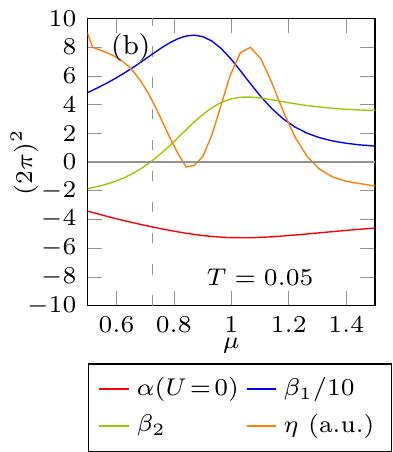}
\caption{Coefficients of the Landau expansion up to sixth order for (a), (b) $T=0.01,0.05$ (corresponding to $T=\SI{30},\SI{150}{K}$ for \chem{Bi_2Te_3}). The
dashed lines mark the selected chemical potential $\mu=\mu_{\text{hex}}$. For this value of chemical potential, the relation between the coefficients $\alpha$, $\beta_{1,2}$,
and $\eta$ favors
the triple-\vec{Q} magnetic phase. $\eta$ is plotted in arbitrary units since only the sign is important. For the numerical analysis we considered only the upper helicity band
contribution.}
\label{fig:landau-higher}
\end{figure}

The sign of $\eta$ depends on the shape of the FS, which is controlled by the chemical potential $\mu$. Our numerical results are depicted in \figref{landau-higher}. We note
that the system can be tuned to a critical chemical potential for which $\eta$ va\-ni\-shes, which leads to an additional degeneracy between the $A_1$ and $A_2$ order
parameters, and $\varPhi_z$ merely represents an additional Goldstone mode. On the other hand, in the vicinity of the value $\mu=\mu_{\text{hex}}=0.725$, correspon\-ding
to a hexagonal FS, we find $\eta>0$. That means the $A_2$ IR of the magnetic order parameter with $\varPhi_z=\pm\pi/2$ is favored. In the next paragraph, we examine further
properties of this favored magnetic ground state.

\subsection{Skyrmion lattices}

\begin{figure}[b]
\includegraphics{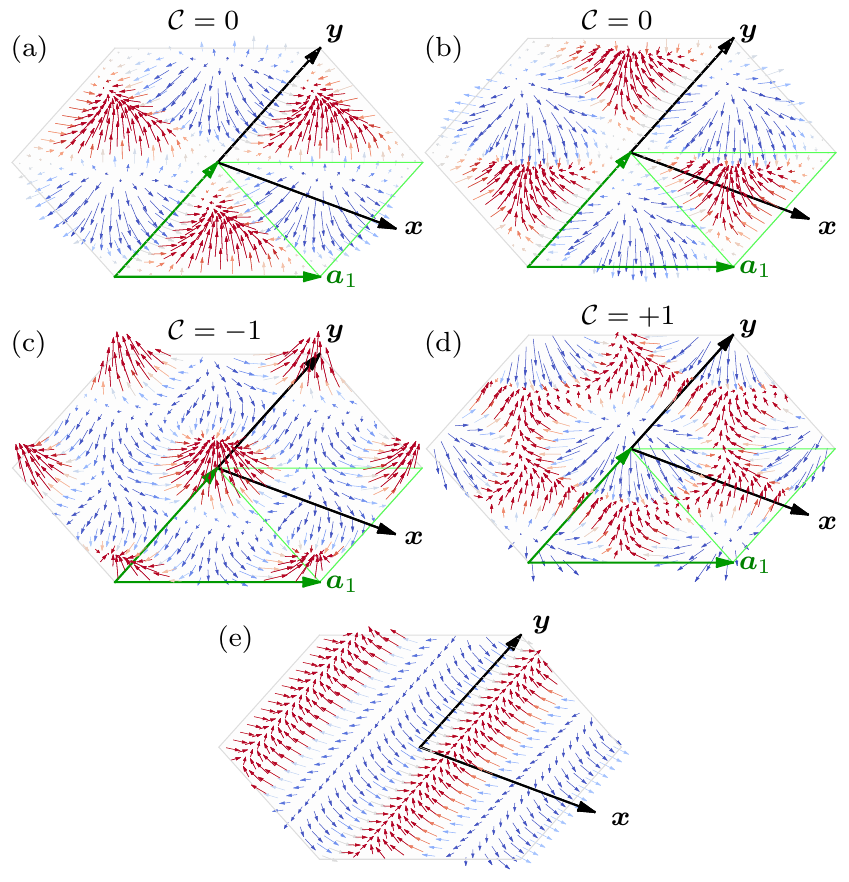}
\caption{Magnetization profile on the surface of a warped TI for $|\vec{r}\cdot\Uvec{Q}_i|\le 2\pi$. The green parallelogram formed by the two triangles is the unit cell.
(a), (b) Triple-\vec{Q} phases belonging to the $A_1$ representation of the group \group{C_{3v}} with skyrmion charge $\OpC=0$: (a) $\varPhi_z=\varPhi_{1,2,3}=0$ and (b)
$\varPhi_z=\varPhi_{1,2,3}=\pi$. (c), (d) Triple-\vec{Q} phases belonging to the $A_2$ representation of the group \group{C_{3v}} with skyrmion charge $\OpC=\pm 1$: (c)
$\varPhi_z=-\pi/2$, $\varPhi_{1,2,3}=\pi/2$ and (d) $\varPhi_z=\pi/2$, $\varPhi_{1,2,3}=-\pi/2$. (e) Single-\vec{Q} phase with order parameter breaking the
\group{C_3} symmetry. For presentation purposes we use in (e) a vector normalization differing from the ones in (a)-(d).}
\label{fig:magnetization}
\end{figure}

Continuing with the convention of \secref{magnetic:landau2} we assume for the rest of the discussion that the $M_x$ component of the order parameter is real and the
$M_{y,z}$ components purely imaginary. With this choice, we find that the magnetic order parameter $\vec{M}(\vec{r})$ belongs for $\varPhi_z=0,\pi$ ($\varPhi_z=\pm\pi/2$) to
the $A_1$ ($A_2$) IR of the point
group \group{C_{3v}}. For a detailed discussion of the accessible magnetic order parameters and their classification according to
the IRs of the \group{C_{3v}} point group, see \appendixref{classification}.

In \figref{magnetization} we depict the resulting magnetization vector in coordinate space
\begin{eqnarray}
\vec{M}(\vec{r})=\sum_{i=1}^3 \vec{M}_{\vec{Q}_i} e^{\imath\vec{Q}_i\cdot\vec{r}} + \ccterm
\end{eqnarray}

\noindent for the values $\varPhi_z=0,\,\pi,\,\pm\pi/2$. One can directly identify the IR to which each magnetic profile belongs. For the $A_1$ representation profiles with
$\varPhi_z=0,\pi$, the magnetization component $M_x(\vec{r})$ is even under the mirror symmetry $x\mapsto-x$, while the $M_{y,z}(\vec{r})$ components change sign. The
opposite situation is realized in the case of the $A_2$ IRs. The magnetization profiles corresponding to $\varPhi_z=\pm\pi/2$ preserve only the \group{C_3} symmetry
and give rise to a skyrmion lattice.

As shown in \figref{magnetization,b,c}, the particular magnetic skyrmion profile exhibits a periodicity that allows us to introduce a unit cell (UC) (enclosed by green lines)
spanned by the vectors
\begin{eqnarray}
  \vec{a}_1 = \frac{2\pi}{\sqrt{3}k_0}\left(\frac{\sqrt{3}}{2},\, \frac{1}{2}\right)\,,\quad
  \vec{a}_2 = \frac{2\pi}{\sqrt{3}k_0}\left(0,\,1\right)\,.\quad
\end{eqnarray}
In order to highlight the unbroken \group{C_3} symmetry we show a hexagon defined by $|\vec{r}\cdot\vec{Q}_i|\le 2\pi$, which consists of three UCs. The periodicity in real
space also suggests the definition of a Brillouin zone (BZ) in \vec{k} space with $|\vec{k}\cdot\Uvec{Q}_i|\le k_0$.

The magnetization vector at opposite edges of the UC is the same, allowing us to compactify the UC. In fact, gluing the opposite edges of the UC yields a manifold
homeomorphic to the flat 2-torus $\TT^2$. On the other hand, the unit vector $\Uvec{M}(\vec{r})$ takes values on the 2-sphere $S^2$. The winding (or first Chern) number of the
mapping from the torus to the sphere is defined as
\begin{eqnarray}
\OpC = \frac{1}{4\pi} \int_{\text{UC}}\difv{r} \Uvec{M}(\vec{r})\cdot\left(\partial_x \Uvec{M}(\vec{r}) \times \partial_y \Uvec{M}(\vec{r})\right)\,.
\end{eqnarray}

\noindent In the present context the Chern number is also called skyrmion charge and takes only integer values. We find that $\OpC=\sign(\varPhi_z)$; thus $\OpC=0$ for the
$A_1$ phases and $\OpC = \pm 1$ for the $A_2$ phases, with $\varPhi_z=\pm\pi/2$. Therefore, in the $A_2$ phases the magnetization is topologically nontrivial with $\pm 1$
skyrmion charge.

The possibility of a topologically nontrivial magnetic ground state has been discussed earlier in \citeref{BaumStern2}. By our detailed analysis we confirmed that the skyrmion
lattice is indeed the true thermodynamic ground state of the system with a hexagonal FS.

\section{Effects of proximity-induced superconductivity}\label{sec:proximity}

One of the prospects of TIs is their use as building blocks for engineering TSCs. A route to this goal relies on bringing the TI in pro\-xi\-mi\-ty to a conventional
superconductor. This requires analyzing the consequences of a proximity-induced superconduc\-ting gap on the magnetic instability and the magnetic surface states. Conversely
the surface magnetism acts as a pair-breaking source for the superconductor and reduces the superconduc\-ting gap near the interface. For the qualitative discussion of the
following section we neglect this effect. However, we will point out that induced supercurrents may contain information about the magnetic structures.

\subsection{Magnetic instability of the surface states in proximity to a superconductor}

On a phenomenological level we expect that the pro\-xi\-mi\-ty induced superconducting gap $\Delta$ introduces additional terms in the Landau free energy expansion, the most
important one for weak $|\Delta|$ being
\begin{eqnarray}
\OpF _{\text{mag,sc}}=c|\Delta|^2\sum_{i=1}^3|M_i|^2\,.
\end{eqnarray}

\noindent Note that the \OpT symmetry violating magnetic phase competes with the spin singlet proximity-induced gap, which implies that $c>0$. The presence of the
coupling term modifies the coefficient $\alpha$ as follows,
\begin{eqnarray}
\alpha(\Delta)=\frac{2}{U}-\chi_{\vec{Q}_1}^1+c|\Delta|^2\,.
\end{eqnarray}

\noindent It can be viewed as a renormalization of the effective interaction. That is, in the presence of $\Delta$, a stronger interaction is required for reaching the magnetic
instability as compared to the situation without. At this level of the Landau theory the presence of superconductivity does not change the nature of the magnetic
order, i.e., our conclusions concerning triple- or single-\vec{Q} phases or the selection between the different triple-\vec{Q} phases remain unchanged. Therefore we assume
for the rest of this section that one of the triple-\vec{Q} phases or the nonmagnetic state is selected.

For a more quantitative analysis of the proximity effects on the magnetic instability we recalculate the quadratic term of the Landau expansion similarly to
\secref{magnetic:landau}, but with $\Delta$ included in the Bogoliubov-de Gennes Hamiltonian (BdG)
\begin{eqnarray}
\HH_{0,\text{sc}}(\vec{k})=\vec{g}(\vec{k})\cdot\tau_z\tvec{\sigma}-\mu\tau_z-\Delta\tau_y\sigma_y\,.
\end{eqnarray}

\noindent We introduced \vec{\tau} Pauli matrices acting in particle-hole space and extended the spin Pauli matrices \vec{\sigma} to particle-hole and spin
space $\tvec{\sigma}\equiv (\tau_z\sigma_x,\,\sigma_y,\,\tau_z\sigma_z)$. Furthermore, we chose a gauge where the proximity-induced superconducting gap $\Delta$ is real and
positive. After introducing the four-component Nambu spinor
\begin{eqnarray}
\HPsi\hc_{\vec{k}}=\big(c\hc_{\vec{k}\up},\,c\hc_{\vec{k}\down},\,c\nohc_{-\vec{k}\up},\,c\nohc_{-\vec{k}\down}\big) \,,
\end{eqnarray}

\noindent we can rewrite the Hamiltonian  as $\OpH_{0,{\text{sc}}}=\nicefrac{1}{2}\int \difv{k}
\HPsi_{\vec{k}}\hc\HH_{0,\text{sc}}(\vec{k})\HPsi_{\vec{k}}$. The mean-field-decoupled mag\-netic contribution then reads
\begin{eqnarray}
  \OpH_{\text{mag}}\nohc &=& \int\difk{q}\Bigg[\frac{|\vec{M}_{\vec{q}}|^2}{U} \nonumber\\
  &&+\frac{1}{2}\int\difk{k}\HPsi_{\vec{k}+\vec{q}/2}\hc\vec{M}_{\vec{q}}\nohc\cdot\tvec{\sigma}\,\HPsi_{\vec{k}-\vec{q}/2}\nohc\Bigg]\,.
  \label{eq:hamiltonian-magnetic}
\end{eqnarray}

\noindent Note the presence of a factor $1/2$ which cancels the double counting of the electronic degrees of freedom.

In order to perform the Landau expansion, we need the modified Green's function
\begin{eqnarray}
  \HG_{0,\text{sc}}(k)&=&\frac{(\imath k_n)^2-\left(|\vec{g}(\vec{k})|^2+\mu^2+\Delta^2\right)-2\mu\vec{g}(\vec{k})\cdot\tvec{\sigma}}
  {\left(k_n^2+\tilde{\epsilon}_{\vec{k},+}^2\right)\left(k_n^2+\tilde{\epsilon}_{\vec{k},-}^2\right)}\nonumber\\
  &\times&\left[\imath k_n+\HH_{0,\text{sc}}(\vec{k})\right]\nonumber\\\nonumber\\
  &=&\sum_{s=\pm} \frac{\widetilde{P}_s(\vec{k})\left[\imath k_n+\HH_{0,\text{sc}}(\vec{k})\right]}
  {\left(\imath k_n-\tilde{\epsilon}_{\vec{k},s}\right)\left(\imath k_n+\tilde{\epsilon}_{\vec{k},s}\right)}.\qquad
\end{eqnarray}

\noindent Here $k_n$ again denotes the fermionic Matsubara frequencies. We introduced the Nambu-space-extended he\-li\-ci\-ty band projectors
$\widetilde{P}_\pm(\vec{k})=\left[\II \pm \uvec{g}(\vec{k})\cdot\tvec{\sigma}\right]/2$ and the eigen\-energies $\tilde{\epsilon}_{\vec{k},\pm} =
\sqrt{\epsilon_{\vec{k},\pm}^2+\Delta^2}$. For the calculation of the modified spin susceptibility we switch to the helicity basis of the Hamiltonian
$\{\ket{e,\vec{k},s}\,,\ket{h,\vec{k},s}\}$, with $s=\pm 1$ denoting the helicities and $e$ ($h$) corresponding to the electron (hole) space:
\begin{eqnarray}
\ket{e,\vec{k},\pm} &=& \begin{pmatrix}1\\0\end{pmatrix} \otimes \ket{\vec{k},\pm}\,,\\
\ket{h,\vec{k},\pm} &=& \Xi\ket{e,\vec{k},\pm}=\tau_x \begin{pmatrix}1\\0\end{pmatrix} \otimes \OpK \ket{\vec{k},\pm}\nonumber\\
&=&\begin{pmatrix}0\\1\end{pmatrix}\otimes\ket{-\vec{k},\pm}^*\,.
\end{eqnarray}

\noindent The charge conjugate partner of $\ket{e,\vec{k},\pm}$ is obtained by acting with the charge conjugation operator $\Xi=\tau_x\OpK$. For instance, for the upper
helicity band we obtain
\begin{eqnarray}
\ket{h,\vec{k},+}
=\begin{pmatrix}0\\1\end{pmatrix}\otimes \begin{pmatrix}
-e^{+\frac{\imath\varphi_{\vec{k}}}{2}}\sin\tfrac{\vartheta_{\vec{k}}}{2}\\
\phantom{-}e^{-\frac{\imath\varphi_{\vec{k}}}{2}}\cos\tfrac{\vartheta_{\vec{k}}}{2}\end{pmatrix}\,.
\end{eqnarray}

\noindent Here we used $\varphi_{-\vec{k}}=\varphi_{\vec{k}}$ and $\vartheta_{-\vec{k}}=\vartheta_{\vec{k}}+\pi$. Notice that the complex conjugation operator \OpK also
acts on momenta $\OpK\vec{k}=-\vec{k}$.

According to the previous sections, the major contribution to the spin susceptibility is provided by the upper helicity band. Therefore, we consider only the upper
helicity band by appropriately projecting the Hamiltonian and Green's function. The projected Hamiltonian, acting now only in Nambu space, reads
\begin{eqnarray}
\HH_{0,\text{sc}}^+(\vec{k})=\epsilon_{\vec{k},+}\tau_z+\Delta\tau_x\,.\label{eq:hamiltonian-helicity-basis}
\end{eqnarray}

\noindent Note that with the given choice of the helicity eigen\-states, the projected superconducting gap is \vec{k}-independent. A diffe\-rent gauge would yield an odd
\vec{k}-dependence, as discussed in \citeref{FuKane2008}.

The correspon\-ding projected Green's function reads
\begin{eqnarray}
\HG_{0,\text{sc}}^+(k)=\frac{\imath k_n+\HH_{0,\text{sc}}^+(\vec{k})}{\left(\imath k_n-\tilde{\epsilon}_{\vec{k},+}\right)\left(\imath
k_n+\tilde{\epsilon}_{\vec{k},+}\right)}\,.
\end{eqnarray}

\noindent Using this Green's function we obtain the spin susceptibility, modified by the effects of the proximity-induced superconducting gap,
\begin{eqnarray}
  \tilde{\chi}^{ab}_{\vec{q},++} &=& -\frac{1}{2} \sum_{\lambda,\lambda'=\pm} \!\!\int\! \difk{k}
  \frac{n_F(\lambda\tilde{\epsilon}_{\vec{k},+}) -
n_F(\lambda'\tilde{\epsilon}_{\vec{k}+\vec{q},+})}{\lambda\tilde{\epsilon}_{\vec{k},+}-\lambda'\tilde{\epsilon}_{\vec{k}+\vec{q},+}} \nonumber\\
  &&\times
 \frac{\tilde{\epsilon}_{\vec{k},+} \tilde{\epsilon}_{\vec{k}+\vec{q},+} + \lambda\lambda' (\epsilon_{\vec{k},+}\epsilon_{\vec{k}+\vec{q},+}+
\Delta^2)}{2 \tilde{\epsilon}_{\vec{k},+}
\tilde{\epsilon}_{\vec{k}+\vec{q},+}} \nonumber\\
 &&\times\braket{\vec{k},+ | \sigma^a | \vec{k}+\vec{q},+} \braket{\vec{k}+\vec{q},+ | \sigma^b |\vec{k},+}\,.
\end{eqnarray}

\begin{figure}[t]
\includegraphics{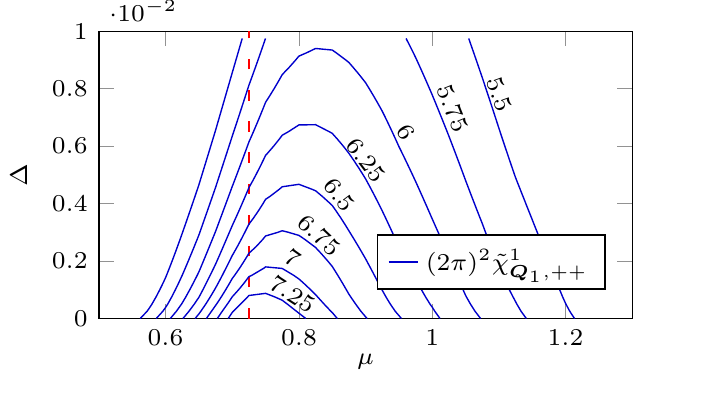}
\caption{Contours of the largest eigenvalue of the spin susceptibility for values between 5.5 and 7.25
(only the contribution of the upper helicity band is shown)
versus the proximity-induced superconducting gap $\Delta$ and the chemical potential $\mu$.
The dashed red line is the chemical potential
$\mu=\mu_{\text{hex}}$ for which the FS is closest to hexagonal.}
\label{fig:deltaisolines}
\end{figure}

In \figref{deltaisolines} we present a contour plot of the spin susceptibility calculated for $T=0$ as a function of the che\-mi\-cal potential $\mu$ and superconduc\-ting gap
$\Delta$. As anticipated the spin susceptibility is reduced by the presence of $\Delta$. For example, if $\Delta=\num{1.93e-3}$ corresponding to \SI{0.5}{meV} for
\chem{Bi_2Te_3}, the spin susceptibility is reduced by approximately 10\%.

\subsection{Low energy model for topological superconductivity}

In this section we derive a low-energy model for the surface electrons of the TI and the magnetic order parameter under the influence of pro\-xi\-mi\-ty-induced
superconductivity. Essentially, the latter model will describe the TSC expected to appear in this hybrid device. It will allow us to establish the topological criterion
related to the emergence of Majorana fermions.

We start from the Hamiltonian \eqref{hamiltonian-helicity-basis} and include the magnetic term of \eqref{hamiltonian-magnetic} in the upper helicity band projection. We can
restrict ourselves to the nesting wave vectors $\pm\vec{Q}_i$, which connect the $\vec{k}\pm\vec{Q}_i/2$ upper he\-li\-ci\-ty states ($i=1,2,3$). Under these conditions, we
obtain a BdG Hamiltonian $\HH_{\text{mag,sc}}(\vec{k})$, with three decoupled blocks $\hh_i(\vec{k})$. Each one of the blocks reads
\begin{eqnarray}
\hh_i(\vec{k})&=&\delta_{i,\vec{k},+}\tau_z+\delta_{i,\vec{k},-}\tau_z\rho_z +\Delta\tau_x
\nonumber\\
&+&Mf^R_i(\vec{k})\rho_x- Mf^I_i(\vec{k})\rho_y
\end{eqnarray}
with
\begin{eqnarray}\delta_{i,\vec{k},\pm}= \frac{\epsilon_{\vec{k}+\vec{Q}_i/2,+}\pm\epsilon_{\vec{k}-\vec{Q}_i/2,+}}{2}\,
\end{eqnarray}
\noindent and $f_i(\vec{k}) \equiv \Uvec{M}_i\cdot\vec{\sigma}^+_i(\vec{k})$ (we introduced $\vec{\sigma}^+_i(\vec{k})$ in \secref{magnetic:landau2} and discuss it further
in \appendixref{landau}). Moreover, we enlarged the spinor space to account for the two vectors $\vec{k}\pm\vec{Q}_i/2$ by introducing  \vec{\rho} Pauli matrices ac\-ting
in the latter $2\times 2$ subspace. The spinors for each of the blocks are defined in  $\vec{\tau}\otimes\vec{\rho}$ space as
\begin{eqnarray}
  \HPsi_{i,\vec{k}}\hc&=&\big(\psi_{\vec{k}+\vec{Q}_i/2,+}\hc\,,\psi_{\vec{k}-\vec{Q}_i/2,+}\hc\,,\nonumber\\
   &&\phantom{(}\psi_{-\vec{k}-\vec{Q}_i/2,+}\,,\psi_{-\vec{k}+\vec{Q}_i/2,+}\big)\,.
\end{eqnarray}
The eigenenergies are given by $\pm E_{i,\vec{k},s}$ with $s=\pm 1$,
\begin{eqnarray}
\hspace*{-4mm} E_{i,\vec{k},s} = \!\sqrt{\delta_{i,\vec{k},+}^2\!+\!\delta_{i,\vec{k},-}^2\!+\!\Delta^2 \!+\! M^2 |f_i(\vec{k})|^2 + s 2 \, T_i(\vec{k})} \quad
\end{eqnarray}
 and
\begin{eqnarray}
\hspace*{-4mm} T_i(\vec{k}) = \!\sqrt{\delta_{i,\vec{k},+}^2 \delta_{i,\vec{k},-}^2\!+\!M^2 |f_i(\vec{k})|^2 (\delta_{i,\vec{k},+}^2\!+\!\Delta^2)}\,.\nonumber
\end{eqnarray}
Note that each one of the above block Hamiltonians can support a single zero-energy Majorana mode when a gap in the dispersion closes at $\vec{k}=\vec{0}$
\cite{FuKane2008,AliceaReview,KSG,*Pekker}. If we use the approximate $\Uvec{M}_i$ given in \eqref{m-vector}, we find that $|f_i(\vec{0})|\approx\sqrt{2}$. Since
$\delta_{i,\vec{0},\pm}=0$, the criterion for the gap closing yielding a single Majorana mode, becomes
\begin{eqnarray}
\Delta=M|f_i(\vec{0})|\,,\label{eq:criterion}
\end{eqnarray}

\noindent i.e., $\Delta \approx \sqrt{2}M$. We conclude that the exact dependence of the magnetic order parameter on the pro\-xi\-mi\-ty-induced gap $\Delta$ is crucial
for deciding whether a TSC is indeed feasible in the particular heterostructure. For this reason, we determine self-consistently the dependence $M=M(\Delta)$ in the following
section.

\subsection{Magnetic gap in the presence of superconductivity -- Feasibility of a TSC}\label{sec:proximity:self-consistent}

In this section, we solve the self-consistency relation within the low-energy model derived in the previous paragraph and determine the influence of the superconduc\-ting gap
$\Delta$ on the surface magnetization, focusing on zero temperature since this is most relevant for Majorana fermion scenarios. The free
energy for each block of the Hamiltonian is given by
\begin{eqnarray}
\OpF_i=\frac{2 M^2}{U}-\frac{1}{2}\sum_{s=\pm} \int \difk{k}E_{i,\vec{k},s} \, ,
\end{eqnarray}
which yields the self-consistency relation
\begin{eqnarray}
M= U \sum_{s=\pm} \int \difk{k} \frac{M |f_i(\vec{k})|^2}{8E_{i,\vec{k},s}} \left[1 + \frac{\Delta^2 + \delta_{i,\vec{k},+}^2}{s T_i(\vec{k})}\right]\,.\qquad
\end{eqnarray}

\begin{figure}[t]
\includegraphics{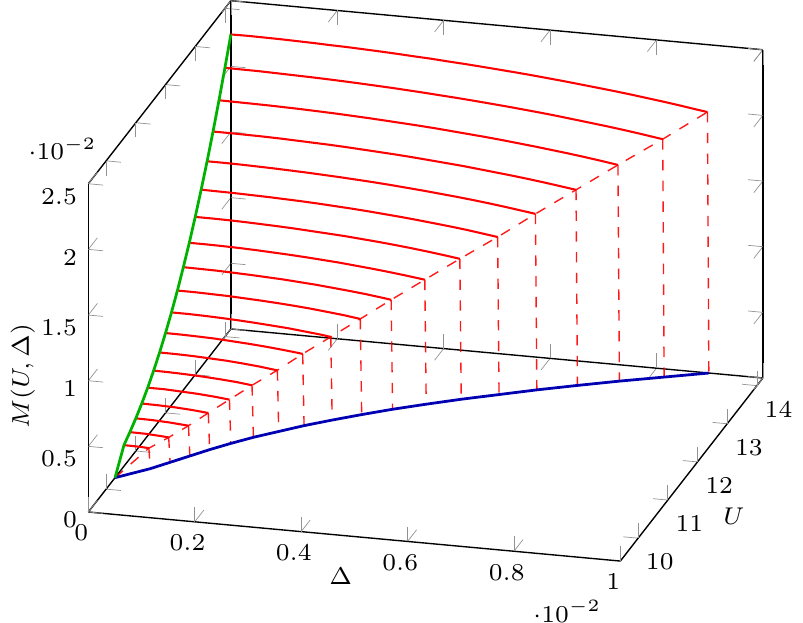}
\caption{Magnetization $M(U,\Delta)$ obtained as a solution of the self-consistency relation. We observe a first-order transition with increasing $\Delta$. When we consider
the TI \chem{Bi_2Te_3}  the interaction strength $U=13$ corresponds to $U=\SI{3.38}{eV}$, a superconducting gap of $\Delta=0.0025$ to $\Delta=\SI{0.65}{meV}$, and the magnetic
gap of $M \approx 0.01$ to $M=\SI{2.6}{meV}$. The blue line coincides with the instability obtained from the Stoner criterion, $U_{\text{crit}}=2/\chi_{\vec{Q}_1,++}^1$, and
the green curve illustrates the increase of the magnetization $M$ with $U$ for $\Delta=0$. Only the upper helicity band has been taken into account.}
\label{fig:selfconsistent}
\end{figure}

For varying strength of the interaction $U$ and $\Delta$, the resulting strength $M(U,\Delta)$ is depicted in \figref{selfconsistent}. Due to imperfect ne\-sting a minimum
strength of the interaction is required to induce a magnetic gap. We also va\-ried the chemical potential and found that the magnetic gap remains almost unchanged for up to
5\% detuning from $\mu_{\text{hex}}$. For larger va\-lues, the FS strongly differs from the hexagonal surface, and additional nesting vectors become important (see
\appendixref{susc-high-symmetry}). Upon increasing $\Delta$ the magnetic order parameter shows a first-order phase transition. This is expected due to the different
pro\-per\-ties of the proximity-induced singlet gap $\Delta$ and the magnetic order parameter $M$ under \OpT. Strictly speaking the jump of $M(U,\Delta)$ marks the limit up to
which a stable or metastable nonzero solution is found. Here we did not compare the corresponding values of the free energy to precisely determine the transition point, but
we do not expect the qualitative picture to be changed. A similar phase transition has been found before for a model with spatially modulated superconducting pairing terms
\cite{Aperis}.

\begin{figure}[t]
\includegraphics{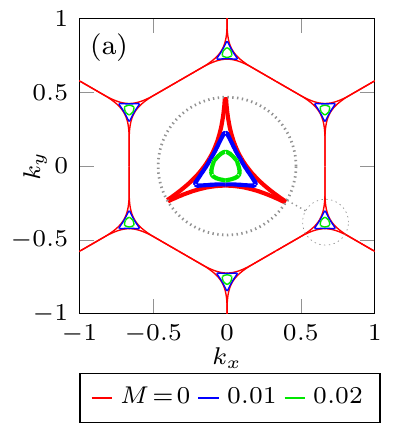}
\includegraphics{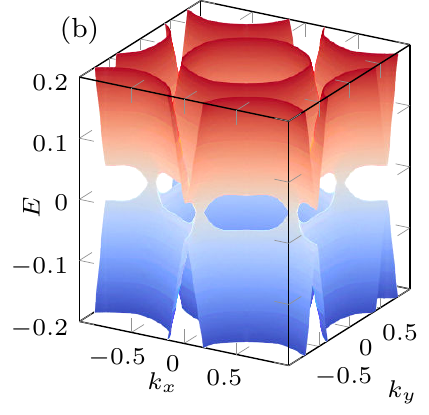}
\caption{(a) Extended scheme of the FS for $\mu=\mu_{\text{hex}}$ and $\Delta=0$ for $M=0,0.01,0.02$ (red, blue, and green, respectively). (b) Partly gapped
band structure for $M=0.02$.}
\label{fig:reconstruction}
\end{figure}

When the superconducting gap is nonzero, $\Delta\ne 0$, and below the critical value given by \eqref{criterion}, the spectrum is fully gapped. In \figref{reconstruction}
we display the reconstructed FS and band structure for $\Delta=0$ but a nonvanishing value of the magnetic gap. We notice that the magnetic ground state is associated with a
FS which has gapped and ungapped parts. The ungapped parts arise as consequence of the imperfect nesting and grow when we tune the chemical potential away from the optimal
value $\mu_{\text{hex}}$ producing the most perfect hexagonal FS. Note that $\vec{Q}_1(k_0)$ depends on $\mu$, i.e. it follows the chemical potential. The features of the
band structure should be obser\-vable in angular resolved photoemission spectroscopy.

To infer whether Majorana fermions may exist in the given hybrid structure we need to examine whether the ine\-qua\-li\-ty $|\Delta|\le\sqrt{2}M(U,\Delta)$ can be satisfied, where
the magnetic gap $M=M(U, \Delta)$ depends on the interaction $U$ and $\Delta$. If we work at $\mu=\mu_{\text{hex}}$ with an induced superconducting gap $\Delta=0.0025$
(corresponding to \SI{0.65}{meV} for \chem{Bi_2Te_3}), we find a required strength of the interaction of $U \gtrsim 13$ (corresponding to $U \gtrsim \SI{3.38}{eV}$ for
\chem{Bi_2Te_3}) and a magnetic gap of $M\approx 0.01$ (\SI{2.6}{meV}). Since both values appear accessible we conclude that the considered
he\-te\-ro\-stru\-ctu\-re can indeed support Majorana fermions and can be used to engineer a \group{C_3}-symmetric TSC
\cite{KotetesClassi,KaneMirror,*Shinsei,*SatoU1,*SatoU2,*TriDirac}.

\section{Control and fingerprints of the magnetic phases}\label{sec:external}

To further study the warping-driven magnetic phases we explore their properties under the influence of external fields. Here we focus on the triple-\vec{Q} phases, which under
the symmetry conditions studied so far are favored compared to the single-\vec{Q} phases. The latter can become re\-le\-vant if \group{C_3} symmetry-breaking fields are
introduced. In \secref{external:tailor} we show that a perpendicular magnetic field $B_z$ provides the most direct way for controlling the skyrmion charge without
violating \group{C_3} symmetry. In \secref{external:detect} we discuss the influence of an imposed supercurrent on the proximity-coupled TI. Based on symmetry arguments we
note a tendency that magnetic skyrmion phases induce distinctive multipolar patterns of supercurrents, which might serve as signatures of the magnetic phases.

\subsection{Tailoring the magnetic phase diagram}\label{sec:external:tailor}

Given the \group{C_{3v}} point group symmetry of the TI surface, we perform a symmetry classification of prominent external fields which can tailor the nature of the magnetic
ground state. In \tableref{fields} we categorize the electric $\vec{E}(\vec{r})$ and magnetic $\vec{B}(\vec{r})$ fields according to the IRs of the point group. We also
included a symmetry classification for the supercurrent vector $\vec{J}(\vec{r})\propto\nabla\phi(\vec{r})$, which is relevant when the TI is in proximity to a bulk
superconductor with order parameter $\Delta(\vec{r})=\Delta e^{\imath\phi(\vec{r})}$. As long as \vec{B} and \vec{J} are static and spatially homogeneous, they can only
influence the magnetic phase diagram by coupling to $\varPhi_z$. In contrast, when a constant electric field \vec{E} is applied, a term of the form
$\varPhi_xE_x+\varPhi_yE_y$ is allowed.

External fields coupling to $\varPhi_z$  determine the favored triple-$\vec{Q}$ phase. The only terms of the Landau expansion of \secref{magnetic:landau} in which $\varPhi_z$
appears  have the form $(M_1M_2M_3)^n=|M_1M_2M_3|e^{\imath n\varPhi_z}$ and $(M_1^*M_2^*M_3^*)^n=|M_1M_2M_3|e^{-\imath n\varPhi_z}$ with $n\in \NN$. As discussed in that
section, as a result of \OpT symmetry only terms with even $n$ are allowed. Consequently the lowest order accessible term  was $2\eta|M_1M_2M_3|^2\cos(2\varPhi_z)$. The
modification of $\eta$, which potentially can arise as consequence of appro\-pria\-te \OpT-symmetric external fields, allows switching between the possible triple-\vec{Q}
phases and thus controlling the skyrmion charge \OpC.

In contrast, by applying \OpT-violating external fields of the types shown in \tableref{combos} one can influence the phase diagram already at lower orders, ren\-de\-ring the
sixth-order terms and the value of $\eta$ irrelevant near the magnetic phase boun\-daries where the order parameters are small. The third-order terms
$|M_1M_2M_3|\cos\varPhi_z$ and $|M_1M_2M_3|\sin\varPhi_z$, which belong to the $A_1$ and $A_2$ representations, respectively, can couple to \OpT symmetry breaking combinations of
external fields. Fields which couple to the term $\cos\varPhi_z$ ($\sin\varPhi_z$) will stabilize the triple-\vec{Q} phases with $\varPhi_z=0,\pi$ ($\varPhi_z=\pm\pi/2$).

\begin{table}[b]
\caption{Symmetry classification of the phase degrees of freedom \vec{\varPhi}, the electric field \vec{E}, the magnetic field \vec{B} and the supercurrent \vec{J}, according
to IRs of \group{C_{3v}} and their behavior under time reversal \OpT.}
  \label{table:fields}
  \begin{ruledtabular}
    \begin{tabular}{>{$}l<{$}|>{$}r<{$}>{$}r<{$}>{$}r<{$}|>{$}l<{$}|>{$}l<{$}}
      \text{IR} & \II & 2C_3 & 3\sigma_v & \OpT=+1 & \OpT=-1\\
      \hline
      A_1&       1 &    \phantom{-}1 &    \phantom{-}1 & E_z &J_z\\
      A_2&       1 &    \phantom{-}1 &              -1 & \varPhi_z & B_z\\
      E &        2 &              -1 &               0 & (\varPhi_x,\varPhi_y),\,(E_x,E_y)&(B_x,B_y),\,(J_x,J_y)\\
    \end{tabular}
  \end{ruledtabular}

\caption{Combinations of magnetic field \vec{B} and supercurrent \vec{J} which couple to $|M_1M_2M_3|\cos\varPhi_z$ or $|M_1M_2M_3|\sin\varPhi_z$. The presence of the
respective terms can select the triple-\vec{Q} magnetic ground state. The $\cos\varPhi_z$ ($\sin\varPhi_z$) establishes the $\varPhi_z=0,\pi$ ($\varPhi_z=\pm\pi/2$)
magnetic ground state.}
  \label{table:combos}
  \begin{ruledtabular}
    \begin{tabular}{>{$}l<{$}|>{$}l<{$}}
      |M_1M_2M_3|\cos\varPhi_z\cdot&|M_1M_2M_3|\sin\varPhi_z\cdot\\
      \hline
      B_x(B_x^2-3B_y^2)           & B_z,\, B_z^3,\, B_z(B_x^2+B_y^2)\\
      J_y(J_y^2-3J_x^2)           & B_y(B_y^2-3B_x^2),\, J_x(J_x^2-3J_y^2)\\
      2J_xJ_yB_y-(J_x^2-J_y^2)B_x & 2J_xJ_yB_x+(J_x^2-J_y^2)B_y\\
      2B_xB_yJ_x+(B_x^2-B_y^2)J_y & 2B_xB_yJ_y-(B_x^2-B_y^2)J_x\\
    \end{tabular}
  \end{ruledtabular}
\end{table}

For instance, an external perpendicular magnetic field $B_z$ only couples to the $A_2$ term, contributing to the Landau expansion the term $|M_1M_2M_3|\sin\varPhi_zB_z$.
At the extrema it reduces to $\OpC B_z$, where $\OpC=\pm 1$. That is, a perpendicular field $B_z$, which is expe\-ri\-men\-tally easily accessible, directly couples to the
skyrmion charge and enables its manipulation. It also stabilizes the topologically nontrivial skyrmion phase, with the important  consequence that near the phase boun\-dary
even a weak field can impose the skyrmion phases with $\OpC=\pm 1$, although the FS may de\-via\-te from the hexagonal shape.

Constant in-plane magnetic fields $B_{x,y}$ or imposed supercurrents $J_{x,y}$ break the \group{C_3} symmetry and favor single-\vec{Q} (stripe) phases.

From a symmetry perspective, accor\-ding to \tableref{combos}, a field distribution with nonzero value of $B_x(B_x^2-3B_y^2)$ or $J_x(J_x^2-3J_y^2)$ would stabilize the $A_1$
or $A_2$ phase with $\OpC=0$ or $\OpC=\pm1$, respectively. Similar conclusions can be reached for the remaining third-order combinations of external fields presented in
\tableref{combos}, which preserve \group{C_3} symmetry but violate \OpT symmetry. While this may be of use in principle, it has probably little practical relevance. It will be
difficult to ge\-ne\-rate such third-order moments of the field distributions without producing also the linear \group{C_3} symmetry breaking fields $B_{x,y}$ or $J_{x,y}$.
This strongly restricts the potential of multipolar field distributions as knobs for manipulations. Nonetheless, as we discuss in the next paragraph, they can be potentially
valuable as means of detecting the magnetic skyrmion phases.

\subsection{Supercurrent signature of skyrmions}\label{sec:external:detect}

The spin-orbit coupling on the TI surface implies that an in-plane supercurrent yields an effective in-plane Zeeman field $\uvec{z}\times\vec{J}$ \cite{KotetesClassi},
while the octupolar moment $J_x(J_x^2-3J_y^2)$ produces a $B_z$ component. This follows from \eqref{Hamiltonian0} after the replacement
$\vec{k}\rightarrow\vec{k}+\vec{J}\tau_z/2$, which leads to
\begin{eqnarray}
&&\vec{B}_{\text{eff}}(\vec{k})=-\frac{J_y}{2}\uvec{x}+\frac{J_x}{2}\uvec{y}+\frac{J_x}{8}(J_x^2-3J_y^2)\uvec{z}\nonumber\\
&&+\frac{J_x}{2}(k_x^2-3k_y^2)\uvec{z}+k_x(k_xJ_x-3k_yJ_y)\uvec{z}\,.\label{eq:Beff}
\end{eqnarray}
This effective magnetic field couples to the magnetic order parameters and allows controlling the magnetic phase as discussed before. However the coupling also has
consequences in the reverse direction. Rather than manipulating the magnetic order by applied fields, we can look for the influence of the magnetic order onto the fields.

In particular, a magnetic order $\vec{M}(\vec{r})$ can induce a supercurrent $\vec{J}(\vec{r})$ which could serve as a probe of the magnetic order. Most significant is the
induced octupolar distribution $\OpJ(\vec{r})\equiv J_x(\vec{r})[J_x^2(\vec{r})-3J_y^2(\vec{r})]$, which is the lowest order term providing information concerning the skyrmion
charge.

\begin{figure}[t]
\includegraphics{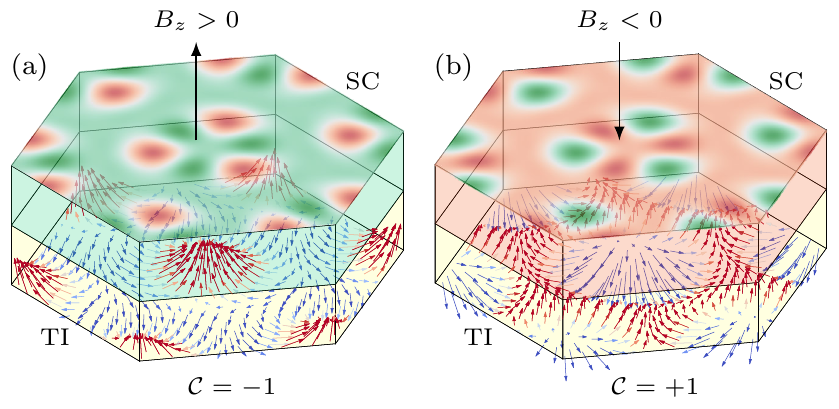}
\caption{Distinctive pattern of the supercurrent distribution $\OpJ(\vec{r})$ which might be induced in a conventional superconductor placed on top of a TI in the magnetic
skyrmion phase. A weak perpendicular magnetic field $B_z>0$ ($<0$) selects the skyrmion charge $\OpC=-1$ ($+1$) as shown in (a) and (b). The quantity
$\OpJ_{\vec{0}}=\int_{\text{UC}}\difv{r}\OpJ(\vec{r})$ is proportional to \OpC and can serve as a signature of the magnetic state. Note that due to scree\-ning effects the
pat\-tern observed in experiments could deviate from  what is plotted. However, smooth modifications would not alter the topological invariant quantity $\OpJ_{\vec{0}}$, as
long as the value of the skyrmion charge remains unchanged.}
\label{fig:octupole}
\end{figure}

At this point we establish the connection between $\OpJ(\vec{r})$ and the skyrmion charge \OpC. The integral of the octupolar supercurrent distribution $\OpJ(\vec{r})$ over
the unit cell, $\OpJ_{\vec{0}}\equiv\int_{\text{UC}}\difv{r}\OpJ(\vec{r})$, becomes nonzero only in the topologically nontrivial skyrmion phases, and in par\-ti\-cu\-lar
$\OpC=\sign\OpJ_{\vec{0}}$. The relation stems from the identical transformation behavior of $\OpJ_{\vec{0}}$ and the skyrmion charge $\OpC$ under
\group{C_{3v}} operations, which allows a coupling of the form $\OpJ_{\vec{0}}\OpC$. Thus \OpC acts as a source term for $\OpJ_{\vec{0}}$; i.e., $\OpJ_{\vec{0}}\propto\OpC$.

The plot in \figref{octupole} illustrates the potential effect of the magnetic skyrmion structure on the profile of $\OpJ(\vec{r})$. It is obtained without taking into account
the unavoi\-dable screening effects and without describing the superconducting properties in proximity to the TI fully and self-consistently. However, we expect a trend in the
direction shown. At lowest order in the magnetization we obtain $\vec{J}(\vec{r})\propto\uvec{z} \times \vec{M}(\vec{r})$, which yields the supercurrent distribution
\begin{eqnarray}
  \OpJ(\vec{r})\propto M_y(\vec{r})\left[M_y^2(\vec{r})-3M_x^2(\vec{r})\right]\,.
\end{eqnarray}

\noindent The real distribution may deviate due to screening effects, but the value $\OpJ_{\vec{0}}$ is a topological invariant and cannot change as long as the magnetic
ground state is not modified.

\section{Conclusion}\label{sec:conclusion}

We have investigated the magnetic instabilities which can spontaneously develop in the electronic surface states of a topological insulator with strong hexagonal war\-ping
effects. The latter can lead to a well-nested hexagonal Fermi surface and magnetic instability towards the formation of a nontrivial magnetic order. Since the nesting is not
perfect the transition only occurs above a critical strength of a Hubbard interaction. By a\-na\-ly\-zing a Landau theory we determined which ones of the compe\-ting magnetic
phases are realized for varying values of the chemical potential, thus exten\-ding earlier work. The Landau expansion at fourth order revealed that the single-\vec{Q} phase is
not favored; it could become favored if the \group{C_3} symmetry is broken, e.g., by applied field. Otherwise the system develops one of two types of triple-\vec{Q}
magnetic phases, which differ in the value of the skyrmion charge \OpC. By tuning the chemical potential one can switch between the different triple-\vec{Q} phases with
$\OpC=0,\pm1$. For the chemical potential with most perfectly hexagonally warped Fermi surface the topo\-lo\-gi\-cal\-ly nontrivial triple-\vec{Q} skyrmion phase with
$\OpC=\pm1$ is favored.

In addition we discussed new perspectives derived from spontaneous magnetic ordering on the surface of the topological insulator. In par\-ti\-cular, we showed that placing the
magnetized surface in proximity to a conventional superconductor allows engineering a \group{C_3}-symmetric topological superconductor, and we demonstrated the feasibility of
such a topological state for realistic values of the pro\-xi\-mi\-ty induced superconducting gap. The presence of the unitary symmetry allows introducing additional
topolo\-gi\-cal invariants and is expected to lead to multiple Majorana fermions trapped at a single defect. Moreover, switching on and off the unitary \group{C_3} symmetry
can assist implemen\-ting topological quantum information proces\-sing protocols \cite{KotetesClassi}.

With an eye on developing functional devices relying on the magnetically ordered topological insulator surface, we further investigated the modification of the magnetic phase
diagram when a perpendicular magnetic field is applied. By exploiting a quantum phase transition between two topologically distinct magnetic phases, we can create a skyrmion
switch, which can be valuable for (quantum) information storage, differing from other experimentally developed platforms \cite{Wiesendanger}. In addition the presence of a
skyrmion may lead to a characteristic pattern of supercurrents which can serve as a signature.

\begin{acknowledgments}

We thank A. Heimes, K. Rogdakis, G. Varelogiannis, A. Aperis, A. Shnirman, L. Molenkamp, E. K\"onig, P. Orth and M. Scheurer for valuable discussions.
\end{acknowledgments}

\bibliography{refs}

\begin{thebibliography}{61}%
\makeatletter
\providecommand \@ifxundefined [1]{%
 \@ifx{#1\undefined}
}%
\providecommand \@ifnum [1]{%
 \ifnum #1\expandafter \@firstoftwo
 \else \expandafter \@secondoftwo
 \fi
}%
\providecommand \@ifx [1]{%
 \ifx #1\expandafter \@firstoftwo
 \else \expandafter \@secondoftwo
 \fi
}%
\providecommand \natexlab [1]{#1}%
\providecommand \enquote  [1]{``#1''}%
\providecommand \bibnamefont  [1]{#1}%
\providecommand \bibfnamefont [1]{#1}%
\providecommand \citenamefont [1]{#1}%
\providecommand \href@noop [0]{\@secondoftwo}%
\providecommand \href [0]{\begingroup \@sanitize@url \@href}%
\providecommand \@href[1]{\@@startlink{#1}\@@href}%
\providecommand \@@href[1]{\endgroup#1\@@endlink}%
\providecommand \@sanitize@url [0]{\catcode `\\12\catcode `\$12\catcode
  `\&12\catcode `\#12\catcode `\^12\catcode `\_12\catcode `\%12\relax}%
\providecommand \@@startlink[1]{}%
\providecommand \@@endlink[0]{}%
\providecommand \url  [0]{\begingroup\@sanitize@url \@url }%
\providecommand \@url [1]{\endgroup\@href {#1}{\urlprefix }}%
\providecommand \urlprefix  [0]{URL }%
\providecommand \Eprint [0]{\href }%
\providecommand \doibase [0]{http://dx.doi.org/}%
\providecommand \selectlanguage [0]{\@gobble}%
\providecommand \bibinfo  [0]{\@secondoftwo}%
\providecommand \bibfield  [0]{\@secondoftwo}%
\providecommand \translation [1]{[#1]}%
\providecommand \BibitemOpen [0]{}%
\providecommand \bibitemStop [0]{}%
\providecommand \bibitemNoStop [0]{.\EOS\space}%
\providecommand \EOS [0]{\spacefactor3000\relax}%
\providecommand \BibitemShut  [1]{\csname bibitem#1\endcsname}%
\let\auto@bib@innerbib\@empty
\bibitem [{\citenamefont {Bernevig}\ \emph {et~al.}(2006)\citenamefont
  {Bernevig}, \citenamefont {Hughes},\ and\ \citenamefont {Zhang}}]{BHZ}%
  \BibitemOpen
  \bibfield  {author} {\bibinfo {author} {\bibfnamefont {B.~A.}\ \bibnamefont
  {Bernevig}}, \bibinfo {author} {\bibfnamefont {T.~L.}\ \bibnamefont
  {Hughes}}, \ and\ \bibinfo {author} {\bibfnamefont {S.-C.}\ \bibnamefont
  {Zhang}},\ }\href {\doibase 10.1126/science.1133734} {\bibfield  {journal}
  {\bibinfo  {journal} {Science}\ }\textbf {\bibinfo {volume} {314}},\ \bibinfo
  {pages} {1757} (\bibinfo {year} {2006})}\BibitemShut {NoStop}%
\bibitem [{\citenamefont {Fu}\ and\ \citenamefont {Kane}(2007)}]{Fu-Kane}%
  \BibitemOpen
  \bibfield  {author} {\bibinfo {author} {\bibfnamefont {L.}~\bibnamefont
  {Fu}}\ and\ \bibinfo {author} {\bibfnamefont {C.~L.}\ \bibnamefont {Kane}},\
  }\href {\doibase 10.1103/physrevb.76.045302} {\bibfield  {journal} {\bibinfo
  {journal} {Phys. Rev. B}\ }\textbf {\bibinfo {volume} {76}},\ \bibinfo
  {pages} {045302} (\bibinfo {year} {2007})}\BibitemShut {NoStop}%
\bibitem [{\citenamefont {Zhang}\ \emph {et~al.}(2009)\citenamefont {Zhang},
  \citenamefont {Liu}, \citenamefont {Qi}, \citenamefont {Dai}, \citenamefont
  {Fang},\ and\ \citenamefont {Zhang}}]{HJZhang}%
  \BibitemOpen
  \bibfield  {author} {\bibinfo {author} {\bibfnamefont {H.}~\bibnamefont
  {Zhang}}, \bibinfo {author} {\bibfnamefont {C.-X.}\ \bibnamefont {Liu}},
  \bibinfo {author} {\bibfnamefont {X.-L.}\ \bibnamefont {Qi}}, \bibinfo
  {author} {\bibfnamefont {X.}~\bibnamefont {Dai}}, \bibinfo {author}
  {\bibfnamefont {Z.}~\bibnamefont {Fang}}, \ and\ \bibinfo {author}
  {\bibfnamefont {S.-C.}\ \bibnamefont {Zhang}},\ }\href {\doibase
  10.1038/nphys1270} {\bibfield  {journal} {\bibinfo  {journal} {Nat. Phys.}\
  }\textbf {\bibinfo {volume} {5}},\ \bibinfo {pages} {438} (\bibinfo {year}
  {2009})}\BibitemShut {NoStop}%
\bibitem [{\citenamefont {K{\"o}nig}\ \emph {et~al.}(2007)\citenamefont
  {K{\"o}nig}, \citenamefont {Wiedmann}, \citenamefont {Brune}, \citenamefont
  {Roth}, \citenamefont {Buhmann}, \citenamefont {Molenkamp}, \citenamefont
  {Qi},\ and\ \citenamefont {Zhang}}]{QSHE}%
  \BibitemOpen
  \bibfield  {author} {\bibinfo {author} {\bibfnamefont {M.}~\bibnamefont
  {K{\"o}nig}}, \bibinfo {author} {\bibfnamefont {S.}~\bibnamefont {Wiedmann}},
  \bibinfo {author} {\bibfnamefont {C.}~\bibnamefont {Brune}}, \bibinfo
  {author} {\bibfnamefont {A.}~\bibnamefont {Roth}}, \bibinfo {author}
  {\bibfnamefont {H.}~\bibnamefont {Buhmann}}, \bibinfo {author} {\bibfnamefont
  {L.~W.}\ \bibnamefont {Molenkamp}}, \bibinfo {author} {\bibfnamefont {X.-L.}\
  \bibnamefont {Qi}}, \ and\ \bibinfo {author} {\bibfnamefont {S.-C.}\
  \bibnamefont {Zhang}},\ }\href {\doibase 10.1126/science.1148047} {\bibfield
  {journal} {\bibinfo  {journal} {Science}\ }\textbf {\bibinfo {volume}
  {318}},\ \bibinfo {pages} {766} (\bibinfo {year} {2007})}\BibitemShut
  {NoStop}%
\bibitem [{\citenamefont {Hsieh}\ \emph {et~al.}(2008)\citenamefont {Hsieh},
  \citenamefont {Qian}, \citenamefont {Wray}, \citenamefont {Xia},
  \citenamefont {Hor}, \citenamefont {Cava},\ and\ \citenamefont
  {Hasan}}]{Hsieh1}%
  \BibitemOpen
  \bibfield  {author} {\bibinfo {author} {\bibfnamefont {D.}~\bibnamefont
  {Hsieh}}, \bibinfo {author} {\bibfnamefont {D.}~\bibnamefont {Qian}},
  \bibinfo {author} {\bibfnamefont {L.}~\bibnamefont {Wray}}, \bibinfo {author}
  {\bibfnamefont {Y.}~\bibnamefont {Xia}}, \bibinfo {author} {\bibfnamefont
  {Y.~S.}\ \bibnamefont {Hor}}, \bibinfo {author} {\bibfnamefont {R.~J.}\
  \bibnamefont {Cava}}, \ and\ \bibinfo {author} {\bibfnamefont {M.~Z.}\
  \bibnamefont {Hasan}},\ }\href {\doibase 10.1038/nature06843} {\bibfield
  {journal} {\bibinfo  {journal} {Nature}\ }\textbf {\bibinfo {volume} {452}},\
  \bibinfo {pages} {970} (\bibinfo {year} {2008})}\BibitemShut {NoStop}%
\bibitem [{\citenamefont {Hsieh}\ \emph
  {et~al.}(2009{\natexlab{a}})\citenamefont {Hsieh}, \citenamefont {Xia},
  \citenamefont {Wray}, \citenamefont {Qian}, \citenamefont {Pal},
  \citenamefont {Dil}, \citenamefont {Osterwalder}, \citenamefont {Meier},
  \citenamefont {Bihlmayer}, \citenamefont {Kane}, \citenamefont {Hor},
  \citenamefont {Cava},\ and\ \citenamefont {Hasan}}]{Hsieh2}%
  \BibitemOpen
  \bibfield  {author} {\bibinfo {author} {\bibfnamefont {D.}~\bibnamefont
  {Hsieh}}, \bibinfo {author} {\bibfnamefont {Y.}~\bibnamefont {Xia}}, \bibinfo
  {author} {\bibfnamefont {L.}~\bibnamefont {Wray}}, \bibinfo {author}
  {\bibfnamefont {D.}~\bibnamefont {Qian}}, \bibinfo {author} {\bibfnamefont
  {A.}~\bibnamefont {Pal}}, \bibinfo {author} {\bibfnamefont {J.~H.}\
  \bibnamefont {Dil}}, \bibinfo {author} {\bibfnamefont {J.}~\bibnamefont
  {Osterwalder}}, \bibinfo {author} {\bibfnamefont {F.}~\bibnamefont {Meier}},
  \bibinfo {author} {\bibfnamefont {G.}~\bibnamefont {Bihlmayer}}, \bibinfo
  {author} {\bibfnamefont {C.~L.}\ \bibnamefont {Kane}}, \bibinfo {author}
  {\bibfnamefont {Y.~S.}\ \bibnamefont {Hor}}, \bibinfo {author} {\bibfnamefont
  {R.~J.}\ \bibnamefont {Cava}}, \ and\ \bibinfo {author} {\bibfnamefont
  {M.~Z.}\ \bibnamefont {Hasan}},\ }\href {\doibase 10.1126/science.1167733}
  {\bibfield  {journal} {\bibinfo  {journal} {Science}\ }\textbf {\bibinfo
  {volume} {323}},\ \bibinfo {pages} {919} (\bibinfo {year}
  {2009}{\natexlab{a}})}\BibitemShut {NoStop}%
\bibitem [{\citenamefont {Chen}\ \emph {et~al.}(2009)\citenamefont {Chen},
  \citenamefont {Analytis}, \citenamefont {Chu}, \citenamefont {Liu},
  \citenamefont {Mo}, \citenamefont {Qi}, \citenamefont {Zhang}, \citenamefont
  {Lu}, \citenamefont {Dai}, \citenamefont {Fang}, \citenamefont {Zhang},
  \citenamefont {Fisher}, \citenamefont {Hussain},\ and\ \citenamefont
  {Shen}}]{Chen}%
  \BibitemOpen
  \bibfield  {author} {\bibinfo {author} {\bibfnamefont {Y.~L.}\ \bibnamefont
  {Chen}}, \bibinfo {author} {\bibfnamefont {J.~G.}\ \bibnamefont {Analytis}},
  \bibinfo {author} {\bibfnamefont {J.-H.}\ \bibnamefont {Chu}}, \bibinfo
  {author} {\bibfnamefont {Z.~K.}\ \bibnamefont {Liu}}, \bibinfo {author}
  {\bibfnamefont {S.-K.}\ \bibnamefont {Mo}}, \bibinfo {author} {\bibfnamefont
  {X.~L.}\ \bibnamefont {Qi}}, \bibinfo {author} {\bibfnamefont {H.~J.}\
  \bibnamefont {Zhang}}, \bibinfo {author} {\bibfnamefont {D.~H.}\ \bibnamefont
  {Lu}}, \bibinfo {author} {\bibfnamefont {X.}~\bibnamefont {Dai}}, \bibinfo
  {author} {\bibfnamefont {Z.}~\bibnamefont {Fang}}, \bibinfo {author}
  {\bibfnamefont {S.~C.}\ \bibnamefont {Zhang}}, \bibinfo {author}
  {\bibfnamefont {I.~R.}\ \bibnamefont {Fisher}}, \bibinfo {author}
  {\bibfnamefont {Z.}~\bibnamefont {Hussain}}, \ and\ \bibinfo {author}
  {\bibfnamefont {Z.-X.}\ \bibnamefont {Shen}},\ }\href {\doibase
  10.1126/science.1173034} {\bibfield  {journal} {\bibinfo  {journal}
  {Science}\ }\textbf {\bibinfo {volume} {325}},\ \bibinfo {pages} {178}
  (\bibinfo {year} {2009})}\BibitemShut {NoStop}%
\bibitem [{\citenamefont {Xia}\ \emph {et~al.}(2009)\citenamefont {Xia},
  \citenamefont {Qian}, \citenamefont {Hsieh}, \citenamefont {Wray},
  \citenamefont {Pal}, \citenamefont {Lin}, \citenamefont {Bansil},
  \citenamefont {Grauer}, \citenamefont {Hor}, \citenamefont {Cava},\ and\
  \citenamefont {Hasan}}]{Xia}%
  \BibitemOpen
  \bibfield  {author} {\bibinfo {author} {\bibfnamefont {Y.}~\bibnamefont
  {Xia}}, \bibinfo {author} {\bibfnamefont {D.}~\bibnamefont {Qian}}, \bibinfo
  {author} {\bibfnamefont {D.}~\bibnamefont {Hsieh}}, \bibinfo {author}
  {\bibfnamefont {L.}~\bibnamefont {Wray}}, \bibinfo {author} {\bibfnamefont
  {A.}~\bibnamefont {Pal}}, \bibinfo {author} {\bibfnamefont {H.}~\bibnamefont
  {Lin}}, \bibinfo {author} {\bibfnamefont {A.}~\bibnamefont {Bansil}},
  \bibinfo {author} {\bibfnamefont {D.}~\bibnamefont {Grauer}}, \bibinfo
  {author} {\bibfnamefont {Y.~S.}\ \bibnamefont {Hor}}, \bibinfo {author}
  {\bibfnamefont {R.~J.}\ \bibnamefont {Cava}}, \ and\ \bibinfo {author}
  {\bibfnamefont {M.~Z.}\ \bibnamefont {Hasan}},\ }\href {\doibase
  10.1038/nphys1274} {\bibfield  {journal} {\bibinfo  {journal} {Nat. Phys.}\
  }\textbf {\bibinfo {volume} {5}},\ \bibinfo {pages} {398} (\bibinfo {year}
  {2009})}\BibitemShut {NoStop}%
\bibitem [{\citenamefont {Br{\"u}ne}\ \emph {et~al.}(2011)\citenamefont
  {Br{\"u}ne}, \citenamefont {Liu}, \citenamefont {Novik}, \citenamefont
  {Hankiewicz}, \citenamefont {Buhmann}, \citenamefont {Chen}, \citenamefont
  {Qi}, \citenamefont {Shen}, \citenamefont {Zhang},\ and\ \citenamefont
  {Molenkamp}}]{BrueneQHEBulkHgTe}%
  \BibitemOpen
  \bibfield  {author} {\bibinfo {author} {\bibfnamefont {C.}~\bibnamefont
  {Br{\"u}ne}}, \bibinfo {author} {\bibfnamefont {C.~X.}\ \bibnamefont {Liu}},
  \bibinfo {author} {\bibfnamefont {E.~G.}\ \bibnamefont {Novik}}, \bibinfo
  {author} {\bibfnamefont {E.~M.}\ \bibnamefont {Hankiewicz}}, \bibinfo
  {author} {\bibfnamefont {H.}~\bibnamefont {Buhmann}}, \bibinfo {author}
  {\bibfnamefont {Y.~L.}\ \bibnamefont {Chen}}, \bibinfo {author}
  {\bibfnamefont {X.~L.}\ \bibnamefont {Qi}}, \bibinfo {author} {\bibfnamefont
  {Z.~X.}\ \bibnamefont {Shen}}, \bibinfo {author} {\bibfnamefont {S.~C.}\
  \bibnamefont {Zhang}}, \ and\ \bibinfo {author} {\bibfnamefont {L.~W.}\
  \bibnamefont {Molenkamp}},\ }\href {\doibase 10.1103/physrevlett.106.126803}
  {\bibfield  {journal} {\bibinfo  {journal} {Phys. Rev. Lett.}\ }\textbf
  {\bibinfo {volume} {106}},\ \bibinfo {pages} {126803} (\bibinfo {year}
  {2011})}\BibitemShut {NoStop}%
\bibitem [{\citenamefont {Souma}\ \emph {et~al.}(2011)\citenamefont {Souma},
  \citenamefont {Kosaka}, \citenamefont {Sato}, \citenamefont {Komatsu},
  \citenamefont {Takayama}, \citenamefont {Takahashi}, \citenamefont {Kriener},
  \citenamefont {Segawa},\ and\ \citenamefont {Ando}}]{AndoSpinTexture}%
  \BibitemOpen
  \bibfield  {author} {\bibinfo {author} {\bibfnamefont {S.}~\bibnamefont
  {Souma}}, \bibinfo {author} {\bibfnamefont {K.}~\bibnamefont {Kosaka}},
  \bibinfo {author} {\bibfnamefont {T.}~\bibnamefont {Sato}}, \bibinfo {author}
  {\bibfnamefont {M.}~\bibnamefont {Komatsu}}, \bibinfo {author} {\bibfnamefont
  {A.}~\bibnamefont {Takayama}}, \bibinfo {author} {\bibfnamefont
  {T.}~\bibnamefont {Takahashi}}, \bibinfo {author} {\bibfnamefont
  {M.}~\bibnamefont {Kriener}}, \bibinfo {author} {\bibfnamefont
  {K.}~\bibnamefont {Segawa}}, \ and\ \bibinfo {author} {\bibfnamefont
  {Y.}~\bibnamefont {Ando}},\ }\href {\doibase 10.1103/physrevlett.106.216803}
  {\bibfield  {journal} {\bibinfo  {journal} {Phys. Rev. Lett.}\ }\textbf
  {\bibinfo {volume} {106}},\ \bibinfo {pages} {216803} (\bibinfo {year}
  {2011})}\BibitemShut {NoStop}%
\bibitem [{\citenamefont {Kane}\ and\ \citenamefont
  {Mele}(2005{\natexlab{a}})}]{KaneMele1}%
  \BibitemOpen
  \bibfield  {author} {\bibinfo {author} {\bibfnamefont {C.~L.}\ \bibnamefont
  {Kane}}\ and\ \bibinfo {author} {\bibfnamefont {E.~J.}\ \bibnamefont
  {Mele}},\ }\href {\doibase 10.1103/physrevlett.95.146802} {\bibfield
  {journal} {\bibinfo  {journal} {Phys. Rev. Lett.}\ }\textbf {\bibinfo
  {volume} {95}},\ \bibinfo {pages} {146802} (\bibinfo {year}
  {2005}{\natexlab{a}})}\BibitemShut {NoStop}%
\bibitem [{\citenamefont {Kane}\ and\ \citenamefont
  {Mele}(2005{\natexlab{b}})}]{KaneMele2}%
  \BibitemOpen
  \bibfield  {author} {\bibinfo {author} {\bibfnamefont {C.~L.}\ \bibnamefont
  {Kane}}\ and\ \bibinfo {author} {\bibfnamefont {E.~J.}\ \bibnamefont
  {Mele}},\ }\href {\doibase 10.1103/physrevlett.95.226801} {\bibfield
  {journal} {\bibinfo  {journal} {Phys. Rev. Lett.}\ }\textbf {\bibinfo
  {volume} {95}},\ \bibinfo {pages} {226801} (\bibinfo {year}
  {2005}{\natexlab{b}})}\BibitemShut {NoStop}%
\bibitem [{\citenamefont {Moore}\ and\ \citenamefont {Balents}(2007)}]{Moore}%
  \BibitemOpen
  \bibfield  {author} {\bibinfo {author} {\bibfnamefont {J.~E.}\ \bibnamefont
  {Moore}}\ and\ \bibinfo {author} {\bibfnamefont {L.}~\bibnamefont
  {Balents}},\ }\href {\doibase 10.1103/physrevb.75.121306} {\bibfield
  {journal} {\bibinfo  {journal} {Phys. Rev. B}\ }\textbf {\bibinfo {volume}
  {75}},\ \bibinfo {pages} {121306} (\bibinfo {year} {2007})}\BibitemShut
  {NoStop}%
\bibitem [{\citenamefont {Qi}\ \emph {et~al.}(2008)\citenamefont {Qi},
  \citenamefont {Hughes},\ and\ \citenamefont {Zhang}}]{ZhangTQFT}%
  \BibitemOpen
  \bibfield  {author} {\bibinfo {author} {\bibfnamefont {X.-L.}\ \bibnamefont
  {Qi}}, \bibinfo {author} {\bibfnamefont {T.~L.}\ \bibnamefont {Hughes}}, \
  and\ \bibinfo {author} {\bibfnamefont {S.-C.}\ \bibnamefont {Zhang}},\ }\href
  {\doibase 10.1103/physrevb.78.195424} {\bibfield  {journal} {\bibinfo
  {journal} {Phys. Rev. B}\ }\textbf {\bibinfo {volume} {78}},\ \bibinfo
  {pages} {195424} (\bibinfo {year} {2008})}\BibitemShut {NoStop}%
\bibitem [{\citenamefont {Roy}(2009)}]{Roy}%
  \BibitemOpen
  \bibfield  {author} {\bibinfo {author} {\bibfnamefont {R.}~\bibnamefont
  {Roy}},\ }\href {\doibase 10.1103/physrevb.79.195321} {\bibfield  {journal}
  {\bibinfo  {journal} {Phys. Rev. B}\ }\textbf {\bibinfo {volume} {79}},\
  \bibinfo {pages} {195321} (\bibinfo {year} {2009})}\BibitemShut {NoStop}%
\bibitem [{\citenamefont {Hasan}\ and\ \citenamefont {Kane}(2010)}]{KaneHasan}%
  \BibitemOpen
  \bibfield  {author} {\bibinfo {author} {\bibfnamefont {M.~Z.}\ \bibnamefont
  {Hasan}}\ and\ \bibinfo {author} {\bibfnamefont {C.~L.}\ \bibnamefont
  {Kane}},\ }\href {\doibase 10.1103/revmodphys.82.3045} {\bibfield  {journal}
  {\bibinfo  {journal} {Rev. Mod. Phys.}\ }\textbf {\bibinfo {volume} {82}},\
  \bibinfo {pages} {3045} (\bibinfo {year} {2010})}\BibitemShut {NoStop}%
\bibitem [{\citenamefont {Qi}\ and\ \citenamefont {Zhang}(2011)}]{ZhangQi}%
  \BibitemOpen
  \bibfield  {author} {\bibinfo {author} {\bibfnamefont {X.-L.}\ \bibnamefont
  {Qi}}\ and\ \bibinfo {author} {\bibfnamefont {S.-C.}\ \bibnamefont {Zhang}},\
  }\href {\doibase 10.1103/revmodphys.83.1057} {\bibfield  {journal} {\bibinfo
  {journal} {Rev. Mod. Phys.}\ }\textbf {\bibinfo {volume} {83}},\ \bibinfo
  {pages} {1057} (\bibinfo {year} {2011})}\BibitemShut {NoStop}%
\bibitem [{\citenamefont {Ando}(2013)}]{Ando}%
  \BibitemOpen
  \bibfield  {author} {\bibinfo {author} {\bibfnamefont {Y.}~\bibnamefont
  {Ando}},\ }\href {\doibase 10.7566/JPSJ.82.102001} {\bibfield  {journal}
  {\bibinfo  {journal} {J. Phys. Soc. Jpn.}\ }\textbf {\bibinfo {volume}
  {82}},\ \bibinfo {pages} {102001} (\bibinfo {year} {2013})}\BibitemShut
  {NoStop}%
\bibitem [{\citenamefont {Akhmerov}\ \emph {et~al.}(2009)\citenamefont
  {Akhmerov}, \citenamefont {Nilsson},\ and\ \citenamefont
  {Beenakker}}]{AkhmerovMFDetection}%
  \BibitemOpen
  \bibfield  {author} {\bibinfo {author} {\bibfnamefont {A.~R.}\ \bibnamefont
  {Akhmerov}}, \bibinfo {author} {\bibfnamefont {J.}~\bibnamefont {Nilsson}}, \
  and\ \bibinfo {author} {\bibfnamefont {C.~W.~J.}\ \bibnamefont {Beenakker}},\
  }\href {\doibase 10.1103/physrevlett.102.216404} {\bibfield  {journal}
  {\bibinfo  {journal} {Phys. Rev. Lett.}\ }\textbf {\bibinfo {volume} {102}},\
  \bibinfo {pages} {216404} (\bibinfo {year} {2009})}\BibitemShut {NoStop}%
\bibitem [{\citenamefont {Linder}\ \emph {et~al.}(2010)\citenamefont {Linder},
  \citenamefont {Tanaka}, \citenamefont {Yokoyama}, \citenamefont {Sudb{\o}},\
  and\ \citenamefont {Nagaosa}}]{Linder}%
  \BibitemOpen
  \bibfield  {author} {\bibinfo {author} {\bibfnamefont {J.}~\bibnamefont
  {Linder}}, \bibinfo {author} {\bibfnamefont {Y.}~\bibnamefont {Tanaka}},
  \bibinfo {author} {\bibfnamefont {T.}~\bibnamefont {Yokoyama}}, \bibinfo
  {author} {\bibfnamefont {A.}~\bibnamefont {Sudb{\o}}}, \ and\ \bibinfo
  {author} {\bibfnamefont {N.}~\bibnamefont {Nagaosa}},\ }\href {\doibase
  10.1103/PhysRevB.81.184525} {\bibfield  {journal} {\bibinfo  {journal} {Phys.
  Rev. B}\ }\textbf {\bibinfo {volume} {81}},\ \bibinfo {pages} {184525}
  (\bibinfo {year} {2010})}\BibitemShut {NoStop}%
\bibitem [{\citenamefont {Fu}\ and\ \citenamefont {Kane}(2008)}]{FuKane2008}%
  \BibitemOpen
  \bibfield  {author} {\bibinfo {author} {\bibfnamefont {L.}~\bibnamefont
  {Fu}}\ and\ \bibinfo {author} {\bibfnamefont {C.~L.}\ \bibnamefont {Kane}},\
  }\href {\doibase 10.1103/physrevlett.100.096407} {\bibfield  {journal}
  {\bibinfo  {journal} {Phys. Rev. Lett.}\ }\textbf {\bibinfo {volume} {100}},\
  \bibinfo {pages} {096407} (\bibinfo {year} {2008})}\BibitemShut {NoStop}%
\bibitem [{\citenamefont {Alicea}(2012)}]{AliceaReview}%
  \BibitemOpen
  \bibfield  {author} {\bibinfo {author} {\bibfnamefont {J.}~\bibnamefont
  {Alicea}},\ }\href {\doibase 10.1088/0034-4885/75/7/076501} {\bibfield
  {journal} {\bibinfo  {journal} {Rep. Prog. Phys.}\ }\textbf {\bibinfo
  {volume} {75}},\ \bibinfo {pages} {076501} (\bibinfo {year}
  {2012})}\BibitemShut {NoStop}%
\bibitem [{\citenamefont {Beenakker}(2013)}]{BeenakkerReview}%
  \BibitemOpen
  \bibfield  {author} {\bibinfo {author} {\bibfnamefont {C.}~\bibnamefont
  {Beenakker}},\ }\href {\doibase 10.1146/annurev-conmatphys-030212-184337}
  {\bibfield  {journal} {\bibinfo  {journal} {Annu. Rev. Con. Mat. Phys.}\
  }\textbf {\bibinfo {volume} {4}},\ \bibinfo {pages} {113} (\bibinfo {year}
  {2013})}\BibitemShut {NoStop}%
\bibitem [{\citenamefont {Kotetes}(2013)}]{KotetesClassi}%
  \BibitemOpen
  \bibfield  {author} {\bibinfo {author} {\bibfnamefont {P.}~\bibnamefont
  {Kotetes}},\ }\href {\doibase 10.1088/1367-2630/15/10/105027} {\bibfield
  {journal} {\bibinfo  {journal} {New J. Phys.}\ }\textbf {\bibinfo {volume}
  {15}},\ \bibinfo {pages} {105027} (\bibinfo {year} {2013})}\BibitemShut
  {NoStop}%
\bibitem [{\citenamefont {Sac{\'e}p{\'e}}\ \emph {et~al.}(2011)\citenamefont
  {Sac{\'e}p{\'e}}, \citenamefont {Oostinga}, \citenamefont {Li}, \citenamefont
  {Ubaldini}, \citenamefont {Couto}, \citenamefont {Giannini},\ and\
  \citenamefont {Morpurgo}}]{Morpurgo}%
  \BibitemOpen
  \bibfield  {author} {\bibinfo {author} {\bibfnamefont {B.}~\bibnamefont
  {Sac{\'e}p{\'e}}}, \bibinfo {author} {\bibfnamefont {J.~B.}\ \bibnamefont
  {Oostinga}}, \bibinfo {author} {\bibfnamefont {J.}~\bibnamefont {Li}},
  \bibinfo {author} {\bibfnamefont {A.}~\bibnamefont {Ubaldini}}, \bibinfo
  {author} {\bibfnamefont {N.~J.}\ \bibnamefont {Couto}}, \bibinfo {author}
  {\bibfnamefont {E.}~\bibnamefont {Giannini}}, \ and\ \bibinfo {author}
  {\bibfnamefont {A.~F.}\ \bibnamefont {Morpurgo}},\ }\href {\doibase
  10.1038/ncomms1586} {\bibfield  {journal} {\bibinfo  {journal} {Nat. Comm.}\
  }\textbf {\bibinfo {volume} {2}},\ \bibinfo {pages} {575} (\bibinfo {year}
  {2011})}\BibitemShut {NoStop}%
\bibitem [{\citenamefont {Williams}\ \emph {et~al.}(2012)\citenamefont
  {Williams}, \citenamefont {Bestwick}, \citenamefont {Gallagher},
  \citenamefont {Hong}, \citenamefont {Cui}, \citenamefont {Bleich},
  \citenamefont {Analytis}, \citenamefont {Fisher},\ and\ \citenamefont
  {Goldhaber-Gordon}}]{Analytis}%
  \BibitemOpen
  \bibfield  {author} {\bibinfo {author} {\bibfnamefont {J.~R.}\ \bibnamefont
  {Williams}}, \bibinfo {author} {\bibfnamefont {A.~J.}\ \bibnamefont
  {Bestwick}}, \bibinfo {author} {\bibfnamefont {P.}~\bibnamefont {Gallagher}},
  \bibinfo {author} {\bibfnamefont {S.~S.}\ \bibnamefont {Hong}}, \bibinfo
  {author} {\bibfnamefont {Y.}~\bibnamefont {Cui}}, \bibinfo {author}
  {\bibfnamefont {A.~S.}\ \bibnamefont {Bleich}}, \bibinfo {author}
  {\bibfnamefont {J.~G.}\ \bibnamefont {Analytis}}, \bibinfo {author}
  {\bibfnamefont {I.~R.}\ \bibnamefont {Fisher}}, \ and\ \bibinfo {author}
  {\bibfnamefont {D.}~\bibnamefont {Goldhaber-Gordon}},\ }\href {\doibase
  10.1103/physrevlett.109.056803} {\bibfield  {journal} {\bibinfo  {journal}
  {Phys. Rev. Lett.}\ }\textbf {\bibinfo {volume} {109}},\ \bibinfo {pages}
  {056803} (\bibinfo {year} {2012})}\BibitemShut {NoStop}%
\bibitem [{\citenamefont {Veldhorst}\ \emph {et~al.}(2012)\citenamefont
  {Veldhorst}, \citenamefont {Snelder}, \citenamefont {Hoek}, \citenamefont
  {Gang}, \citenamefont {Guduru}, \citenamefont {Wang}, \citenamefont
  {Zeitler}, \citenamefont {van~der Wiel}, \citenamefont {Golubov},
  \citenamefont {Hilgenkamp},\ and\ \citenamefont {Brinkman}}]{Brinkman}%
  \BibitemOpen
  \bibfield  {author} {\bibinfo {author} {\bibfnamefont {M.}~\bibnamefont
  {Veldhorst}}, \bibinfo {author} {\bibfnamefont {M.}~\bibnamefont {Snelder}},
  \bibinfo {author} {\bibfnamefont {M.}~\bibnamefont {Hoek}}, \bibinfo {author}
  {\bibfnamefont {T.}~\bibnamefont {Gang}}, \bibinfo {author} {\bibfnamefont
  {V.~K.}\ \bibnamefont {Guduru}}, \bibinfo {author} {\bibfnamefont {X.~L.}\
  \bibnamefont {Wang}}, \bibinfo {author} {\bibfnamefont {U.}~\bibnamefont
  {Zeitler}}, \bibinfo {author} {\bibfnamefont {W.~G.}\ \bibnamefont {van~der
  Wiel}}, \bibinfo {author} {\bibfnamefont {A.~A.}\ \bibnamefont {Golubov}},
  \bibinfo {author} {\bibfnamefont {H.}~\bibnamefont {Hilgenkamp}}, \ and\
  \bibinfo {author} {\bibfnamefont {A.}~\bibnamefont {Brinkman}},\ }\href
  {\doibase 10.1038/nmat3255} {\bibfield  {journal} {\bibinfo  {journal} {Nat.
  Mat.}\ }\textbf {\bibinfo {volume} {11}},\ \bibinfo {pages} {417} (\bibinfo
  {year} {2012})}\BibitemShut {NoStop}%
\bibitem [{\citenamefont {Qu}\ \emph {et~al.}(2012)\citenamefont {Qu},
  \citenamefont {Yang}, \citenamefont {Shen}, \citenamefont {Ding},
  \citenamefont {Chen}, \citenamefont {Ji}, \citenamefont {Liu}, \citenamefont
  {Fan}, \citenamefont {Jing}, \citenamefont {Yang},\ and\ \citenamefont
  {Lu}}]{Lu}%
  \BibitemOpen
  \bibfield  {author} {\bibinfo {author} {\bibfnamefont {F.}~\bibnamefont
  {Qu}}, \bibinfo {author} {\bibfnamefont {F.}~\bibnamefont {Yang}}, \bibinfo
  {author} {\bibfnamefont {J.}~\bibnamefont {Shen}}, \bibinfo {author}
  {\bibfnamefont {Y.}~\bibnamefont {Ding}}, \bibinfo {author} {\bibfnamefont
  {J.}~\bibnamefont {Chen}}, \bibinfo {author} {\bibfnamefont {Z.}~\bibnamefont
  {Ji}}, \bibinfo {author} {\bibfnamefont {G.}~\bibnamefont {Liu}}, \bibinfo
  {author} {\bibfnamefont {J.}~\bibnamefont {Fan}}, \bibinfo {author}
  {\bibfnamefont {X.}~\bibnamefont {Jing}}, \bibinfo {author} {\bibfnamefont
  {C.}~\bibnamefont {Yang}}, \ and\ \bibinfo {author} {\bibfnamefont
  {L.}~\bibnamefont {Lu}},\ }\href {\doibase 10.1038/srep00339} {\bibfield
  {journal} {\bibinfo  {journal} {Sci. Rep.}\ }\textbf {\bibinfo {volume}
  {2}},\ \bibinfo {pages} {339} (\bibinfo {year} {2012})}\BibitemShut {NoStop}%
\bibitem [{\citenamefont {Knez}\ \emph {et~al.}(2012)\citenamefont {Knez},
  \citenamefont {Du},\ and\ \citenamefont {Sullivan}}]{Du}%
  \BibitemOpen
  \bibfield  {author} {\bibinfo {author} {\bibfnamefont {I.}~\bibnamefont
  {Knez}}, \bibinfo {author} {\bibfnamefont {R.-R.}\ \bibnamefont {Du}}, \ and\
  \bibinfo {author} {\bibfnamefont {G.}~\bibnamefont {Sullivan}},\ }\href
  {\doibase 10.1103/PhysRevLett.109.186603} {\bibfield  {journal} {\bibinfo
  {journal} {Phys. Rev. Lett.}\ }\textbf {\bibinfo {volume} {109}},\ \bibinfo
  {pages} {186603} (\bibinfo {year} {2012})}\BibitemShut {NoStop}%
\bibitem [{\citenamefont {Oostinga}\ \emph {et~al.}(2013)\citenamefont
  {Oostinga}, \citenamefont {Maier}, \citenamefont {Sch{\"u}ffelgen},
  \citenamefont {Knott}, \citenamefont {Ames}, \citenamefont {Br{\"u}ne},
  \citenamefont {Tkachov}, \citenamefont {Buhmann},\ and\ \citenamefont
  {Molenkamp}}]{Molenkamp}%
  \BibitemOpen
  \bibfield  {author} {\bibinfo {author} {\bibfnamefont {J.~B.}\ \bibnamefont
  {Oostinga}}, \bibinfo {author} {\bibfnamefont {L.}~\bibnamefont {Maier}},
  \bibinfo {author} {\bibfnamefont {P.}~\bibnamefont {Sch{\"u}ffelgen}},
  \bibinfo {author} {\bibfnamefont {D.}~\bibnamefont {Knott}}, \bibinfo
  {author} {\bibfnamefont {C.}~\bibnamefont {Ames}}, \bibinfo {author}
  {\bibfnamefont {C.}~\bibnamefont {Br{\"u}ne}}, \bibinfo {author}
  {\bibfnamefont {G.}~\bibnamefont {Tkachov}}, \bibinfo {author} {\bibfnamefont
  {H.}~\bibnamefont {Buhmann}}, \ and\ \bibinfo {author} {\bibfnamefont
  {L.~W.}\ \bibnamefont {Molenkamp}},\ }\href {\doibase
  10.1103/physrevx.3.021007} {\bibfield  {journal} {\bibinfo  {journal} {Phys.
  Rev. X}\ }\textbf {\bibinfo {volume} {3}},\ \bibinfo {pages} {021007}
  (\bibinfo {year} {2013})}\BibitemShut {NoStop}%
\bibitem [{\citenamefont {Hart}\ \emph {et~al.}(2014)\citenamefont {Hart},
  \citenamefont {Ren}, \citenamefont {Wagner}, \citenamefont {Leubner},
  \citenamefont {Mühlbauer}, \citenamefont {Brüne}, \citenamefont {Buhmann},
  \citenamefont {Molenkamp},\ and\ \citenamefont {Yacoby}}]{Yacoby}%
  \BibitemOpen
  \bibfield  {author} {\bibinfo {author} {\bibfnamefont {S.}~\bibnamefont
  {Hart}}, \bibinfo {author} {\bibfnamefont {H.}~\bibnamefont {Ren}}, \bibinfo
  {author} {\bibfnamefont {T.}~\bibnamefont {Wagner}}, \bibinfo {author}
  {\bibfnamefont {P.}~\bibnamefont {Leubner}}, \bibinfo {author} {\bibfnamefont
  {M.}~\bibnamefont {Mühlbauer}}, \bibinfo {author} {\bibfnamefont
  {C.}~\bibnamefont {Brüne}}, \bibinfo {author} {\bibfnamefont
  {H.}~\bibnamefont {Buhmann}}, \bibinfo {author} {\bibfnamefont {L.~W.}\
  \bibnamefont {Molenkamp}}, \ and\ \bibinfo {author} {\bibfnamefont
  {A.}~\bibnamefont {Yacoby}},\ }\href {\doibase 10.1038/nphys3036} {\bibfield
  {journal} {\bibinfo  {journal} {Nat. Phys.}\ }\textbf {\bibinfo {volume}
  {10}},\ \bibinfo {pages} {638} (\bibinfo {year} {2014})}\BibitemShut
  {NoStop}%
\bibitem [{\citenamefont {Fu}(2009)}]{FuHex}%
  \BibitemOpen
  \bibfield  {author} {\bibinfo {author} {\bibfnamefont {L.}~\bibnamefont
  {Fu}},\ }\href {\doibase 10.1103/physrevlett.103.266801} {\bibfield
  {journal} {\bibinfo  {journal} {Phys. Rev. Lett.}\ }\textbf {\bibinfo
  {volume} {103}},\ \bibinfo {pages} {266801} (\bibinfo {year}
  {2009})}\BibitemShut {NoStop}%
\bibitem [{\citenamefont {Liu}\ \emph {et~al.}(2010)\citenamefont {Liu},
  \citenamefont {Qi}, \citenamefont {Zhang}, \citenamefont {Dai}, \citenamefont
  {Fang},\ and\ \citenamefont {Zhang}}]{ModelHamTI}%
  \BibitemOpen
  \bibfield  {author} {\bibinfo {author} {\bibfnamefont {C.-X.}\ \bibnamefont
  {Liu}}, \bibinfo {author} {\bibfnamefont {X.-L.}\ \bibnamefont {Qi}},
  \bibinfo {author} {\bibfnamefont {H.-J.}\ \bibnamefont {Zhang}}, \bibinfo
  {author} {\bibfnamefont {X.}~\bibnamefont {Dai}}, \bibinfo {author}
  {\bibfnamefont {Z.}~\bibnamefont {Fang}}, \ and\ \bibinfo {author}
  {\bibfnamefont {S.-C.}\ \bibnamefont {Zhang}},\ }\href {\doibase
  10.1103/PhysRevB.82.045122} {\bibfield  {journal} {\bibinfo  {journal} {Phys.
  Rev. B}\ }\textbf {\bibinfo {volume} {82}},\ \bibinfo {pages} {045122}
  (\bibinfo {year} {2010})}\BibitemShut {NoStop}%
\bibitem [{\citenamefont {Alpichshev}\ \emph {et~al.}(2010)\citenamefont
  {Alpichshev}, \citenamefont {Analytis}, \citenamefont {Chu}, \citenamefont
  {Fisher}, \citenamefont {Chen}, \citenamefont {Shen}, \citenamefont {Fang},\
  and\ \citenamefont {Kapitulnik}}]{STMofHex}%
  \BibitemOpen
  \bibfield  {author} {\bibinfo {author} {\bibfnamefont {Z.}~\bibnamefont
  {Alpichshev}}, \bibinfo {author} {\bibfnamefont {J.~G.}\ \bibnamefont
  {Analytis}}, \bibinfo {author} {\bibfnamefont {J.-H.}\ \bibnamefont {Chu}},
  \bibinfo {author} {\bibfnamefont {I.~R.}\ \bibnamefont {Fisher}}, \bibinfo
  {author} {\bibfnamefont {Y.~L.}\ \bibnamefont {Chen}}, \bibinfo {author}
  {\bibfnamefont {Z.~X.}\ \bibnamefont {Shen}}, \bibinfo {author}
  {\bibfnamefont {A.}~\bibnamefont {Fang}}, \ and\ \bibinfo {author}
  {\bibfnamefont {A.}~\bibnamefont {Kapitulnik}},\ }\href {\doibase
  10.1103/physrevlett.104.016401} {\bibfield  {journal} {\bibinfo  {journal}
  {Phys. Rev. Lett.}\ }\textbf {\bibinfo {volume} {104}},\ \bibinfo {pages}
  {016401} (\bibinfo {year} {2010})}\BibitemShut {NoStop}%
\bibitem [{\citenamefont {Wang}\ \emph {et~al.}(2011)\citenamefont {Wang},
  \citenamefont {Hsieh}, \citenamefont {Pilon}, \citenamefont {Fu},
  \citenamefont {Gardner}, \citenamefont {Lee},\ and\ \citenamefont
  {Gedik}}]{GedikHex}%
  \BibitemOpen
  \bibfield  {author} {\bibinfo {author} {\bibfnamefont {Y.~H.}\ \bibnamefont
  {Wang}}, \bibinfo {author} {\bibfnamefont {D.}~\bibnamefont {Hsieh}},
  \bibinfo {author} {\bibfnamefont {D.}~\bibnamefont {Pilon}}, \bibinfo
  {author} {\bibfnamefont {L.}~\bibnamefont {Fu}}, \bibinfo {author}
  {\bibfnamefont {D.~R.}\ \bibnamefont {Gardner}}, \bibinfo {author}
  {\bibfnamefont {Y.~S.}\ \bibnamefont {Lee}}, \ and\ \bibinfo {author}
  {\bibfnamefont {N.}~\bibnamefont {Gedik}},\ }\href {\doibase
  10.1103/physrevlett.107.207602} {\bibfield  {journal} {\bibinfo  {journal}
  {Phys. Rev. Lett.}\ }\textbf {\bibinfo {volume} {107}},\ \bibinfo {pages}
  {207602} (\bibinfo {year} {2011})}\BibitemShut {NoStop}%
\bibitem [{\citenamefont {Henk}\ \emph {et~al.}(2012)\citenamefont {Henk},
  \citenamefont {Flieger}, \citenamefont {Maznichenko}, \citenamefont {Mertig},
  \citenamefont {Ernst}, \citenamefont {Eremeev},\ and\ \citenamefont
  {Chulkov}}]{Chulkov}%
  \BibitemOpen
  \bibfield  {author} {\bibinfo {author} {\bibfnamefont {J.}~\bibnamefont
  {Henk}}, \bibinfo {author} {\bibfnamefont {M.}~\bibnamefont {Flieger}},
  \bibinfo {author} {\bibfnamefont {I.~V.}\ \bibnamefont {Maznichenko}},
  \bibinfo {author} {\bibfnamefont {I.}~\bibnamefont {Mertig}}, \bibinfo
  {author} {\bibfnamefont {A.}~\bibnamefont {Ernst}}, \bibinfo {author}
  {\bibfnamefont {S.~V.}\ \bibnamefont {Eremeev}}, \ and\ \bibinfo {author}
  {\bibfnamefont {E.~V.}\ \bibnamefont {Chulkov}},\ }\href {\doibase
  10.1103/physrevlett.109.076801} {\bibfield  {journal} {\bibinfo  {journal}
  {Phys. Rev. Lett.}\ }\textbf {\bibinfo {volume} {109}},\ \bibinfo {pages}
  {076801} (\bibinfo {year} {2012})}\BibitemShut {NoStop}%
\bibitem [{\citenamefont {Kuroda}\ \emph {et~al.}(2010)\citenamefont {Kuroda},
  \citenamefont {Arita}, \citenamefont {Miyamoto}, \citenamefont {Ye},
  \citenamefont {Jiang}, \citenamefont {Kimura}, \citenamefont {Krasovskii},
  \citenamefont {Chulkov}, \citenamefont {Iwasawa}, \citenamefont {Okuda},
  \citenamefont {Shimada}, \citenamefont {Ueda}, \citenamefont {Namatame},\
  and\ \citenamefont {Taniguchi}}]{KurodaHexWarp1sArpes}%
  \BibitemOpen
  \bibfield  {author} {\bibinfo {author} {\bibfnamefont {K.}~\bibnamefont
  {Kuroda}}, \bibinfo {author} {\bibfnamefont {M.}~\bibnamefont {Arita}},
  \bibinfo {author} {\bibfnamefont {K.}~\bibnamefont {Miyamoto}}, \bibinfo
  {author} {\bibfnamefont {M.}~\bibnamefont {Ye}}, \bibinfo {author}
  {\bibfnamefont {J.}~\bibnamefont {Jiang}}, \bibinfo {author} {\bibfnamefont
  {A.}~\bibnamefont {Kimura}}, \bibinfo {author} {\bibfnamefont {E.~E.}\
  \bibnamefont {Krasovskii}}, \bibinfo {author} {\bibfnamefont {E.~V.}\
  \bibnamefont {Chulkov}}, \bibinfo {author} {\bibfnamefont {H.}~\bibnamefont
  {Iwasawa}}, \bibinfo {author} {\bibfnamefont {T.}~\bibnamefont {Okuda}},
  \bibinfo {author} {\bibfnamefont {K.}~\bibnamefont {Shimada}}, \bibinfo
  {author} {\bibfnamefont {Y.}~\bibnamefont {Ueda}}, \bibinfo {author}
  {\bibfnamefont {H.}~\bibnamefont {Namatame}}, \ and\ \bibinfo {author}
  {\bibfnamefont {M.}~\bibnamefont {Taniguchi}},\ }\href {\doibase
  10.1103/physrevlett.105.076802} {\bibfield  {journal} {\bibinfo  {journal}
  {Phys. Rev. Lett.}\ }\textbf {\bibinfo {volume} {105}},\ \bibinfo {pages}
  {076802} (\bibinfo {year} {2010})}\BibitemShut {NoStop}%
\bibitem [{\citenamefont {Jiang}\ and\ \citenamefont {Wu}(2011)}]{JiangStripe}%
  \BibitemOpen
  \bibfield  {author} {\bibinfo {author} {\bibfnamefont {J.-H.}\ \bibnamefont
  {Jiang}}\ and\ \bibinfo {author} {\bibfnamefont {S.}~\bibnamefont {Wu}},\
  }\href {\doibase 10.1103/physrevb.83.205124} {\bibfield  {journal} {\bibinfo
  {journal} {Phys. Rev. B}\ }\textbf {\bibinfo {volume} {83}},\ \bibinfo
  {pages} {205124} (\bibinfo {year} {2011})}\BibitemShut {NoStop}%
\bibitem [{\citenamefont {Baum}\ and\ \citenamefont
  {Stern}(2012{\natexlab{a}})}]{BaumStern1}%
  \BibitemOpen
  \bibfield  {author} {\bibinfo {author} {\bibfnamefont {Y.}~\bibnamefont
  {Baum}}\ and\ \bibinfo {author} {\bibfnamefont {A.}~\bibnamefont {Stern}},\
  }\href {\doibase 10.1103/physrevb.85.121105} {\bibfield  {journal} {\bibinfo
  {journal} {Phys. Rev. B}\ }\textbf {\bibinfo {volume} {85}},\ \bibinfo
  {pages} {121105} (\bibinfo {year} {2012}{\natexlab{a}})}\BibitemShut
  {NoStop}%
\bibitem [{\citenamefont {Baum}\ and\ \citenamefont
  {Stern}(2012{\natexlab{b}})}]{BaumStern2}%
  \BibitemOpen
  \bibfield  {author} {\bibinfo {author} {\bibfnamefont {Y.}~\bibnamefont
  {Baum}}\ and\ \bibinfo {author} {\bibfnamefont {A.}~\bibnamefont {Stern}},\
  }\href {\doibase 10.1103/physrevb.86.195116} {\bibfield  {journal} {\bibinfo
  {journal} {Phys. Rev. B}\ }\textbf {\bibinfo {volume} {86}},\ \bibinfo
  {pages} {195116} (\bibinfo {year} {2012}{\natexlab{b}})}\BibitemShut
  {NoStop}%
\bibitem [{\citenamefont {Li}\ and\ \citenamefont
  {Carbotte}(2013{\natexlab{a}})}]{CarbotteCond1}%
  \BibitemOpen
  \bibfield  {author} {\bibinfo {author} {\bibfnamefont {Z.}~\bibnamefont
  {Li}}\ and\ \bibinfo {author} {\bibfnamefont {J.~P.}\ \bibnamefont
  {Carbotte}},\ }\href {\doibase 10.1103/physrevb.87.155416} {\bibfield
  {journal} {\bibinfo  {journal} {Phys. Rev. B}\ }\textbf {\bibinfo {volume}
  {87}},\ \bibinfo {pages} {155416} (\bibinfo {year}
  {2013}{\natexlab{a}})}\BibitemShut {NoStop}%
\bibitem [{\citenamefont {Li}\ and\ \citenamefont
  {Carbotte}(2013{\natexlab{b}})}]{CarbotteCond2}%
  \BibitemOpen
  \bibfield  {author} {\bibinfo {author} {\bibfnamefont {Z.}~\bibnamefont
  {Li}}\ and\ \bibinfo {author} {\bibfnamefont {J.~P.}\ \bibnamefont
  {Carbotte}},\ }\href {\doibase 10.1103/physrevb.88.045414} {\bibfield
  {journal} {\bibinfo  {journal} {Phys. Rev. B}\ }\textbf {\bibinfo {volume}
  {88}},\ \bibinfo {pages} {045414} (\bibinfo {year}
  {2013}{\natexlab{b}})}\BibitemShut {NoStop}%
\bibitem [{\citenamefont {Xiao}\ and\ \citenamefont {Wen}(2013)}]{OptCond}%
  \BibitemOpen
  \bibfield  {author} {\bibinfo {author} {\bibfnamefont {X.}~\bibnamefont
  {Xiao}}\ and\ \bibinfo {author} {\bibfnamefont {W.}~\bibnamefont {Wen}},\
  }\href {\doibase 10.1103/physrevb.88.045442} {\bibfield  {journal} {\bibinfo
  {journal} {Phys. Rev. B}\ }\textbf {\bibinfo {volume} {88}},\ \bibinfo
  {pages} {045442} (\bibinfo {year} {2013})}\BibitemShut {NoStop}%
\bibitem [{\citenamefont {Choy}\ \emph {et~al.}(2011)\citenamefont {Choy},
  \citenamefont {Edge}, \citenamefont {Akhmerov},\ and\ \citenamefont
  {Beenakker}}]{Choy}%
  \BibitemOpen
  \bibfield  {author} {\bibinfo {author} {\bibfnamefont {T.-P.}\ \bibnamefont
  {Choy}}, \bibinfo {author} {\bibfnamefont {J.~M.}\ \bibnamefont {Edge}},
  \bibinfo {author} {\bibfnamefont {A.~R.}\ \bibnamefont {Akhmerov}}, \ and\
  \bibinfo {author} {\bibfnamefont {C.~W.~J.}\ \bibnamefont {Beenakker}},\
  }\href {\doibase 10.1103/physrevb.84.195442} {\bibfield  {journal} {\bibinfo
  {journal} {Phys. Rev. B}\ }\textbf {\bibinfo {volume} {84}},\ \bibinfo
  {pages} {195442} (\bibinfo {year} {2011})}\BibitemShut {NoStop}%
\bibitem [{\citenamefont {Nakosai}\ \emph {et~al.}(2013)\citenamefont
  {Nakosai}, \citenamefont {Tanaka},\ and\ \citenamefont
  {Nagaosa}}]{Nakosai2013}%
  \BibitemOpen
  \bibfield  {author} {\bibinfo {author} {\bibfnamefont {S.}~\bibnamefont
  {Nakosai}}, \bibinfo {author} {\bibfnamefont {Y.}~\bibnamefont {Tanaka}}, \
  and\ \bibinfo {author} {\bibfnamefont {N.}~\bibnamefont {Nagaosa}},\ }\href
  {\doibase 10.1103/physrevb.88.180503} {\bibfield  {journal} {\bibinfo
  {journal} {Phys. Rev. B}\ }\textbf {\bibinfo {volume} {88}},\ \bibinfo
  {pages} {180503} (\bibinfo {year} {2013})}\BibitemShut {NoStop}%
\bibitem [{\citenamefont {P{\"o}yh{\"o}nen}\ \emph {et~al.}(2014)\citenamefont
  {P{\"o}yh{\"o}nen}, \citenamefont {Weststr{\"o}m}, \citenamefont
  {R{\"o}ntynen},\ and\ \citenamefont {Ojanen}}]{Ojanen}%
  \BibitemOpen
  \bibfield  {author} {\bibinfo {author} {\bibfnamefont {K.}~\bibnamefont
  {P{\"o}yh{\"o}nen}}, \bibinfo {author} {\bibfnamefont {A.}~\bibnamefont
  {Weststr{\"o}m}}, \bibinfo {author} {\bibfnamefont {J.}~\bibnamefont
  {R{\"o}ntynen}}, \ and\ \bibinfo {author} {\bibfnamefont {T.}~\bibnamefont
  {Ojanen}},\ }\href {\doibase 10.1103/physrevb.89.115109} {\bibfield
  {journal} {\bibinfo  {journal} {Phys. Rev. B}\ }\textbf {\bibinfo {volume}
  {89}},\ \bibinfo {pages} {115109} (\bibinfo {year} {2014})}\BibitemShut
  {NoStop}%
\bibitem [{\citenamefont {Sedlmayr}\ \emph {et~al.}(2015)\citenamefont
  {Sedlmayr}, \citenamefont {Aguiar-Hualde},\ and\ \citenamefont
  {Bena}}]{Bena}%
  \BibitemOpen
  \bibfield  {author} {\bibinfo {author} {\bibfnamefont {N.}~\bibnamefont
  {Sedlmayr}}, \bibinfo {author} {\bibfnamefont {J.~M.}\ \bibnamefont
  {Aguiar-Hualde}}, \ and\ \bibinfo {author} {\bibfnamefont {C.}~\bibnamefont
  {Bena}},\ }\href {\doibase 10.1103/PhysRevB.91.115415} {\bibfield  {journal}
  {\bibinfo  {journal} {Phys. Rev. B}\ }\textbf {\bibinfo {volume} {91}},\
  \bibinfo {pages} {115415} (\bibinfo {year} {2015})}\BibitemShut {NoStop}%
\bibitem [{\citenamefont {Kotetes}\ \emph {et~al.}(2013)\citenamefont
  {Kotetes}, \citenamefont {Sch{\"o}n},\ and\ \citenamefont {Shnirman}}]{KSG}%
  \BibitemOpen
  \bibfield  {author} {\bibinfo {author} {\bibfnamefont {P.}~\bibnamefont
  {Kotetes}}, \bibinfo {author} {\bibfnamefont {G.}~\bibnamefont {Sch{\"o}n}},
  \ and\ \bibinfo {author} {\bibfnamefont {A.}~\bibnamefont {Shnirman}},\
  }\href {\doibase 10.3938/jkps.62.1558} {\bibfield  {journal} {\bibinfo
  {journal} {J. Korean Phys. Soc.}\ }\textbf {\bibinfo {volume} {62}},\
  \bibinfo {pages} {1558} (\bibinfo {year} {2013})}\BibitemShut {NoStop}%
\bibitem [{\citenamefont {Jiang}\ \emph {et~al.}(2013)\citenamefont {Jiang},
  \citenamefont {Pekker}, \citenamefont {Alicea}, \citenamefont {Refael},
  \citenamefont {Oreg}, \citenamefont {Brataas},\ and\ \citenamefont {von
  Oppen}}]{Pekker}%
  \BibitemOpen
  \bibfield  {author} {\bibinfo {author} {\bibfnamefont {L.}~\bibnamefont
  {Jiang}}, \bibinfo {author} {\bibfnamefont {D.}~\bibnamefont {Pekker}},
  \bibinfo {author} {\bibfnamefont {J.}~\bibnamefont {Alicea}}, \bibinfo
  {author} {\bibfnamefont {G.}~\bibnamefont {Refael}}, \bibinfo {author}
  {\bibfnamefont {Y.}~\bibnamefont {Oreg}}, \bibinfo {author} {\bibfnamefont
  {A.}~\bibnamefont {Brataas}}, \ and\ \bibinfo {author} {\bibfnamefont
  {F.}~\bibnamefont {von Oppen}},\ }\href {\doibase 10.1103/physrevb.87.075438}
  {\bibfield  {journal} {\bibinfo  {journal} {Phys. Rev. B}\ }\textbf {\bibinfo
  {volume} {87}},\ \bibinfo {pages} {075438} (\bibinfo {year}
  {2013})}\BibitemShut {NoStop}%
\bibitem [{\citenamefont {Zhang}\ \emph {et~al.}(2010)\citenamefont {Zhang},
  \citenamefont {He}, \citenamefont {Chang}, \citenamefont {Song},
  \citenamefont {Wang}, \citenamefont {Chen}, \citenamefont {Jia},
  \citenamefont {Fang}, \citenamefont {Dai}, \citenamefont {Shan},
  \citenamefont {Shen}, \citenamefont {Niu}, \citenamefont {Qi}, \citenamefont
  {Zhang}, \citenamefont {Ma},\ and\ \citenamefont {Xue}}]{He}%
  \BibitemOpen
  \bibfield  {author} {\bibinfo {author} {\bibfnamefont {Y.}~\bibnamefont
  {Zhang}}, \bibinfo {author} {\bibfnamefont {K.}~\bibnamefont {He}}, \bibinfo
  {author} {\bibfnamefont {C.-Z.}\ \bibnamefont {Chang}}, \bibinfo {author}
  {\bibfnamefont {C.-L.}\ \bibnamefont {Song}}, \bibinfo {author}
  {\bibfnamefont {L.-L.}\ \bibnamefont {Wang}}, \bibinfo {author}
  {\bibfnamefont {X.}~\bibnamefont {Chen}}, \bibinfo {author} {\bibfnamefont
  {J.-F.}\ \bibnamefont {Jia}}, \bibinfo {author} {\bibfnamefont
  {Z.}~\bibnamefont {Fang}}, \bibinfo {author} {\bibfnamefont {X.}~\bibnamefont
  {Dai}}, \bibinfo {author} {\bibfnamefont {W.-Y.}\ \bibnamefont {Shan}},
  \bibinfo {author} {\bibfnamefont {S.-Q.}\ \bibnamefont {Shen}}, \bibinfo
  {author} {\bibfnamefont {Q.}~\bibnamefont {Niu}}, \bibinfo {author}
  {\bibfnamefont {X.-L.}\ \bibnamefont {Qi}}, \bibinfo {author} {\bibfnamefont
  {S.-C.}\ \bibnamefont {Zhang}}, \bibinfo {author} {\bibfnamefont {X.-C.}\
  \bibnamefont {Ma}}, \ and\ \bibinfo {author} {\bibfnamefont {Q.-K.}\
  \bibnamefont {Xue}},\ }\href {\doibase 10.1038/nphys1689} {\bibfield
  {journal} {\bibinfo  {journal} {Nat. Phys.}\ }\textbf {\bibinfo {volume}
  {6}},\ \bibinfo {pages} {584} (\bibinfo {year} {2010})}\BibitemShut {NoStop}%
\bibitem [{\citenamefont {Pan}\ \emph {et~al.}(2011)\citenamefont {Pan},
  \citenamefont {Vescovo}, \citenamefont {Fedorov}, \citenamefont {Gardner},
  \citenamefont {Lee}, \citenamefont {Chu}, \citenamefont {Gu},\ and\
  \citenamefont {Valla}}]{VallaFullSpinPol}%
  \BibitemOpen
  \bibfield  {author} {\bibinfo {author} {\bibfnamefont {Z.-H.}\ \bibnamefont
  {Pan}}, \bibinfo {author} {\bibfnamefont {E.}~\bibnamefont {Vescovo}},
  \bibinfo {author} {\bibfnamefont {A.~V.}\ \bibnamefont {Fedorov}}, \bibinfo
  {author} {\bibfnamefont {D.}~\bibnamefont {Gardner}}, \bibinfo {author}
  {\bibfnamefont {Y.~S.}\ \bibnamefont {Lee}}, \bibinfo {author} {\bibfnamefont
  {S.}~\bibnamefont {Chu}}, \bibinfo {author} {\bibfnamefont {G.~D.}\
  \bibnamefont {Gu}}, \ and\ \bibinfo {author} {\bibfnamefont {T.}~\bibnamefont
  {Valla}},\ }\href {\doibase 10.1103/physrevlett.106.257004} {\bibfield
  {journal} {\bibinfo  {journal} {Phys. Rev. Lett.}\ }\textbf {\bibinfo
  {volume} {106}},\ \bibinfo {pages} {257004} (\bibinfo {year}
  {2011})}\BibitemShut {NoStop}%
\bibitem [{\citenamefont {Hsieh}\ \emph
  {et~al.}(2009{\natexlab{b}})\citenamefont {Hsieh}, \citenamefont {Xia},
  \citenamefont {Qian}, \citenamefont {Wray}, \citenamefont {Dil},
  \citenamefont {Meier}, \citenamefont {Osterwalder}, \citenamefont {Patthey},
  \citenamefont {Checkelsky}, \citenamefont {Ong}, \citenamefont {Fedorov},
  \citenamefont {Lin}, \citenamefont {Bansil}, \citenamefont {Grauer},
  \citenamefont {Hor}, \citenamefont {Cava},\ and\ \citenamefont
  {Hasan}}]{HsiehTunableTI}%
  \BibitemOpen
  \bibfield  {author} {\bibinfo {author} {\bibfnamefont {D.}~\bibnamefont
  {Hsieh}}, \bibinfo {author} {\bibfnamefont {Y.}~\bibnamefont {Xia}}, \bibinfo
  {author} {\bibfnamefont {D.}~\bibnamefont {Qian}}, \bibinfo {author}
  {\bibfnamefont {L.}~\bibnamefont {Wray}}, \bibinfo {author} {\bibfnamefont
  {J.}~\bibnamefont {Dil}}, \bibinfo {author} {\bibfnamefont {F.}~\bibnamefont
  {Meier}}, \bibinfo {author} {\bibfnamefont {J.}~\bibnamefont {Osterwalder}},
  \bibinfo {author} {\bibfnamefont {L.}~\bibnamefont {Patthey}}, \bibinfo
  {author} {\bibfnamefont {J.}~\bibnamefont {Checkelsky}}, \bibinfo {author}
  {\bibfnamefont {N.}~\bibnamefont {Ong}}, \bibinfo {author} {\bibfnamefont
  {A.}~\bibnamefont {Fedorov}}, \bibinfo {author} {\bibfnamefont
  {H.}~\bibnamefont {Lin}}, \bibinfo {author} {\bibfnamefont {A.}~\bibnamefont
  {Bansil}}, \bibinfo {author} {\bibfnamefont {D.}~\bibnamefont {Grauer}},
  \bibinfo {author} {\bibfnamefont {Y.}~\bibnamefont {Hor}}, \bibinfo {author}
  {\bibfnamefont {R.}~\bibnamefont {Cava}}, \ and\ \bibinfo {author}
  {\bibfnamefont {M.}~\bibnamefont {Hasan}},\ }\href {\doibase
  10.1038/nature08234} {\bibfield  {journal} {\bibinfo  {journal} {Nature}\
  }\textbf {\bibinfo {volume} {460}},\ \bibinfo {pages} {1101} (\bibinfo {year}
  {2009}{\natexlab{b}})}\BibitemShut {NoStop}%
\bibitem [{\citenamefont {Gr{\"u}ner}(1994)}]{GruenerSpin}%
  \BibitemOpen
  \bibfield  {author} {\bibinfo {author} {\bibfnamefont {G.}~\bibnamefont
  {Gr{\"u}ner}},\ }\href {\doibase 10.1103/revmodphys.66.1} {\bibfield
  {journal} {\bibinfo  {journal} {Rev. Mod. Phys.}\ }\textbf {\bibinfo {volume}
  {66}},\ \bibinfo {pages} {1} (\bibinfo {year} {1994})}\BibitemShut {NoStop}%
\bibitem [{\citenamefont {Schulz}(1990)}]{Schulz}%
  \BibitemOpen
  \bibfield  {author} {\bibinfo {author} {\bibfnamefont {H.~J.}\ \bibnamefont
  {Schulz}},\ }\href {\doibase 10.1103/physrevlett.64.1445} {\bibfield
  {journal} {\bibinfo  {journal} {Phys. Rev. Lett.}\ }\textbf {\bibinfo
  {volume} {64}},\ \bibinfo {pages} {1445} (\bibinfo {year}
  {1990})}\BibitemShut {NoStop}%
\bibitem [{\citenamefont {Aperis}\ \emph {et~al.}(2010)\citenamefont {Aperis},
  \citenamefont {Varelogiannis},\ and\ \citenamefont {Littlewood}}]{Aperis}%
  \BibitemOpen
  \bibfield  {author} {\bibinfo {author} {\bibfnamefont {A.}~\bibnamefont
  {Aperis}}, \bibinfo {author} {\bibfnamefont {G.}~\bibnamefont
  {Varelogiannis}}, \ and\ \bibinfo {author} {\bibfnamefont {P.~B.}\
  \bibnamefont {Littlewood}},\ }\href {\doibase 10.1103/physrevlett.104.216403}
  {\bibfield  {journal} {\bibinfo  {journal} {Phys. Rev. Lett.}\ }\textbf
  {\bibinfo {volume} {104}},\ \bibinfo {pages} {216403} (\bibinfo {year}
  {2010})}\BibitemShut {NoStop}%
\bibitem [{\citenamefont {Zhang}\ \emph {et~al.}(2013)\citenamefont {Zhang},
  \citenamefont {Kane},\ and\ \citenamefont {Mele}}]{KaneMirror}%
  \BibitemOpen
  \bibfield  {author} {\bibinfo {author} {\bibfnamefont {F.}~\bibnamefont
  {Zhang}}, \bibinfo {author} {\bibfnamefont {C.~L.}\ \bibnamefont {Kane}}, \
  and\ \bibinfo {author} {\bibfnamefont {E.~J.}\ \bibnamefont {Mele}},\ }\href
  {\doibase 10.1103/physrevlett.111.056403} {\bibfield  {journal} {\bibinfo
  {journal} {Phys. Rev. Lett.}\ }\textbf {\bibinfo {volume} {111}},\ \bibinfo
  {pages} {056403} (\bibinfo {year} {2013})}\BibitemShut {NoStop}%
\bibitem [{\citenamefont {Chiu}\ \emph {et~al.}(2013)\citenamefont {Chiu},
  \citenamefont {Yao},\ and\ \citenamefont {Ryu}}]{Shinsei}%
  \BibitemOpen
  \bibfield  {author} {\bibinfo {author} {\bibfnamefont {C.-K.}\ \bibnamefont
  {Chiu}}, \bibinfo {author} {\bibfnamefont {H.}~\bibnamefont {Yao}}, \ and\
  \bibinfo {author} {\bibfnamefont {S.}~\bibnamefont {Ryu}},\ }\href {\doibase
  10.1103/physrevb.88.075142} {\bibfield  {journal} {\bibinfo  {journal} {Phys.
  Rev. B}\ }\textbf {\bibinfo {volume} {88}},\ \bibinfo {pages} {075142}
  (\bibinfo {year} {2013})}\BibitemShut {NoStop}%
\bibitem [{\citenamefont {Ueno}\ \emph {et~al.}(2013)\citenamefont {Ueno},
  \citenamefont {Yamakage}, \citenamefont {Tanaka},\ and\ \citenamefont
  {Sato}}]{SatoU1}%
  \BibitemOpen
  \bibfield  {author} {\bibinfo {author} {\bibfnamefont {Y.}~\bibnamefont
  {Ueno}}, \bibinfo {author} {\bibfnamefont {A.}~\bibnamefont {Yamakage}},
  \bibinfo {author} {\bibfnamefont {Y.}~\bibnamefont {Tanaka}}, \ and\ \bibinfo
  {author} {\bibfnamefont {M.}~\bibnamefont {Sato}},\ }\href {\doibase
  10.1103/physrevlett.111.087002} {\bibfield  {journal} {\bibinfo  {journal}
  {Phys. Rev. Lett.}\ }\textbf {\bibinfo {volume} {111}},\ \bibinfo {pages}
  {087002} (\bibinfo {year} {2013})}\BibitemShut {NoStop}%
\bibitem [{\citenamefont {Shiozaki}\ and\ \citenamefont {Sato}(2014)}]{SatoU2}%
  \BibitemOpen
  \bibfield  {author} {\bibinfo {author} {\bibfnamefont {K.}~\bibnamefont
  {Shiozaki}}\ and\ \bibinfo {author} {\bibfnamefont {M.}~\bibnamefont
  {Sato}},\ }\href {\doibase 10.1103/physrevb.90.165114} {\bibfield  {journal}
  {\bibinfo  {journal} {Phys. Rev. B}\ }\textbf {\bibinfo {volume} {90}},\
  \bibinfo {pages} {165114} (\bibinfo {year} {2014})}\BibitemShut {NoStop}%
\bibitem [{\citenamefont {Fang}\ \emph {et~al.}(2014)\citenamefont {Fang},
  \citenamefont {Bernevig},\ and\ \citenamefont {Gilbert}}]{TriDirac}%
  \BibitemOpen
  \bibfield  {author} {\bibinfo {author} {\bibfnamefont {C.}~\bibnamefont
  {Fang}}, \bibinfo {author} {\bibfnamefont {B.}~\bibnamefont {Bernevig}}, \
  and\ \bibinfo {author} {\bibfnamefont {M.}~\bibnamefont {Gilbert}},\
  }\href@noop {} {\bibfield  {journal} {\bibinfo  {journal} {ArXiv e-prints}\ }
  (\bibinfo {year} {2014})},\ \Eprint {http://arxiv.org/abs/1401.1823}
  {arXiv:1401.1823} \BibitemShut {NoStop}%
\bibitem [{\citenamefont {Romming}\ \emph {et~al.}(2013)\citenamefont
  {Romming}, \citenamefont {Hanneken}, \citenamefont {Menzel}, \citenamefont
  {Bickel}, \citenamefont {Wolter}, \citenamefont {von Bergmann}, \citenamefont
  {Kubetzka},\ and\ \citenamefont {Wiesendanger}}]{Wiesendanger}%
  \BibitemOpen
  \bibfield  {author} {\bibinfo {author} {\bibfnamefont {N.}~\bibnamefont
  {Romming}}, \bibinfo {author} {\bibfnamefont {C.}~\bibnamefont {Hanneken}},
  \bibinfo {author} {\bibfnamefont {M.}~\bibnamefont {Menzel}}, \bibinfo
  {author} {\bibfnamefont {J.~E.}\ \bibnamefont {Bickel}}, \bibinfo {author}
  {\bibfnamefont {B.}~\bibnamefont {Wolter}}, \bibinfo {author} {\bibfnamefont
  {K.}~\bibnamefont {von Bergmann}}, \bibinfo {author} {\bibfnamefont
  {A.}~\bibnamefont {Kubetzka}}, \ and\ \bibinfo {author} {\bibfnamefont
  {R.}~\bibnamefont {Wiesendanger}},\ }\href {\doibase 10.1126/science.1240573}
  {\bibfield  {journal} {\bibinfo  {journal} {Science}\ }\textbf {\bibinfo
  {volume} {341}},\ \bibinfo {pages} {636} (\bibinfo {year}
  {2013})}\BibitemShut {NoStop}%
\end{thebibliography}%

\appendix

\begin{widetext}

\section{Susceptibility for momentum transfer along high-symmetry lines}\label{appendix:susc-high-symmetry}

\begin{figure}[H]
\includegraphics{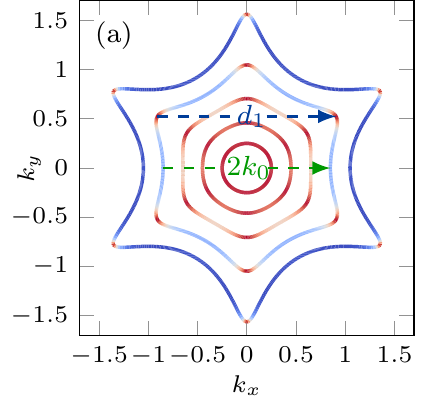}
\hspace*{3em}
\includegraphics{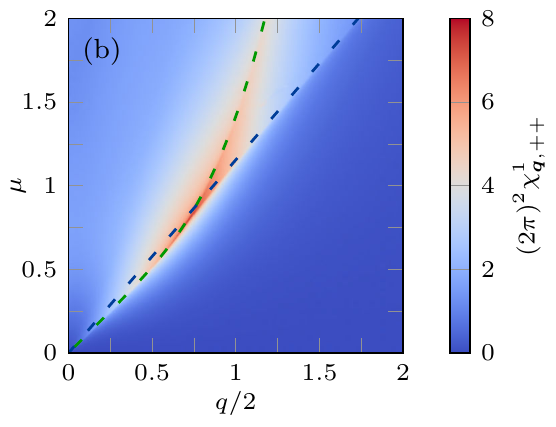}
\hspace*{1em}
\includegraphics{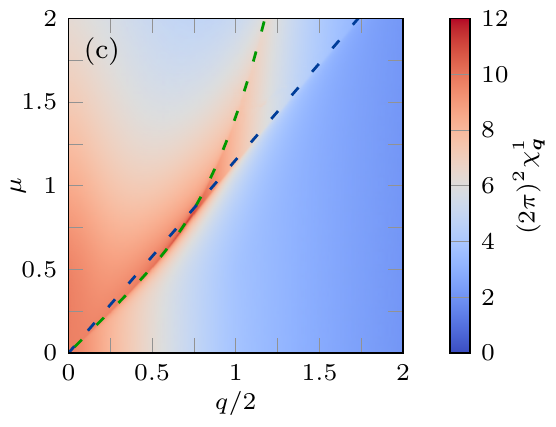}

\includegraphics{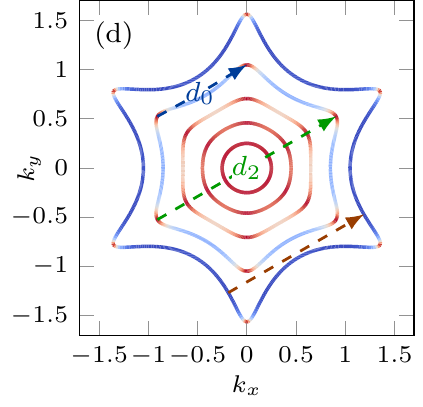}
\hspace*{3em}
\includegraphics{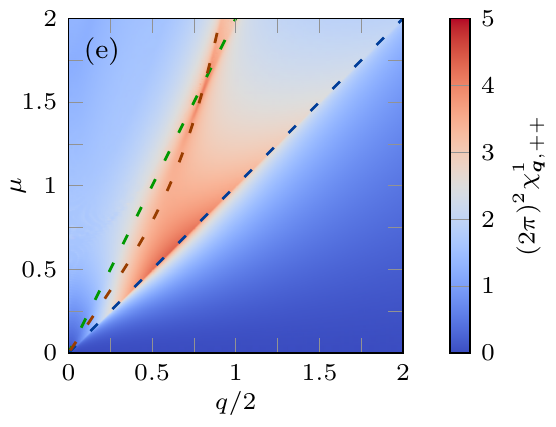}
\hspace*{1em}
\includegraphics{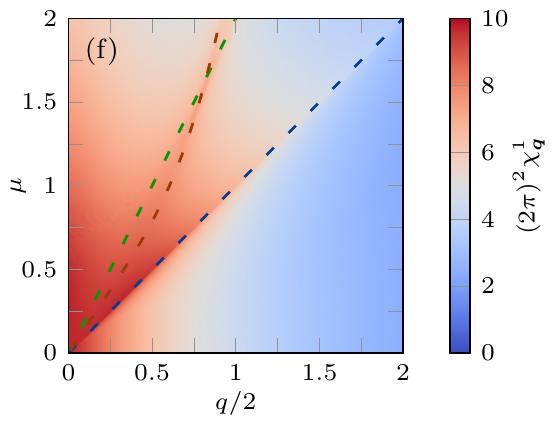}
\caption{Susceptibility for momentum transfers along high-symmetry lines. See the discussion below.}
\label{fig:susc-high-symmetry}

\vspace*{2em}

\includegraphics{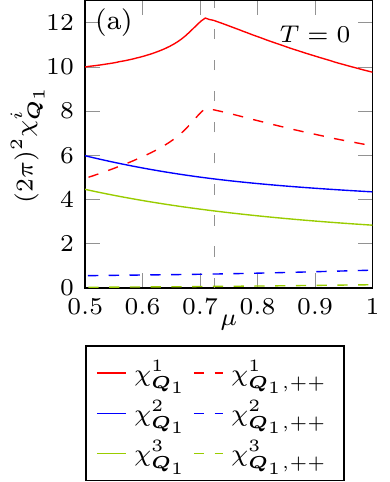}
\includegraphics{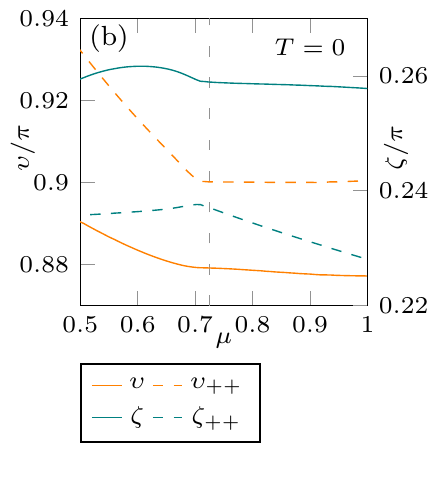}
\includegraphics{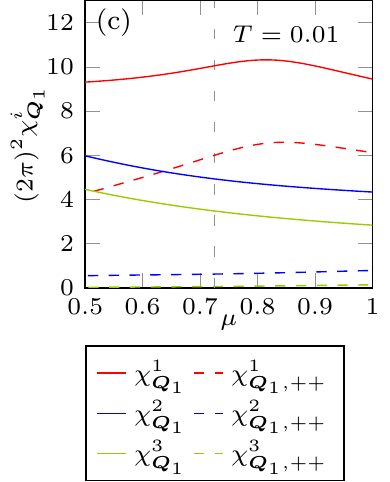}
\includegraphics{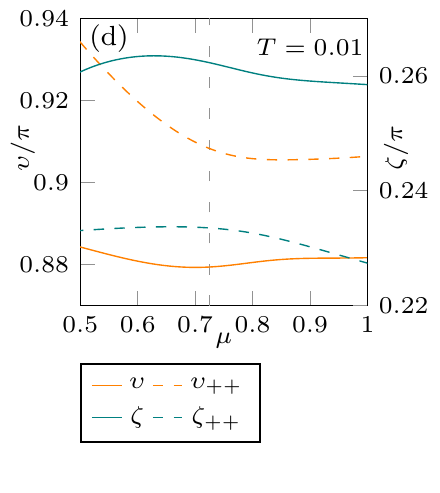}
\caption{Susceptibility eigenvalues and spherical angles of the magnetization direction. See the discussion below.}
\label{fig:landau}
\end{figure}

In this appendix we study the behavior of the susceptibility in a wider range of the chemical potential, thus covering the regimes from circular via hexagonal to
snowflake-like. We focus on momentum transfers along high-symmetry lines.

The distance between two nested sides of the Fermi surface, for a given $\mu$, is given by $2k_0$. Using \eqref{fermi} we find the distances $d_{0,1,2}$ between the corners of
the FS, shown in \figref{susc-high-symmetry,a,d}, to be given by $d_0=2\mu\sin(\pi/6)=\mu$, $d_1 = 2\mu \sin(\pi/3)=\sqrt{3} \mu$, and $d_2 = 2 k(\pi/6,\mu)=2 \mu$. We
investigate the spin susceptibility for $T=0$ as a function of $\mu$ and the mo\-du\-lus of two vectors $\vec{q}=q(1,0)$ and $\vec{q'}=q'(\sqrt{3}/2,1/2)$, both running along
high-symmetry lines in momentum space. The simultaneous scanning of these parameters allows us to determine the optimal relations, $\mu(q)$ and $\mu(q')$, that maximize the
spin susceptibility. In the plots of \figref{susc-high-symmetry,b,c,e,f} we observe different ridges where the susceptibility has a maximum maximal. These ridges can be
attributed to scattering between sides of the FS and corners with high DOS. As shown in \figref{susc-high-symmetry,b,c}, we find a dominant ridge in the susceptibility  along
$\mu = (q/2)\sqrt{1 +(q/2)^4}$ with $q=2k_0$, which is due to the nesting vector connecting the approximately flat parts of the FS. There is also a subdominant ridge in the
susceptibility, which goes along $\mu = q/\sqrt{3} = d_1/\sqrt{3}$ and arises due to the nesting of the high-DOS corners of the FS connected with the mirror symmetries
$(x,y)\mapsto(x,-y)$. In \figref{susc-high-symmetry,e,f} we find one linear ridge along $\mu = q' = d_0$, originating from nested neighbo\-ring corners of the FS. There is
another subleading linear ridge along $\mu=q'/2=d_2/2$ due to nesting between opposite corners connected by $(x,y)\mapsto (-x,-y)$. The arising nonlinear ridge is related to
the nesting of FS edges which bend outwards for higher values of $\mu$.

In \figref{landau,a,c} we present the three spin suscepti\-bi\-li\-ty eigenvalues for the most dominant ridge $\mu=k_0\sqrt{1+k_0^4}$ and the nesting vector
$\vec{Q}_1=2k_0(1,0)$. The plots of (a) and (c) correspond to $T=0$ and $T=0.01$, respectively. For \chem{Bi_2Te_3} the temperature value $T=0.01$ corresponds to $\SI{30}{K}$.
The solid curves show the contribution of all bands and the dashed lines correspond only to the upper helicity intraband contribution. Note that the lower helicity intraband
contribution is  negligible. We observe that the largest spin susceptibility eigenvalue is two or three times larger that the remaining two. The dashed vertical
line shows the case for the hexagonal FS with the chemical potential $\mu=\mu_{\text{hex}}$. In spite of the fact that the susceptibility decreases monotonically,
the magnetization direction is stable for higher temperatures as can be seen in \figref{landau,d}.

\section{Calculation of the Landau theory coefficients}\label{appendix:landau}

The free energy up to fourth order can be written as follows:
\begin{eqnarray}
\OpF^{(4)} = \alpha \sum_{i=1}^3 |M_i|^2 + \frac{\beta_1}{2} \sum_{i=1}^3 |M_i|^4 + \beta_2 \sum_{i<j} |M_i|^2 |M_j|^2\,.
\end{eqnarray}

\noindent We obtain for the above coefficients and $\eta$
\begin{alignat}{4}
  \alpha(U=0)  &= \left.\frac{\partial^2 \OpF}{\partial M\nohc_1 \partial M_1^*}\right|_{M_{1,2,3}=0} &=&\quad \Pi^{(2)}(\vec{Q_1},-\vec{Q_1})\,,\\
  2  \beta_1 &= \left.\frac{\partial^4 \OpF}{(\partial M\nohc_1 \partial M_1^*)^2}\right|_{M_{1,2,3}=0} &=&\quad \frac{1}{4} \sum_{\sigma \in S_4}
  \Pi^{(4)}\left(\sigma(\vec{Q}_1,-\vec{Q}_1,\vec{Q}_1,-\vec{Q}_1)\right)\,,\\
  \beta_2 &= \left.\frac{\partial^4 \OpF}{\partial M\nohc_1 \partial M_1^* \partial M\nohc_2 \partial M_2^*}\right|_{M_{1,2,3}=0} &=&\quad \frac{1}{4} \sum_{\sigma \in S_4}
  \Pi^{(4)}\left(\sigma(\vec{Q}_1,-\vec{Q}_1,\vec{Q}_2,-\vec{Q}_2)\right)\,,\\
  8 \eta  &= \left.\frac{\partial^6 \OpF}{(\partial M_1 \partial M_2 \partial M_3)^2}\right|_{M_{1,2,3}=0} &=&\quad \frac{1}{6} \sum_{\sigma \in S_6}
  \Pi^{(6)}\left(\sigma(\vec{Q}_1,\vec{Q}_2,\vec{Q}_3,\vec{Q}_1,\vec{Q}_2,\vec{Q}_3)\right)\,,
\end{alignat}

\noindent with the correlation functions
\begin{eqnarray}
  \Pi^{(2)}(\vec{q}_1,\vec{q}_2) &=& \sum_{s_1,s_2} \int \difk{k} S^{(2)}(\vec{\epsilon}) \tr\left\{ P_{s_1}(\vec{k}) \sigma_{\vec{q}_1} P_{s_2}(\vec{k}+\vec{q}_2)
    \sigma_{\vec{q}_2} \right\}\,,\\
  \Pi^{(4)}(\vec{q}_1,\ldots,\vec{q}_4) &=& \sum_{s_1,\ldots, s_4} \int \difk{k} S^{(4)}(\vec{\epsilon}) \tr\left\{ P_{s_1}(\vec{k}) \sigma_{\vec{q}_1}
    P_{s_2}(\vec{k}\!-\!\vec{q}_1) \sigma_{\vec{q}_2}  P_{s_3}(\vec{k}\!-\!\vec{q}_1\!-\!\vec{q}_2) \sigma_{\vec{q}_3}  P_{s_4}(\vec{k}\!+\!\vec{q}_4)
    \sigma_{\vec{q}_4} \right\}\,,\\
  \Pi^{(6)}(\vec{q}_1,\ldots,\vec{q}_6) &=& \sum_{s_1,\ldots, s_6} \int \difk{k} S^{(6)}(\vec{\epsilon}) \tr\Big\{ P_{s_1}(\vec{k}) \sigma_{\vec{q}_1}
  P_{s_2}(\vec{k}\!-\!\vec{q}_1) \sigma_{\vec{q}_2}  P_{s_3}(\vec{k}\!-\!\vec{q}_1\!-\!\vec{q}_2) \sigma_{\vec{q}_3} \nonumber\\
  &&\hspace*{40mm}P_{s_4}(\vec{k}\!+\!\vec{q}_4\!+\!\vec{q}_5\!+\!\vec{q}_6) \sigma_{\vec{q}_4} P_{s_5}(\vec{k}\!+\!\vec{q}_5\!+\!\vec{q}_6) \sigma_{\vec{q}_5}
  P_{s_6}(\vec{k}\!+\!\vec{q}_6) \sigma_{\vec{q}_6} \Big\}\,,
\end{eqnarray}

\noindent where $\sigma_{\vec{q}} \equiv \Uvec{M}_{\vec{q}} \cdot \vec{\sigma}$. We used the Matsubara sums $S^{(2,4,6)}(\vec{\epsilon})$ which are in general given by
\begin{eqnarray}
  S(m_1 \times \epsilon_1,\ldots,m_\mu \times \epsilon_\mu) = \frac{1}{\beta}\sum_{k_n} \prod_{\nu=1}^\mu\frac{1}{(\imath k_n-\epsilon_\nu)^{m_\nu}} = \sum_{\nu=1}^\mu
  \frac{1}{(m_\nu-1)!}\frac{\partial^{m_\nu-1}}{\partial \epsilon_\nu^{m_\nu-1}} \left[ n_F(\epsilon_\nu) \prod_{\rho\ne\nu} \frac{1}{(\epsilon_\nu-\epsilon_\rho)^{m_\rho}}
\right]\,,
\end{eqnarray}

\noindent where $\epsilon_i\ne\epsilon_j$ are pairwise different and $m_\nu$ denotes the multiplicities of the poles. In the case of single poles ($m_\nu=1$) we obtain
\begin{eqnarray}
  S(\epsilon_1,\ldots,\epsilon_\mu) = \sum_{\nu=1}^\mu n_F(\epsilon_\nu) \prod_{\rho\ne\nu} \frac{1}{\epsilon_\nu-\epsilon_\rho}\,.
\end{eqnarray}

For the second-order correlation function we find the susceptibility in the explicit form
\begin{eqnarray}
\chi^{ab}_{\vec{q},s,s'}&=&-\int \difk{k}
\frac{n_F(\epsilon_{\vec{k},s})-n_F(\epsilon_{\vec{k}+\vec{q},s'})}{\epsilon_{\vec{k},s}-\epsilon_{\vec{k}+\vec{q},s'}}
\braket{\vec{k},s|\sigma^a|\vec{k}+\vec{q},s'}\braket{\vec{k}+\vec{q},s'| \sigma^b|\vec{k},s}\nonumber\\
&=&-\frac{1}{2}\int \difk{k}
\frac{n_F(\epsilon_{\vec{k},s})-n_F(\epsilon_{\vec{k}+\vec{q},s'})}{\epsilon_{\vec{k},s}-\epsilon_{\vec{k}+\vec{q},s'}}
\bigg\{\delta^{ab}\big[1 - ss'\hg^m(\vec{k})\hg^m(\vec{k}+\vec{q})\big]
+\imath\varepsilon^{abm}\big[s\hg^m(\vec{k})-s'\hg^m(\vec{k}+\vec{q})\big]\nonumber\\
&&+ss'\big[\hg^a(\vec{k})\hg^b(\vec{k}+\vec{q})+\hg^a(\vec{k}+\vec{q})\hg^b(\vec{k})\big]\bigg\}\,.\quad
\end{eqnarray}

The relevant contributions to the susceptibility come from the upper helicity band. Thus we derive the upper helicity projection of \vec{\sigma} for the wave vector \vec{q}
\begin{eqnarray}
  &&\vec{\sigma}^+_{\vec{q}}(\vec{k})=\braket{\vec{k}+\vec{q}/2,+|\vec{\sigma}|\vec{k}-\vec{q}/2,+} =\\
  &&\bigg(
  \sin\frac{\vartheta_{\vec{k}+\vec{q}/2}}{2}\cos\frac{\vartheta_{\vec{k}-\vec{q}/2}}{2}
  \exp\left\{-\imath\frac{\varphi_{\vec{k}+\vec{q}/2}\!+\!\varphi_{\vec{k}-\vec{q}/2}}{2}\right\}+
  \cos\frac{\vartheta_{\vec{k}+\vec{q}/2}}{2}\sin\frac{\vartheta_{\vec{k}-\vec{q}/2}}{2}
  \exp\left\{\imath\frac{\varphi_{\vec{k}+\vec{q}/2}\!+\!\varphi_{\vec{k}-\vec{q}/2}}{2}\right\}
  \,,\nonumber\\
  &&\phantom{=\bigg(}
  \imath\sin\frac{\vartheta_{\vec{k}+\vec{q}/2}}{2}\cos\frac{\vartheta_{\vec{k}-\vec{q}/2}}{2}
  \exp\left\{-\imath\frac{\varphi_{\vec{k}+\vec{q}/2}\!+\!\varphi_{\vec{k}-\vec{q}/2}}{2}\right\}-\imath
  \cos\frac{\vartheta_{\vec{k}+\vec{q}/2}}{2}\sin\frac{\vartheta_{\vec{k}-\vec{q}/2}}{2}
  \exp\left\{\imath\frac{\varphi_{\vec{k}+\vec{q}/2}\!+\!\varphi_{\vec{k}-\vec{q}/2}}{2}\right\}
  \,,\nonumber\\
  &&\phantom{=\bigg(}
  \cos\frac{\vartheta_{\vec{k}+\vec{q}/2}}{2}\cos\frac{\vartheta_{\vec{k}-\vec{q}/2}}{2}
  \exp\left\{\imath\frac{\varphi_{\vec{k}+\vec{q}/2}\!-\!\varphi_{\vec{k}-\vec{q}/2}}{2}\right\}-
  \sin\frac{\vartheta_{\vec{k}+\vec{q}/2}}{2}\sin\frac{\vartheta_{\vec{k}-\vec{q}/2}}{2}
  \exp\left\{-\imath\frac{\varphi_{\vec{k}+\vec{q}/2}\!-\!\varphi_{\vec{k}-\vec{q}/2}}{2}\right\}
  \bigg)\,.\nonumber
\end{eqnarray}

\section{Symmetry classification of magnetic order parameters}\label{appendix:classification}

We perform a symmetry classification of the complex magnetic order parameters $\vec{M}_{\vec{q}}$ under the point group \group{C_{3v}}. The transformation behavior of the
axial vector $\vec{M}_{\vec{q}}$ under a group operation $\OpG \in \group{C_{3v}}$ is given by $\OpG \vec{M}_{\vec{q}} \equiv \HD_{\OpG}^-\vec{M}_{\OpG^{-1}\vec{q}}$. In the
case of a hexagonal FS and three nesting vectors $\vec{Q}_{1,2,3}$ we can classify linear combinations of the magnetic order parameters $\vec{M}\nohc_{\vec{Q}_\lambda} =
\vec{M}^*_{-\vec{Q}_\lambda} \equiv (M_{\lambda x},\,M_{\lambda y},\,M_{\lambda z})\trans$ ($\lambda=1,2,3$) which transform under different representations $A_1$, $A_2$ and
$E$ of the point group \group{C_{3v}}. At first we define the 18-dimensional basis in terms of the original order parameters
\begin{eqnarray}
  \vec{M}\trans &\equiv& \begin{pmatrix}
    \vec{M}_{\vec{Q}_1}\trans,&
    \vec{M}_{-\vec{Q}_1}\trans,&
    \vec{M}_{\vec{Q}_2}\trans,&
    \vec{M}_{-\vec{Q}_2}\trans,&
    \vec{M}_{\vec{Q}_3}\trans,&
    \vec{M}_{-\vec{Q}_3}\trans
  \end{pmatrix} \equiv
  \begin{pmatrix}
    M\nohc_{1x},&
    M\nohc_{1y},&
    M\nohc_{1z},&
    M^*_{1x},&
    M^*_{1y},&
    M^*_{1z},&
    \ldots
  \end{pmatrix}\,,\\
  \vec{M}\hc &=&\left[\II_3 \otimes \rho_x \otimes \II_3 \right] \cdot \vec{M}\trans =
  \begin{pmatrix}
    M^*_{1x},&
    M^*_{1y},&
    M^*_{1z},&
    M\nohc_{1x},&
    M\nohc_{1y},&
    M\nohc_{1z},&
    \ldots
  \end{pmatrix}\,,
\end{eqnarray}

\noindent where $\rho_x$ is a Pauli matrix acting in $\pm\vec{Q}_i$ space. For the symmetry classification we consider the action of the point group operations $\OpG \in
\group{C_{3v}}$ on the magnetic order parameters. We write down the representations $\HD_{\vec{M}}(\OpG)$ acting in the basis \vec{M},
\begin{alignat}{2}
  \HD_{\vec{M}}(C_3) &= \HD_\lambda(C_3) \otimes \II_2 \otimes \HD_{C_3}^-\quad\text{and}\quad
  \HD_{\vec{M}}(\sigma_v) &= \HD_\lambda(\sigma_v) \otimes \rho_x \otimes \HD_{\sigma_v}^-\,,
\end{alignat}

\noindent where the transformations in $\lambda$ space are given by
\begin{eqnarray}
  \HD_\lambda(C_3) =
  \begin{pmatrix}
    0 & 0 & 1\\
    1 & 0 & 0\\
    0 & 1 & 0\\
  \end{pmatrix}\quad\text{and}\quad
  \HD_\lambda(\sigma_v) =
  \begin{pmatrix}
    1 & 0 & 0\\
    0 & 0 & 1\\
    0 & 1 & 0\\
  \end{pmatrix}
\end{eqnarray}

\noindent since the rotation $C_3$ rotates the wave vector $\vec{Q}_1$ to $\vec{Q}_2$, etc., and the reflection at the $yz$ plane reflects $\vec{Q}_1 \to -\vec{Q}_1$ and
$\vec{Q}_{2,3} \to -\vec{Q}_{3,2}$. In order to make a connection to the Hamiltonian formalism, we introduce the real order parameter basis
\begin{eqnarray}
  \begin{pmatrix}
    \vec{M}^R_{\vec{q}}\\
    \vspace*{-3mm}\\
    \vec{M}^I_{\vec{q}}
  \end{pmatrix} \equiv
  \frac{1}{\sqrt{2}}\begin{pmatrix}
    1 & 1\\
    -i & i\\
  \end{pmatrix} \cdot
  \begin{pmatrix}
    \vec{M}_{\vec{q}}\\
    \vec{M}_{-\vec{q}}
  \end{pmatrix}\,.
\end{eqnarray}

Linear combinations of the above order parameters transform according to respective representations of the point group. The two-dimensional $E$ representations
are formed by $(\OpM_{x,i},\OpM_{y,i})$. The representations $i=1,2,3$ are chosen is such a way that they transform like a $(k_x,k_y)$ vector. The representations $i=4,5,6$
are chosen to transform like $(-k_y,k_x)$.
\begin{align}
  \mathcal{M}_{A_1,1}&=\nicefrac{1}{\sqrt{12}}\left(-2 M_{1x}^R+M_{2x}^R+M_{3x}^R-\sqrt{3} M_{2y}^R+\sqrt{3} M_{3y}^R\right)\,,\\
  \mathcal{M}_{A_1,2}&=\nicefrac{1}{\sqrt{12}}\left(-\sqrt{3} M_{2x}^I+\sqrt{3} M_{3x}^I+2 M_{1y}^I-M_{2y}^I-M_{3y}^I\right)\,,\\
  \mathcal{M}_{A_1,3}&=\nicefrac{1}{\sqrt{3}}\left(M_{1z}^I+M_{2z}^I+M_{3z}^I\right)\,,\\
  \mathcal{M}_{A_2,1}&=\nicefrac{1}{\sqrt{12}}\left(2 M_{1x}^I-M_{2x}^I-M_{3x}^I+\sqrt{3} M_{2y}^I-\sqrt{3} M_{3y}^I\right)\,,\\
  \mathcal{M}_{A_2,2}&=\nicefrac{1}{\sqrt{12}}\left(-\sqrt{3} M_{2x}^R+\sqrt{3} M_{3x}^R+2 M_{1y}^R-M_{2y}^R-M_{3y}^R\right)\,,\\
  \mathcal{M}_{A_2,3}&=\nicefrac{1}{\sqrt{3}}\left(M_{1z}^R+M_{2z}^R+M_{3z}^R\right)\,,\\
  \mathcal{M}_{x,1}&=\nicefrac{1}{\sqrt{24}}\left(-\sqrt{3} M_{2x}^R+\sqrt{3} M_{3x}^R+3 M_{2y}^R+3 M_{3y}^R\right)\,,\\
  \mathcal{M}_{y,1}&=\nicefrac{1}{\sqrt{24}}\left(-4 M_{1x}^R-M_{2x}^R-M_{3x}^R+\sqrt{3} M_{2y}^R-\sqrt{3} M_{3y}^R\right)\,,\\
  \mathcal{M}_{x,2}&=\nicefrac{1}{\sqrt{24}}\left(3 M_{2x}^I+3 M_{3x}^I+\sqrt{3} M_{2y}^I-\sqrt{3} M_{3y}^I\right)\,,\\
  \mathcal{M}_{y,2}&=\nicefrac{1}{\sqrt{24}}\left(\sqrt{3} M_{2x}^I-\sqrt{3} M_{3x}^I+4 M_{1y}^I+M_{2y}^I+M_{3y}^I\right)\,,\\
  \mathcal{M}_{x,3}&=\nicefrac{1}{\sqrt{2}}\left(M_{3z}^I-M_{2z}^I\right)\,,\\
  \mathcal{M}_{y,3}&=\nicefrac{1}{\sqrt{6}}\left(2 M_{1z}^I-M_{2z}^I-M_{3z}^I\right)\,,\\
  \mathcal{M}_{x,4}&=\nicefrac{1}{\sqrt{24}}\left(\sqrt{3} M_{2x}^I-\sqrt{3} M_{3x}^I-3 M_{2y}^I-3 M_{3y}^I\right)\,,\\
  \mathcal{M}_{y,4}&=\nicefrac{1}{\sqrt{24}}\left(4 M_{1x}^I+M_{2x}^I+M_{3x}^I-\sqrt{3} M_{2y}^I+\sqrt{3} M_{3y}^I\right)\,,\\
  \mathcal{M}_{x,5}&=\nicefrac{1}{\sqrt{24}}\left(3 M_{2x}^R+3 M_{3x}^R+\sqrt{3} M_{2y}^R-\sqrt{3} M_{3y}^R\right)\,,\\
  \mathcal{M}_{y,5}&=\nicefrac{1}{\sqrt{24}}\left(\sqrt{3} M_{2x}^R-\sqrt{3} M_{3x}^R+4 M_{1y}^R+M_{2y}^R+M_{3y}^R\right)\,,\\
  \mathcal{M}_{x,6}&=\nicefrac{1}{\sqrt{2}}\left(M_{3z}^R-M_{2z}^R\right)\,,\\
  \mathcal{M}_{y,6}&=\nicefrac{1}{\sqrt{6}}\left(2 M_{1z}^R-M_{2z}^R-M_{3z}^R\right)\,.
\end{align}

\noindent We point out that the order parameters consist purely of the real (imaginary) parts of the respective original order parameters.

\end{widetext}

\end{document}